\documentclass[12pt]{iopart}

\pdfoutput=1

\usepackage{amsfonts}
\usepackage{amssymb}
\usepackage{graphicx}
\usepackage{mathptmx}
\usepackage{url}

\newcommand{\vect}[1]{\mathbf{#1}}
\newcommand{\qq}{\vect{q}}
\newcommand{\pp}{\vect{p}}
\newcommand{\cc}{\vect{c}}
\newcommand{\nn}{\vect{n}}

\newcommand{\RR}{\mathbb{R}}

\newcommand{\defeq}{\equiv}

\begin{document}

\paper[Higher-dimensional convex billiards with cylindrical shape]
  {Stable and unstable regimes in higher-dimensional convex billiards with
  cylindrical shape}   
\author{Thomas Gilbert\textdagger, David P.~Sanders\textdaggerdbl}
\address{
  \textdagger Center for Nonlinear Phenomena and Complex Systems,
  Universit\'e Libre  de Bruxelles, C.~P.~231, Campus Plaine, B-1050
  Brussels, Belgium}
\address{\textdaggerdbl Departamento de F\'isica, Facultad de Ciencias, Universidad
  Nacional Aut\'onoma de M\'exico,  Ciudad Universitaria, Delegaci\'on Coyoac\'an,
  04510 M\'exico D.F., Mexico}
\ead{thomas.gilbert@ulb.ac.be, dps@fciencias.unam.mx}

\begin{abstract}
  We introduce a class of convex, higher-dimensional billiard models which
  generalise stadium billiards. These models correspond to the free motion
  of a point-particle in a region bounded by cylinders cut by
  planes. They are motivated by models of particles interacting via
  a string-type mechanism, and confined by hard walls. The combination of
  these elements may give rise to a defocusing mechanism, similar to that
  in two dimensions, which allows large chaotic regions in phase space. The
  remaining part of phase space is associated with marginally stable
  behaviour. In fact periodic orbits in these systems generically come in
  continuous parametric families, associated with a pair of parabolic
  eigen-directions: the periodic orbits are unstable in the presence of a
  defocusing mechanism, but marginally stable otherwise. By
  performing the stability analysis of families of periodic orbits at a
  nonlinear level, we establish the conditions under which 
    families are nonlinearly stable or unstable. As a result, we
  identify regions in the parameter space of the models which admit
  non-linearly stable oscillations in the form of whispering gallery
  modes. Where no families of periodic orbits are stable, the billiards are
  completely chaotic, i.e.\ the Lyapunov exponents of the billiard map are
  non-zero.
\end{abstract}

\submitto{\NJP}
\maketitle

\section{Introduction \label{sec.intro}}

Billiard models, in which a point particle moves uniformly until it
undergoes abrupt elastic collisions with a fixed boundary, are the
playground of statistical physicists and 
mathematicians alike, whose work focuses on the interplay between their
dynamical and statistical properties \cite{Katok:2005p890}. 

There are two main categories of \emph{chaotic} billiards. The
better-known ones are (semi-)\emph{dispersing} billiards, of which the hard-sphere
gas is the prototypical example. The motion of hard spheres in a bounded
region, undergoing  elastic collisions with each other, is equivalent to
that of a billiard model -- a point particle which moves uniformly in the
exterior of a collection of spherical cylinders in a high-dimensional phase
space, with specular collisions at the boundary \cite{Szasz:1993p996}. The
Sinai billiard \cite{Sinai:1970p970} and the periodic Lorentz gas
\cite{Bunimovich:1981p479} are two-dimensional examples of dispersing
billiards, in which a particle moves outside a disk on the $2$-torus or a
periodic configuration of them on the plane. The latter is a useful model
of tagged particle diffusion in a binary mixture \cite{Gaspard:2003p298};
it enjoys fast decay of correlations and thus converges to a Brownian
motion \cite{Bunimovich:1991p47}. 
The mechanism giving rise to chaos in
these billiards, as well as in hard sphere gases, is that of dispersion,
where nearby trajectories separate exponentially fast as a function of the
number of collisions with surfaces of positive curvature, implying that the
system has a positive Lyapunov exponent \cite{Sinai:1991p978} and is thus
chaotic. Indeed, hard-sphere gases are essentially the only systems of
interacting particles for which rigorous results, ranging from
hyperbolicity to exponential decay of correlations, have been firmly
established \cite{Simanyi:1999p796, Szasz:2000book, Szasz:2008p35}. 

The second category is that of \emph{defocusing} billiards, the paradigm of
which is the Bunimovich stadium \cite{Bunimovich:1974p930,
  Bunimovich:1979p105}. Here, chaos is due to a mechanism different from
dispersion, namely the defocusing mechanism. As opposed to the Sinai
billiard, the boundary of the stadium curves inwards with respect to the
particle motion, i.e.\ it is \emph{convex}. 
Although nearby trajectories
initially focus after a collision with this boundary, if the distance to 
the next collision is longer than the distance to the focal
point then they eventually defocus even more. This mechanism thus leads to
an overall expansion in phase space, again measured by 
a positive Lyapunov exponent \cite{Chernov:2006p683}.

Since its discovery, the mechanism of defocusing has attracted much
attention in the physics community, particularly in connection
to quantum chaos \cite{Stockmann:1999p979}, acoustic experiments in closed
chaotic cavities \cite{Draeger:1997p917}, optical microcavity laser
experiments \cite{Friedman:2001p900, Harayama:2003p901}, or quantum
conductance experiments \cite{Marcus:1992p977} to name but a few. 

In 
spite of its potential appeal to a broad range of physical applications, it
has, however, remained a difficult problem to establish the conditions under
which the defocusing mechanism can be extended to higher dimensions
\cite{BunManyDimStadium, Wojtkowski:1990p931}. There are still only 
a few models of higher-dimensional stadia based on three- and 
higher-dimensional cavities which have lent themselves to a systematic
study and are known to be fully chaotic \cite{Bunimovich:1997p729,
  Bunimovich:1998p421,  Bunimovich:1998p164,Bunimovich:2006p213,
  Bunimovich:2008p1377}. 

There are arguably two difficulties which must be dealt with in order to
obtain chaotic behaviour: (i) two curved dimensions, as in a spherical cap,
often generate stable oscillations; and (ii) three-dimensional billiard
domains with a single curved   dimension, such as a cylindrical
surface, have flat components which complicate the stability analysis.

The first of these two difficulties is related to the fact that one cannot
simply construct a three-dimensional surface of revolution by rotating, for
example, a two-dimensional stadium to obtain a chaotic three-dimensional 
cavity, since angular momentum is then conserved. Furthermore, as shown by
Wojtkowski \cite{Wojtkowski:1990p931}, 
intersecting three-dimensional spherical caps and flat components may
produce stable 
periodic orbits, giving rise to billiards with mixed phase space. The models
studied by Bunimovich and Rehacek \cite{Bunimovich:1997p729,
  Bunimovich:1998p421, Bunimovich:1998p164} provide an exception to this
principle and, in this sense, are rather special. They are furthermore
non-convex.

A natural alternative to spherical caps is to use boundary components
made out of cylindrical caps and flat planes. It is clear that
``extruding'' or extending 
the two-dimensional stadium to a three-dimensional cylindrical-type stadium 
(a cylindrical shape whose cross section is a stadium) by a perpendicular
translation does not suffice to 
produce a fully chaotic billiard in three dimensions, since the motion in
this perpendicular direction is trivial. 
Thus Papenbrock \cite{Papenbrock:2000p293} proposed
to consider a three-dimensional  stadium with half-cylindrical caps along
perpendicular axes at both ends of a cuboidal shape.  This construction 
gives rise to a billiard which is fully chaotic, as was recently shown
rigorously  
by Bunimovich and Del Magno \cite{Bunimovich:2006p213, 
  Bunimovich:2008p1377}, and is a particular case of more general
constructions discussed by 
Wojtkowski \cite{Wojtkowski:2007p195}. In this respect, it is the first
known example of 
a three-dimensional billiard in a convex domain with this property. 

The specificity of the Papenbrock three-dimensional stadium is that its
cylindrical caps are placed opposite each other in such a way that periodic
orbits visiting the caps are isolated. That is, even though each
cylindrical component is curved along a single direction, , there is no
degree of freedom with which  
to move the periodic orbits around, since the axes of the two cylinders are
mutually transverse. 

The Papenbrock stadium thus avoids the second of the two difficulties we
alluded to above, which are typical of three-dimensional cavities
constructed out of cylindrical components and  flat planes: namely, that
their periodic orbits generically belong to  \emph{continuous parametric
  families}. Take, for example, the three-dimensional 
cylindrical-type stadium discussed above and break its translational
symmetry by cutting it with an oblique plane. A periodic orbit of this
system can be shifted  along the direction of the cylinder axis,
remaining basically unchanged,  thus giving rise to a one-parameter
family. This has the consequence that its stability analysis yields, out of
the four phase-space dimensions of the billiard map, two parabolic 
directions (with unit eigenvalue), corresponding to motion along the family.

Another example of a three-dimensional cavity whose
periodic orbits belong to continuous parametric families is the
three-dimensional billiard whose domain consists of the intersection of a
sphere with a cuboid \cite{Bunimovich:1996p302}. Though this billiard was
numerically found to have four non-zero Lyapunov exponents in a large
interval of its parameter, this property is yet unproven as it does
not lend itself to the kind of analysis performed in
refs~\cite{Bunimovich:1997p729, Bunimovich:1998p421,
  Bunimovich:1998p164,Bunimovich:2006p213, Bunimovich:2008p1377}.  

It is the goal of this paper to consider billiards whose phase space
combines neutral directions with curved regions, associated with
stable and unstable behaviors, and explore the conditions under which they
can display hyperbolic regimes, i.e.\ such that all pairs of opposite
Lyapunov exponents are non-zero. To this end, we introduce a class of
higher-dimensional convex 
billiards, in which the particle motion occurs inside cylindrical 
domains bounded by oblique planes; we call these \emph{cylindrical stadium
  billiards}. The reasons for using the term stadium are
manifold. As we shall show, these billiards share some of the essential
properties of the two-dimensional stadium billiard. In particular, and
foremost, the mechanism 
that drives instability is of the defocusing type; other common
features of specific interest to us, and which will be discussed below,
are the existence of classes of periodic orbits similar to the 
bouncing-ball and whispering-gallery modes of the stadium billiard
\cite{Tanner:1997p950}.

The motivation for such a general class of models arises from models of
particles which interact via 
virtual strings, as introduced
by Papenbrock to model self-bound nuclei \cite{Papenbrock:2000p765,
  Papenbrock:2000p166} and which were subsequently modified by Gilbert and
Lefevere to model heat conduction by gas particles
trapped in a nano-porous solid matrix  \cite{Gilbert:2008p354}. The
interaction between pairs of neighbouring particles is a sort of hard-core
interaction at a distance, and works as follows: particles move freely
until they reach a critical 
relative distance, at which point they interchange the longitudinal
components of their momenta. This can be thought of as an interaction
mediated by a string, which affects the particles only once the string
becomes fully extended, at which point it applies an instantaneous impulse
to each particle so they flip directions. A chain of particles with such
interactions were considered in \cite{Gilbert:2008p354}, but placed inside
a rectangular domain divided up into square cells, with each particle
confined to its own cell. As their study shows, the combination of specular
reflections along flat walls and circular arcs gives rise to chaotic motion
in this system. Thus, given a system of $N\geq2$ particles, they find
$2N - 1$ pairs of non-zero Lyapunov exponents opposite each other (the
single pair of vanishing exponents corresponds to the energy conservation
and associated time translation symmetry).

As we shall shortly describe further, models of particles with such a
string-type interaction, and confined by hard walls, are equivalent to
billiards in  higher-dimensional configuration spaces inside convex  
regions bounded by cylinders and planes. This motivates the introduction of
a general class of convex billiards bounded by cylinders and planes, which
includes such models. Note that this is the opposite of hard-sphere gases,
which are equivalent to billiards outside cylinders.

The presence of a flat direction along the axes of the cylinders in such
systems implies that periodic orbits come in continuous parametric
families, with each periodic orbit displaying marginal stability along
the direction of the family (each dimension associated with a flat
direction actually gives rise to a pair of parabolic eigenvalues along the
corresponding direction and conjugate momentum). Although the existence of
such a family was noted in ref.~\cite{Papenbrock:2000p196}, its detailed
analysis is notably absent in the models previously considered.  
Indeed, the stability of such families of periodic orbits must be studied
 at a nonlinear level in order to determine whether or
not part of the family can or cannot be stabilised -- a daunting task.

As a first step towards a comprehensive understanding of the dynamics of
such systems, we present in this paper a detailed analysis of the dynamical
regimes occurring in one of the simplest such systems, which
consists of a cylinder in three dimensions with a flat bottom and cut by an
oblique plane at its top. 
This analysis will be performed in terms of the stability of periodic
orbits, and comes in two parts. At a linear level, stability is
understood in terms of the lengths of the  
segments of the orbit, and, specifically, that of paths corresponding to
two collisions with the cylindrical surface, separated by a single
collision with the oblique plane. As we shall show, when this length is large
enough, defocusing takes place and the periodic orbit is
unstable. If on the contrary, this length remains small, then the orbit may
remain stable. However, because such periodic orbits have a pair of parabolic 
eigenvalues, the effect of \emph{nonlinear}
perturbations along the neutral directions must be taken into account,
which is the second part of the stability analysis.

There are essentially two classes of periodic orbits of cylindrical stadium
billiards whose segments' lengths can remain small enough that they are
potentially stable. We identify these two classes as \emph{planar} and
\emph{helical} periodic orbits. The former are similar to bouncing-ball
orbits in the stadium billiard, and the latter to whispering-gallery
modes. By analysing in detail the existence conditions and linear and
nonlinear stability properties of these two classes of orbits, we show, at
the linear level, the existence of a bifurcation from marginally-stable to
unstable (hyperbolic) regimes, and, at a nonlinear level, the existence of a
``restoring force'' which, in specific regions of the parameter space, is
able to stabilise the motion by restricting it to a certain part of the  
parameter space where the family of periodic orbits remains marginally
stable, thus preventing the periodic orbit from escaping to a hyperbolic
region through the bifurcation point.  The parameter regions where none 
of the periodic orbit families can be stabilised correspond to chaotic
regimes of the system.

The paper is organised as follows. In \sref{sec.ipart}, we provide a
detailed description of particles interacting through a string-type
mechanism in terms of cylindrical billiard models. In \sref{sec.Not} we
introduce the particular three-dimensional cylindrical 
stadium billiard described above, and we offer a general characterisation of
its periodic orbits in \sref{sec.PO}. The two classes of planar and
helical periodic orbits are studied in further details in sections
\ref{sec.PPO} and \ref{sec.HPO}, respectively, where we identify parametric
regions of non-linear stability of these periodic orbits. A summary and
discussion of our results are presented in \sref{sec.Con}. Two appendices,
\ref{app.PPO} and \ref{app.HPO}, provide exact forms obtained for the
linear and non-linear analysis of some of the periodic orbits studied in
sections~\ref{sec.PPO} and \ref{sec.HPO}.

\section{Interacting-particle models and cylindrical stadium billiards}
\label{sec.ipart}

We start by relating interacting-particle models with string-type
interactions and billiard models, and then introduce the general class of
cylindrical stadium billiards. 

\subsection{Particles with string-type interactions and flat walls}

Consider $N \ge 2$ particles 
interacting via a string-type interaction \cite{Papenbrock:2000p765,
Papenbrock:2000p166, Gilbert:2008p354}.
The particles are joined by strings in some configuration, and 
have the following dynamics. Two particles which are joined by a string
move freely until they reach a mutual distance $r_{ij} = \ell$,
at which point the string becomes taut (fully extended), and exerts an
instantaneous impulse on each particle, which interchanges the components of
their momenta along the line between them. The interaction potential
between the particles due to the string, as a function of the distance $r$,
is thus  $v(r) = 0$ for $r < \ell$ and $v(r)=\infty$ for $r>\ell$. 

The dynamics of the $N$-particle system corresponds to the motion of a
point  $\Gamma \defeq (\qq, \pp) \defeq (\qq_1, \ldots, \qq_N, \pp_1, \dots,
\pp_N)$
in a  $2dN$-dimensional phase space, where $d$ is the spatial dimension of the
system, so that $\qq_i \in \RR^d$ and $\pp_i \in \RR^d$.

If particles $i$ and $j$ are joined by a string of length $\ell$, then they
are  constrained 
by the
inequality
\begin{equation}
 \| \qq_i - \qq_j \|^2 \leq \ell^2,
\end{equation}
which specifies a hyper-spherical region $S \in \RR^{2d}$ in the dimensions
spanned by the coordinates $\qq_i$ and $\qq_j$.
The other coordinates are free, so that 
this corresponds to an allowed cylindrical region $S \times \RR^{d(N-2)}$ in
configuration space.
This is analogous to the case of hard-sphere interactions
\cite{Szasz:1993p996}, for which the constraint 
has the opposite inequality, $ \| \qq_i - \qq_j \|^2 \geq \ell^2$. 

\begin{figure}[tbh]
  \centering 
  \includegraphics[width=0.5\textwidth]{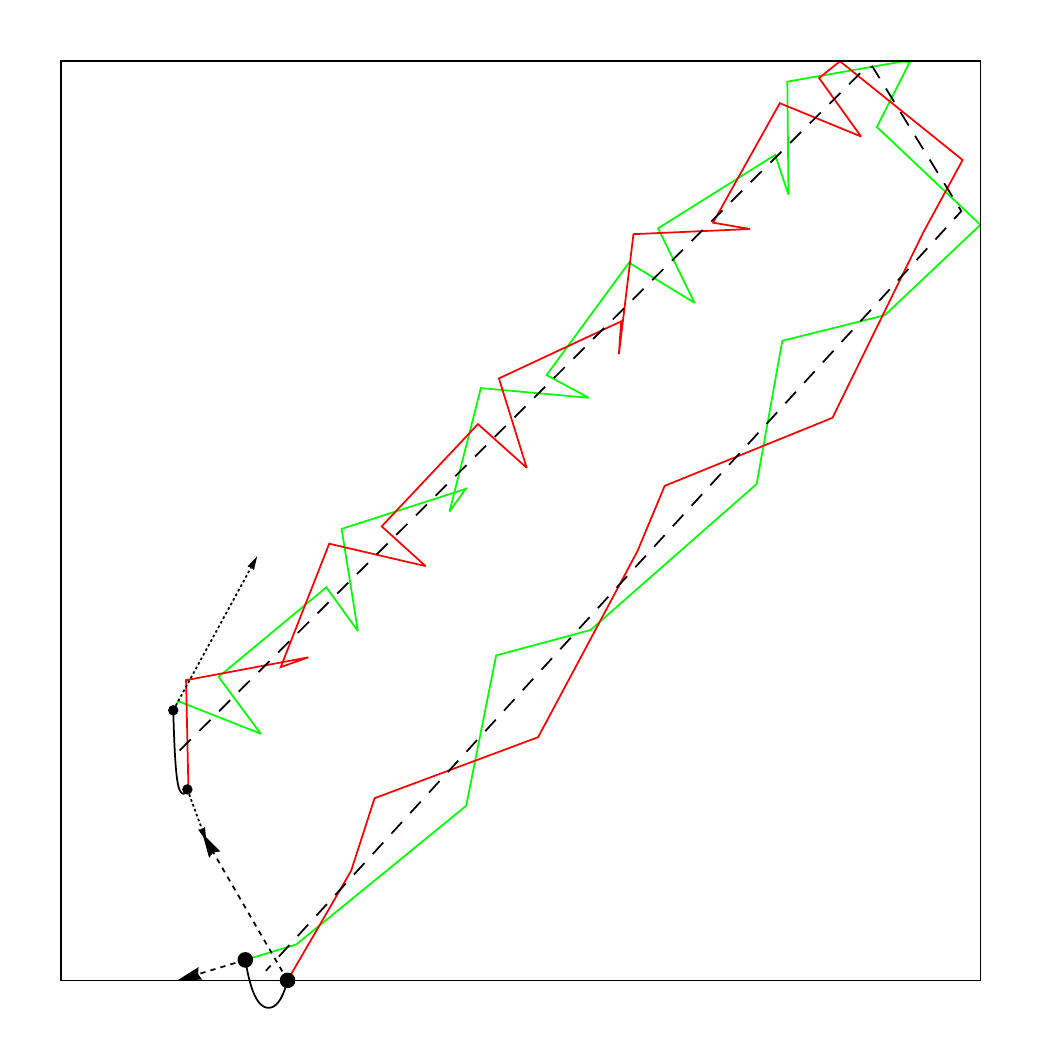}
  \caption{Two interacting particles joined by a string and trapped in a
    square box. Their relative positions are constrained to a disk of
    radius $\ell$, while each particle is reflected elastically by the
    horizontal and vertical walls. The two continuous lines show the
    trajectories of each particle as they interact and collide with the
    walls. The motion of the center of mass (dashed line) is uniform in
    between collisions of either particle with the walls. The dotted arrows
  indicate the directions of the velocity vectors at the initial and final
  positions.}
  \label{fig.2partbox}
\end{figure}
The particles may also be confined by hard walls such as shown in
\fref{fig.2partbox}, where two particles joined by a string are trapped
to a square box.  Each wall imposes 
a further constraint on the position $\qq_i$ of each particle, namely $(\qq_i -
\cc) \cdot \nn \le 0$, where
$\cc$ is a point on the wall and $\nn$ the outgoing normal vector.  Each wall
thus imposes $N$ constraints on the 
total configuration-space position vector $\qq$, which are of the form $(\qq -
\tilde{\cc}) \cdot \tilde{\nn} \le 0$, where
$\tilde{\nn} \defeq (\vect{0}, \ldots, \nn, \ldots, \vect{0})$, with $\nn$
appearing in the $i$th position, and similarly for
 $\tilde{\cc}$; these correspond to half-spaces bounded by
 hyper-planes. 

The configuration space vector $\qq$ is thus confined to
the intersection of the cylinders coming from the string-type interactions,
and the half-spaces corresponding to the bounding hard walls; this region
is the intersection of convex sets, and  is thus itself convex. 
 Due to the fact that the interactions are hard-core, the motion in phase space
is in fact equivalent to a \emph{semi-focusing billiard} inside
this convex set \cite{Bunimovich:2006p213}; this should be compared to the 
semi-dispersing billiards which arise from hard-sphere gases
\cite{Szasz:1993p996}. 

\subsection{Simplest interacting-particle model} 

In order to get a grasp on the types of phenomena which can occur in such
models, it is instructive to look at the simplest model in this class,
which consists of a two-dimensional string-bound diatomic
molecule, with string length (maximum separation) $\ell$, confined in a
channel formed by two parallel, infinitely long, hard walls at $y = \pm
\frac{d}{2}$, i.e.\ similar to the example shown in \fref{fig.2partbox},
but with the vertical walls removed and horizontal 
walls infinitely long.  We denote the particle positions and velocities by 
$\qq_i = (x_i, y_i)$ and $\pp_i = (u_i, v_i)$, respectively, for $i=1,2$.
The string constrains the particles by $\| \qq_1 - \qq_2 \|^2 \le \ell^2$, 
and the above argument shows that the hard walls give 4 allowed
half-spaces, given by extending the inequalities
$-\frac{d}{2} \le y_i \le \frac{d}{2}$ to the configuration space vector
$\qq$. 

\begin{figure}[tbh]
  \centering 
  \includegraphics[width=0.5\textwidth]{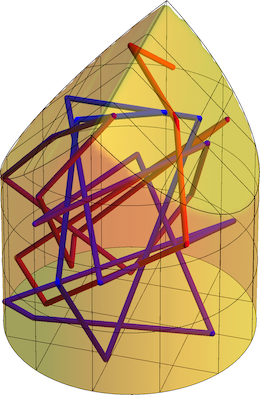}
  \caption{The motion of two interacting particles joined by a string in a
    two-dimensional infinite channel is equivalent to the motion of a point
    particle in a three-dimensional cylindrical domain chopped by four
    planes, or, equivalently, a cylindrical domain with a flat bottom and
    topped by two interesting planes at $z \pm y = h$, as shown above. This
    billiard has a single parameter, the height of the oblique planes,
    given by the ratio of the width of the channel to the length of the
    string of the interacting di-atomic model.} 
  \label{fig.2plstad3d}
\end{figure}
Although the phase space of this system is a priori 8-dimensional,
translational symmetry along the channel implies that the horizontal component 
of  linear momentum of the centre of mass is conserved, so that we may
choose a reference frame in which it vanishes, $u_1 + u_2 = 0$, and
in which we can also fix its position, $x_1 + x_2 = 0$. It is  convenient
to rotate the 
coordinate system by angles $\pi/4$ and normalise them according to
$\ell$, setting $X \equiv (x_1 - x_2)/\ell$, 
$Y \equiv (y_1 - y_2)/\ell$, and $Z \equiv (y_1 + y_2)/\ell$.  
This gives an equivalent billiard system, consisting of a point particle which
bounces elastically inside the region bounded by the cylinder $X^2 +
Y^2 \le 1$, and chopped by the four planes $Y \pm Z = \pm  h$, with $h
\equiv d / \ell$. This, in turn, is 
equivalent to a billiard in a domain bounded below by the plane $Z=0$ and
above by the two planes $Z \pm Y = h$, shown in \fref{fig.2plstad3d}.

Similar constructions give rise to cylindrical stadia in higher
dimensions. For example, in the case of the diatomic molecule inside a
2-dimensional square box with hard walls shown in \fref{fig.2partbox},
there is no longer conservation of momentum, so that we 
obtain a billiard inside a cylinder with two curved and two flat 
directions in $4$ dimensions, bounded by eight hyperplanes. 

\subsection{Cylindrical stadium billiards}

The examples of particles interacting via string-type interactions motivate our
introduction of a large class of
\emph{cylindrical stadium billiards}. Informally, these consist of one or
more cylinders which are cut by (hyper-)planes. More formally, they are the
intersection of one or more infinite cylindrical region(s) with 
two or more half spaces, such that the resulting billiard is bounded 
and such that at least part of the boundary of
the resulting region is curved (that is, comes from a cylinder). This last
condition is required since otherwise the region is polyhedral, and 
the dynamics is known to be non-chaotic
\cite{Chernov:2006p683}.   

Note that these billiards do
  not share the property of the Bunimovich and Papenbrock stadia
  that the cylindrical and flat portions of the boundary meet
  tangentially ($C^1$ boundaries). Nonetheless, they are higher-dimensional
convex billiards with large chaotic regions, and in 
this sense are a possible generalisation of the stadium which are of intrinsic
interest.  Note also that
the stadium is only one of a large class of 2D billiards proved to be chaotic,
many of which do not share
its property of having $C^1$ boundaries
\cite{Chernov:2006p683}. 

\section{A three-dimensional cylindrical stadium billiard\label{sec.Not}}

In order to investigate in more detail the types of mechanisms which are
present and which give rise to chaotic and/or stable motions in cylindrical
stadium billiards, we henceforth restrict attention  to one among the
simplest possible such systems, which is a simplified version of  the 3D
cylinder found in the previous section. 

\begin{figure}[tbh]
  \centering 
  (a)
  \includegraphics[width=0.29\textwidth]{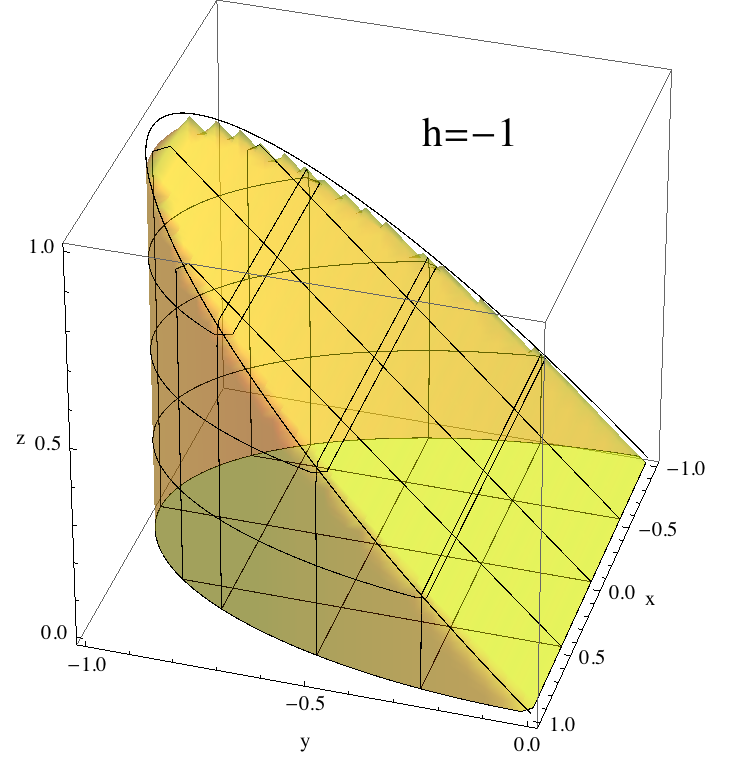}
  \hfill
  (b)
  \includegraphics[width=0.29\textwidth]{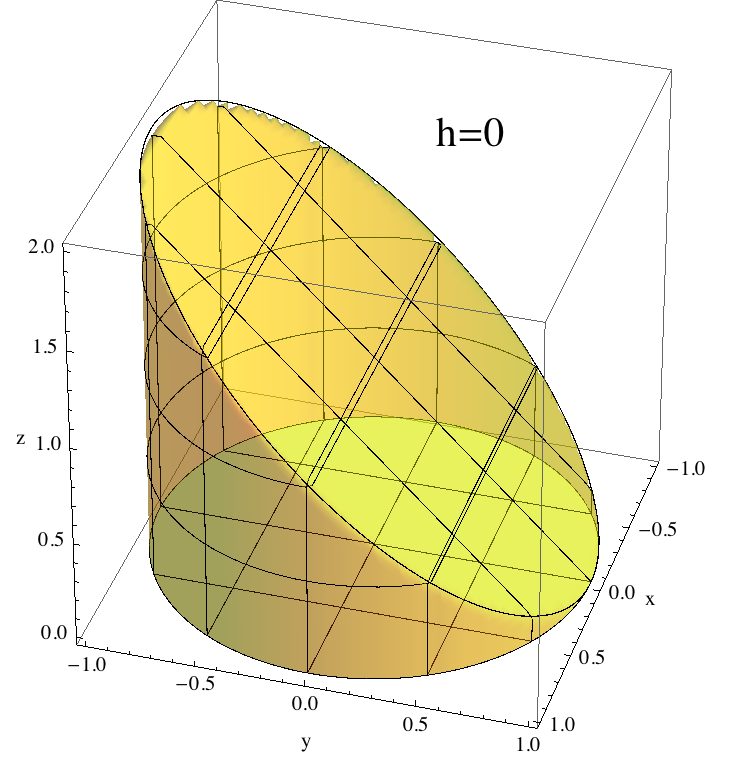}
  \hfill
  (c)
  \includegraphics[width=0.29\textwidth]{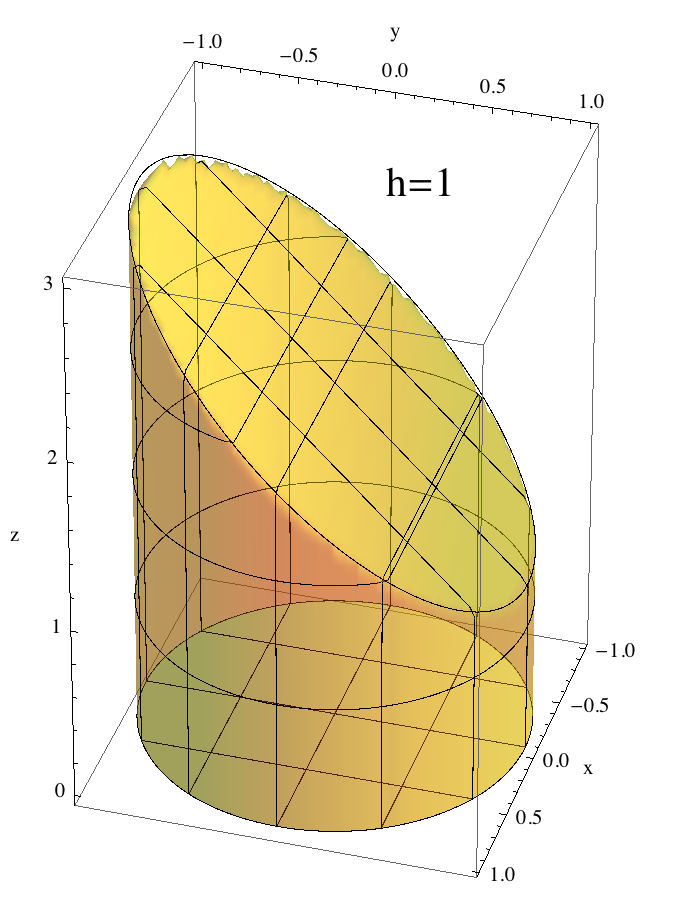}
  \caption{Examples of the cylindrical billiard domains we study: (a) $h =
    -1$; (b) $h = 0$; (c) $h = 1$.}
  \label{fig.cylinder}
\end{figure}

We consider a cylindrical stadium billiard in $\RR^3$ formed by the cylinder
$x^2 + y^2 \le 1$ with unit radius and  vertical axis, and bounded below by
the plane $z=0$, which is perpendicular to the axis, and an oblique plane
that is inclined away from perpendicular. Here we focus on a one-parameter
family of such billiards, where the inclined plane is placed at an angle  
$\pi/4$ with respect to the cylinder axis, and is parameterised by
its height $h$, with equation $y + z = 1+h$. The parameter $h \ge -2$
thus defined measures the minimal height of the cylinder at $y=1$
(``above'' its base); $h < 0$ corresponds to the case where the inclined
plane cuts the flat 
plane inside the cylinder. The convex billiard domain then consists of the
intersection of
the cylinder with the two half-spaces $z \ge 0$ and $y+z \le 1 + h$ -- see
\fref{fig.cylinder}. 

\begin{figure}[tbh]
  \centering
  (a)
   \includegraphics[width=0.29\textwidth]{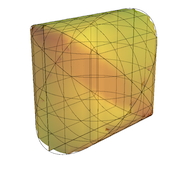}
  \hfill
  (b)
   \includegraphics[width=0.29\textwidth]{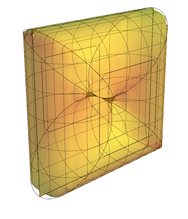}
  \hfill
  (c)
   \includegraphics[width=0.29\textwidth]{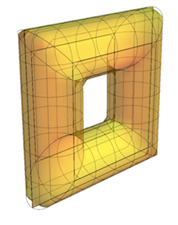}
  \caption{Squared cylindrical stadia obtained by repeatedly reflecting the
    cylindrical billiard domains displayed in \fref{fig.cylinder} in
    their bottom and top planes. (a) $h = -1$; (b) $h = 0$; (c) $h = 1$.} 
  \label{fig.sqcylinder}
\end{figure}

We view this convex billiard as the elementary cell
of an expanded square-shaped cylindrical billiard, or more simply squared
cylindrical stadium, obtained by  ``unfolding'' the elementary cell, that
is, repeatedly reflecting it in its two planes. This cavity is thus made
out of the union of four sections of cylinders, which intersect pairwise at
right angles -- see \fref{fig.sqcylinder}. The only relevant
parameter is the height of each of these 
cylinders; their half-height is the parameter $h$ introduced above, which we
will refer to as the \emph{geometrical parameter}.
It is bounded below by $h = -2$, but is not bounded above.

Considering the unfolded billiard, we will refer to the plane generated by
the axes of the four cylinders as  the \emph{plane of the billiard}. A
transverse plane refers to any plane perpendicular to that plane. These
include the planes perpendicular to the cylinders' axes, as well as the
oblique planes in which they intersect.

Choosing a Cartesian coordinate system whose center is at the center of
symmetry of the square, we take the plane $x=0$ to be the plane of the
billiard. The cylinder axes are then located at $x = 0$
and $y = \pm (1 + h)$, parallel to the $z$-axis, and at $x = 0$ and $z = \pm
(1 + h)$, parallel to the $y$-axis.

The unfolded billiard undergoes significant topological changes at $h = -1$
and $h = 0$. When $h = -1$, each pair of parallel axes coalesces, so that the
unfolded billiard is a convex cavity for $-2 \leq h \leq -1$. When $h = 0$,
two  parallel cylinders intersect along a single line on their
surfaces -- they intersect for $h<0$ and separate for $h>1$. The unfolded
billiard thus has genus $0$ for $h < 0$ and genus $1$ for $h > 0$.

\section{Periodic orbit families\label{sec.PO}}

Much in the same way that one distinguishes bouncing-ball and
whispering-gallery periodic orbits in the two-dimensional stadium billiard,
we will distinguish two specific types of periodic orbits in the
three-dimensional square cylindrical stadium.  The first type consists of
planar periodic orbits, which are analogous to the bouncing-ball periodic
orbits: they bounce from one side of the cavity to the opposite one, within
the plane of the billiard. The second type consists of helical orbits,
which whirl around the surfaces of the cylinders, going smoothly from one
cylinder to the next across their intersection. As we shall show, the
analysis of these two classes of periodic orbits is sufficient to
determine the parametric regions where stable oscillations in cylindrical
stadium billiards occur.

The specificity of  cylindrical billiards is that all their surfaces
have a flat component. This  has the consequence that every periodic
orbit has at least two \emph{parabolic} eigen-directions, i.e.\ whose
corresponding eigenvalues are unity. These are associated to motion
parallel to the axes of the cylinders.

Since the squared cylindrical stadium gives rise to a four-dimensional
symplectic billiard map, a periodic orbit also has another pair of
eigenvalues, in addition to these two parabolic ones, whose product is also
unity. This 
pair is associated to the motion in planes transverse to the cylinders'
axes. It can be either hyperbolic, when the two eigenvalues are real and
are inverses of each other, or elliptic, when they are complex conjugates
with unit modulus. The case with four parabolic eigenvalues is irrelevant.

The consequences of the existence of a pair of parabolic eigenvalues for
each periodic orbit are twofold. First, a periodic orbit is unstable when
the second pair of eigenvalues is hyperbolic, but only marginally stable
when this pair is elliptic. In other words, a given periodic orbit can never
be linearly stable by itself. The second consequence is that periodic
orbits exist in continuous one-parameter families, which can be
parameterised by a displacement along the cylinders' axes. We refer to this
displacement as the \emph{dynamical parameter}, which we will denote in
general by $\epsilon$, and which is defined suitably to parameterise each
family of orbits. It is not to be confused with the height of the
cylinders, which is the model's geometrical parameter $h$. 

If one were to limit the stability analysis to linear terms, as has
previously been the case,  then the naive conclusion
would be that a family of orbits is stable on condition that every
single orbit of this family admits a pair of elliptic eigenvalues, and
provided that there are no geometric constraints which prevent this family
from being complete because of a discontinuity. The reasoning is that the
parabolic eigen-directions behave neutrally, so that given a periodic orbit
and a small perturbation along these directions, the trajectory will move
along them as it oscillates in the plane spanned by the elliptic
eigenvectors, sweeping  through the whole family of associated periodic
orbits, parameterised by different values of $\epsilon$, but left unscathed
in the absence of a destabilizing mechanism. If, on the other hand, the
family of periodic orbits undergoes a transition from elliptic to
hyperbolic regimes as $\epsilon$ varies -- as happens in most cases -- then
the initially stable oscillations in the elliptic regime would become
unstable at the bifurcation, when a value of $\epsilon$ is reached at which
the family crosses over from the elliptic to the hyperbolic regions, and
thus the motion would destabilise.

As it turns out, when elliptic regions occur for a given family of
periodic orbits, transitions from elliptic to hyperbolic regimes are
always found as soon as the value of the maximal height $2+h$ of the
cylinders in the elementary cell is larger than the cylinders' radii, i.e.\
when $h>-1$. The reason is essentially that beyond a bifurcation value of
the dynamical parameter associated to a given family of periodic orbits,
the distance between successive collisions is, at some point on the orbit,
large enough to induce defocusing. According to the above arguments, one
would then be led to believe that no stable oscillations could persist for
square cylindrical stadia with heights larger than this bound. 

However, interestingly, the story does not stop there. Extending 
the stability analysis to the next order beyond linear order, we find that,
under some
assumptions on the initial conditions, nonlinearities are sometimes able to
stabilise a (partial) \emph{family} of periodic orbits in a limited region of the
dynamical
parameter $\epsilon$, leaving the orbit confined to the elliptic
region, 
oscillating back and forth between two values of this parameter. Remarkably,
the nonlinear perturbations are quadratic in the coordinates associated to
the elliptic eigenvalues. As we show below, the key to understanding
whether there exists a nonlinear stabilizing mechanism lies in a qualitative
analysis of the eigenvalue spectrum of these quadratic forms: the signs of
their eigenvalues tell 
us whether or not the oscillations in the plane spanned by the elliptic
eigenvectors are able to stabilise the oscillations along the parabolic
directions.

As our analysis will demonstrate, the conditions for stable oscillations
are met only for a restricted set of values of the geometrical parameter $h$,
consisting of a disjoint union of intervals and isolated values. For
values of the parameter in that set, the phase space is typically mixed -- 
different stable regimes may coexist, together with unstable ones. On the
other hand, away from these values, the square cylindrical stadia are
fully chaotic. 

\section{Planar orbits\label{sec.PPO}}

By planar orbits, we mean those that remain in the plane of the billiard
($x=0$), corresponding to all the orbits whose initial conditions are in
that plane, with velocity components along the $y$- and $z$-axes
only\footnote{There are other planar orbits in planes transverse to
  the cylinders axes, such as a period-four orbit consisting of a bowtie
  figure between two opposite circlular arcs and passing through the center
  of symmetry of the billiard. Such orbits exist for $-1<h<0$ and can be
  found near the $y=0$ and $z=0$ planes, but are not significant since they
  are always unstable.}

For $h \leq 0$, the periodic orbits in that plane are readily identified as
the periodic orbits of a square billiard with sides of length
$2(2+h)$. Given the 
relative prime integers $m$ and $n$, the point $\qq =\{0,y,z\}$ is
periodic provided the  velocity is given by $\pp = \{0, \cos \omega_{m,n},
\sin \omega_{m,n}\}$, where $ \omega_{m,n} = \arctan(n/m)$. Over a
period, the corresponding orbit makes exactly $m$ collisions on each of
the horizontal walls, and $n$ on each of the vertical walls, in a time
$\tau_{m,n} = 4(2+h)\sqrt{m^2+n^2}\,\mathrm{max}(m/n, n/m)$. 

Furthermore, for any pair of relative primes $\{m,n\}$, there is a
continuous family of orbits, indexed by these integers, obtained by varying
the position of the collision impacts on the square sides. Still assuming
$h < 0$, this family extends over all the possible values of $y$ and $z$,
viz. $-(2+h) \le y,\,z\leq 2+h$. 

For $h > 0$, the periodic planar orbits are those of a two-dimensional
square billiard of side length $2(2+h)$ with a square of side length $h$
removed in the centre. The 
periodic orbits of this billiard include some that do not collide with the
central square, which are common to the $h \leq 0$ case, as well as others
that do collide with the central square and which are in general more
difficult to classify.	

\begin{figure}[htb]
  \centering 
  \includegraphics[width=0.3\textwidth]{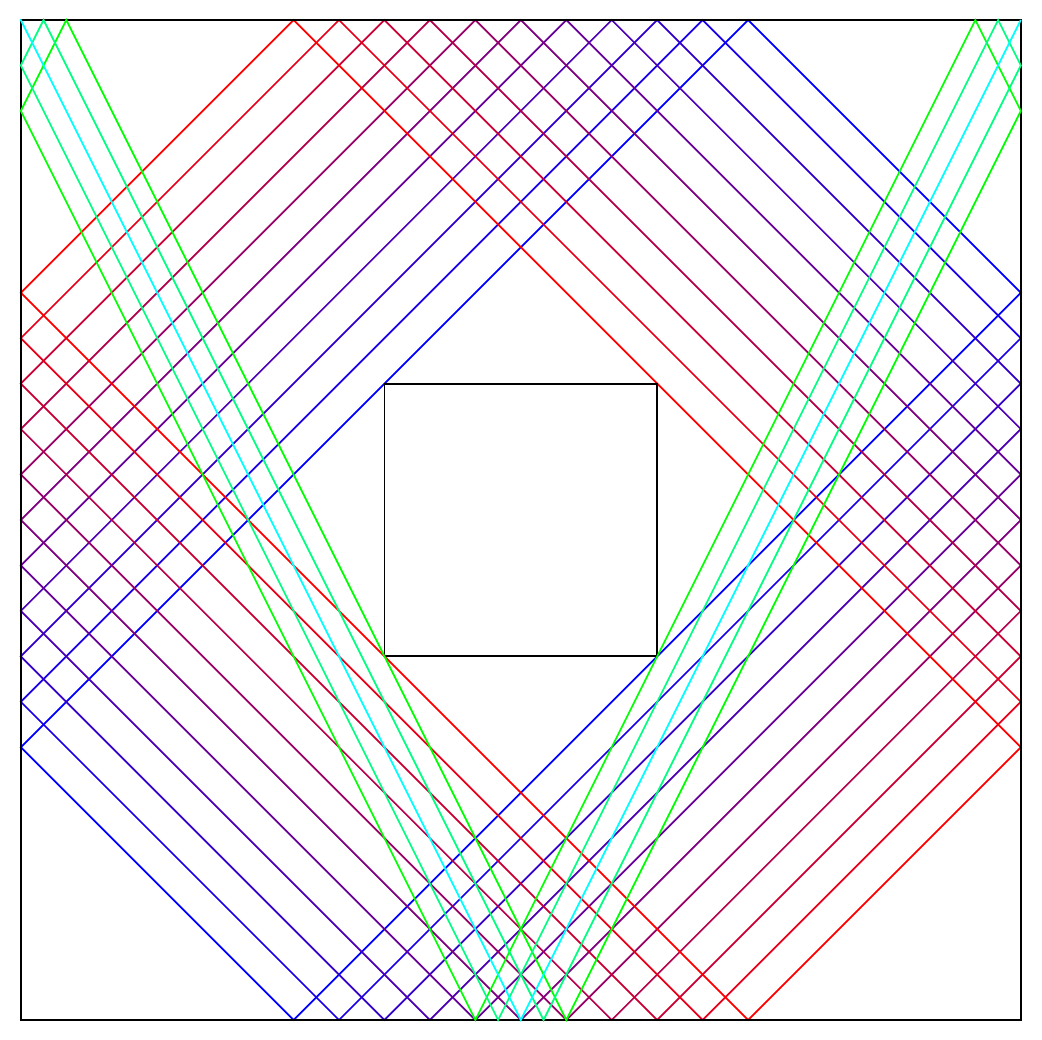}
  \caption{The periodic orbits on a square domaine are specified by the
    direction of the velocity vectors with slopes $\tan(n/m)$, where $m, n
    \in \mathbb{N}$. Several orbits of the $(1,1)_0$ and $(1,2)_0$ families
    are shown, corresponding to the height parameter $h = 3/4$.}
  \label{fig.posquare}
\end{figure}

We will call $(m,n)_0$ planar periodic orbits (PPO) those
periodic orbits of the plane that make collisions only with the outer
surfaces of the cylinders. Examples are shown in 
\fref{fig.posquare}. Except for $n=0$ and $m=n=1$, we will assume $m < n$. 

\subsection{$(1,0)_0$ planar periodic orbits 
  \label{sec.pp1-0_0}}

\subsubsection{Existence \label{sec.pp1-0_0Exs}}

The simplest class of planar orbits are period-2 orbits which bounce back
and forth between two opposite cylinders with velocity perpendicular to the
cylinder axes. Assuming they propagate along the $y$ axis, we identify them
by the phase point with Cartesian coordinates
\begin{equation}
  \qq_0 = \left(
    \begin{array}{c}
      x_0\\
      y_0\\
      z_0
    \end{array}
  \right)
  =
  \left(
    \begin{array}{c}
      0\\
      -2 - h\\
      \epsilon
    \end{array}
  \right),
  \qquad
  \pp_0 = \left(
    \begin{array}{c}
      u_0\\
      v_0\\
      w_0
    \end{array}
  \right)
  =
  \left(
    \begin{array}{c}
      0\\
      1\\
      0
    \end{array}
  \right).
  \label{pp1-0_0-ic}
\end{equation}
  
\begin{figure}[htb]
  \centering 
  (a)
  \includegraphics[width=0.29\textwidth]{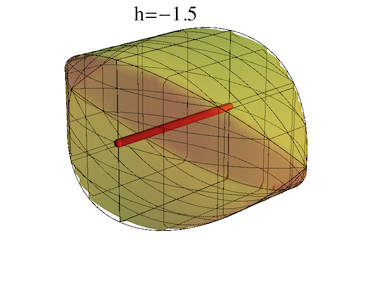}
  \hfill
  (b)
  \includegraphics[width=0.29\textwidth]{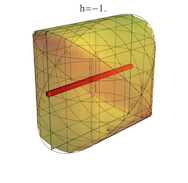}
  \hfill
  (c)
  \includegraphics[width=0.29\textwidth]{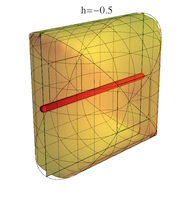}
  \caption{Period-2 $(1,0)_0$ PPOs of the square cylindrical stadium. 
    (a) $h = -3/2$, (b) $h = -1$, (c) $h = -1/2$. These orbits correspond
    to $\epsilon=0$ in \eref{pp1-0_0-ic}, but similar orbits
    exist of all possible values of $\epsilon$ along the vertical axis.} 
  \label{fig.pp1-0_traj}
\end{figure}

These orbits exist everywhere along the axis of the cylinder
so long as $h \leq 0$, i.e.\ $-2-h \leq \epsilon \leq 2+h$, or away from the
center when $h >0$, i.e.\ $h < |\epsilon| \leq 2 + h$. 
Specific examples are displayed in \fref{fig.pp1-0_traj}. 

\subsubsection{Linear stability \label{sec.pp1-0_0Sta}}

The $(1,0)_0$ PPOs are stable only in the interval $-2\le h < -1$. This
result is a straightforward transposition of the stability of periodic
orbits bouncing back and forth between two opposite circular arcs of
identical curvatures: $h = -1$ corresponds to the case where the distance
between the arcs is equal to their diameter. Thus the case (a) in
\fref{fig.pp1-0_traj} is stable, case (b) marginal and case (c)
unstable. 

\paragraph{Cartesian coordinates.} 
It is useful to go through the stability analysis explicitly. To do so, we 
compute the Jacobian of the periodic orbit by following the transformation
rules of the tangent vectors $\vect{\delta q}_i$ and $\vect{\delta p}_i$
along the orbit, where $i = x,y,z$ and each of these vectors has six
dimensions.  

These transformation rules are given according to the two phases of
propagation and collision  by \cite{Dellago:1996p91} 
\begin{equation}
  \begin{array}{lcl}
    \vect{\delta q}_i &\mapsto&  \vect{\delta q}_i 
    + \tau \vect{\delta p}_i\,,\\ 
    \vect{\delta p}_i &\mapsto&   \vect{\delta p}_i\,,
  \end{array}
  \hfill(\mathrm{propagation\,by}\,\tau)
  \label{dqdpt}
\end{equation}
and
\begin{equation}
  \begin{array}{lcl}
    \vect{\delta q}_i &\mapsto&  \vect{\delta q}_i 
    -2\Big(\vect{\delta q}_i\cdot \hat{\mathbf{n}}\Big)\hat{\mathbf{n}}
    \,,\\ 
    \vect{\delta p}_i &\mapsto&   \vect{\delta p}_i
    -2\Big(\vect{\delta p}_i\cdot \hat{\mathbf{n}}\Big)\hat{\mathbf{n}}
    -2 \Big(\pp\cdot \vect{\delta n}_i\Big)\hat{\mathbf{n}}
    - 2 \Big(\pp\cdot \hat{\mathbf{n}}\Big)\vect{\delta n}_i
    \,,
  \end{array}
  \hfill(\mathrm{collision})
  \label{dqdpc}
\end{equation}
where $\hat{\mathbf{n}}$ is the unit vector normal to the surface,
and $\vect{\delta n}$ is a triplet of six component
vectors which depends on the curvature of the surface, 
\begin{equation}
  \vect{\delta n}_i = \left(\frac{\partial \hat{\mathbf{n}}}{\partial
      \vect{q}}\right)_{i,j}
  \left[\vect{\delta q}_j - p_j 
    \frac{
      \vect{\delta q}_k \hat{n}_k}
    {\vect{p}\cdot\hat{\mathbf{n}}}\right]\,.
  \label{dn}
\end{equation}
Given a cylinder with axis in the $y$- or $z$-directions and a
trajectory in the plane of the billiard (the $x$ components of the positions
and velocities vanish), this reduces to
\begin{eqnarray}
  \vect{\delta n}_x &=& \vect{\delta q}_x\,,\\
  \vect{\delta n}_y &=& \vect{\delta n}_z = 0\,.
\end{eqnarray}

The orbit starting at the coordinates \eref{pp1-0_0-ic} bounces back and
forth from $\qq_0$, $\pp_0$ to 
\begin{equation}
  \qq_1 = (0, 2+h, \epsilon), 
  \qquad   
  \pp_1 = (0, -1, 0).
\end{equation}

Correspondingly, letting $\tau = 2(2 + h)$ denote the time between
successive collisions, the tangent vectors are mapped according to
equations \eref{dqdpt}--\eref{dqdpc} as follows, following the
propagation (prop.) and collision (coll.) 
phases:
\begin{eqnarray}
  \fl
  \left(
    \begin{array}{cccccc}
      1&0&0&0&0&0\\
      0&1&0&0&0&0\\
      0&0&1&0&0&0\\
      0&0&0&1&0&0\\
      0&0&0&0&1&0\\
      0&0&0&0&0&1
    \end{array}
  \right)
  &\stackrel{\mapsto}{\mathrm{prop.}}&
  \left(
    \begin{array}{cccccc}
      1&0&0&\tau&0&0\\
      0&1&0&0&\tau&0\\
      0&0&1&0&0&\tau\\
      0&0&0&1&0&0\\
      0&0&0&0&1&0\\
      0&0&0&0&0&1
    \end{array}
  \right)
  \nonumber\\
  &\stackrel{\mapsto}{\mathrm{coll.}}&
  \left(
    \begin{array}{cccccc}
      1&0&0&\tau&0&0\\
      0&-1&0&0&-\tau&0\\
      0&0&1&0&0&\tau\\
      -2&0&0&1-2\tau&0&0\\
      0&0&0&0&-1&0\\
      0&0&0&0&0&1
    \end{array}
  \right)
  \nonumber\\
  &\stackrel{\mapsto}{\mathrm{prop.}}&
  \left(
    \begin{array}{cccccc}
      1-2\tau&0&0&2\tau(1-\tau)&0&0\\
      0&-1&0&0&-2\tau&0\\
      0&0&1&0&0&2\tau\\
      -2&0&0&1-2\tau&0&0\\
      0&0&0&0&-1&0\\
      0&0&0&0&0&1
    \end{array}
  \right)
  \nonumber
  \\
  &\stackrel{\mapsto}{\mathrm{coll.}}&
  \left(
    \begin{array}{cccccc}
      1-2\tau&0&0&2\tau(1-\tau)&0&0\\
      0&1&0&0&2\tau&0\\
      0&0&1&0&0&2\tau\\
      -4(1-\tau)&0&0&(1-2\tau)^2 - 2\tau&0&0\\
      0&0&0&0&1&0\\
      0&0&0&0&0&1
    \end{array}
  \right).
  \label{pp1-0_0-jacobian}
\end{eqnarray}

This matrix has six eigenvalues. Two pairs are trivial, the first
corresponding to the conservation of energy and the associated
time-translation symmetry, and the second being the pair of parabolic
eigenvalues which are due to the flat 
components of the cylinders' surfaces. The only pair of non-trivial
eigenvalues of this matrix are then
\begin{equation}
  \lambda_{1,2} = 8h(3 + h) + 17 \pm (3+2h) \sqrt{(1+h)(2+h)}.
  \label{pp1-0_0-eigenvalues}
\end{equation}
They are elliptic for $-2< h < -1$ and hyperbolic otherwise, as
expected. 

We note that a similar conclusion (with simpler eigenvalues) can be reached
by considering only half of the orbit. This is because the corresponding
periodic orbit in the elementary cell involves only a single collision
with the cylindrical surface. However, because the parity of the
number of reflections on the transverse planes can be odd in the elementary
cell, considering the periodic orbits in the expanded billiard is better
suited to the non-linear stability analysis discussed below.

\paragraph{Birkhoff coordinates.}
The symplectic structure of the collision map is best recovered
by considering, instead of Cartesian coordinates, the Birkhoff
coordinates of the billiard map, i.e.\ the coordinates on the constant
energy surface in which the map is volume-preserving. They are given by:
(i) $\theta$, the position angle, measured in the plane transverse to the
cylinders; (ii) $\xi$, the sine of the associated velocity, measured with
respect to the normal to the cylinder's surface; and (iii)--(iv) the pair $z$,
$w$ of the position and velocity measured along the cylinder axis:
\numparts
\begin{eqnarray}
  \theta_0 &=& \arctan\frac{y_0 + 1 + h}{x_0} = 
  -\frac{\pi}{2}
  \label{pp1-0_0-theta};\\
  \xi_0 &=& \sin \left(\arctan\frac{v_0}{u_0} - \theta_0\right)
  = 0
  \label{pp1-0_0-xi};\\
  z_0 &=& \epsilon
  \label{pp1-0_0-z};\\
  w_0 &=& 0
  \label{pp1-0_0-w}.
\end{eqnarray}
\endnumparts
In terms of the Birkhoff coordinates, the linear map
\eref{pp1-0_0-jacobian} for the perturbations $\delta\theta$, $\delta\xi$,
$\delta w$ and $\delta z$ reduces to 
\begin{equation}
  \left(
    \begin{array}{c}
      \delta\theta\\
      \delta\xi\\
      \delta w\\
      \delta z
    \end{array}
  \right)
  \mapsto
  \left(
    \begin{array}{cccc}
      m_{\theta\theta}&m_{\theta\xi}&0&0\\
      m_{\xi\theta}&m_{\xi\xi}&0&0\\
      0&0&1&0\\
      0&0&m_{zw}&1
    \end{array}
  \right)
  \left(
    \begin{array}{c}
      \delta\theta\\
      \delta\xi\\
      \delta w\\
      \delta z
    \end{array}
  \right).
  \label{pp1-n_0-linearmap}
\end{equation}
It turns out that the structure of this linear map is common to all planar
periodic orbits; the only non-trivial coefficients are the four matrix
elements corresponding to the motion in the $\theta$--$\xi$ plane and the
element of axial velocity contribution to the displacement along the
cylinder axis. In the case of the $(1,0)_0$ PPO, the matrix of elements 
$m_{\alpha \beta}$ is given by 
\begin{equation}
  \fl
  \left(
    \begin{array}{cccc}
      m_{\theta\theta}&m_{\theta\xi}&0&0\\
      m_{\xi\theta}&m_{\xi\xi}&0&0\\
      0&0&1&0\\
      0&0&m_{zw}&1
    \end{array}
  \right)
  =
  \left(
    \begin{array}{cc@{\hspace{-.2cm}}cc}
      17 + 8 h (3 + h)&  -4 (2 + h) (3 + 2 h)&0&0\\
      -4 (1 + h) (3 + 2 h)&17 + 8 h (3 + h)&0&0\\
      0&0&1&0\\
      0&0&4 (2 + h)&1
    \end{array}
  \right).
  \label{pp1-0_0-linearmap}
\end{equation}
The four eigenvalues of this matrix form the pair of parabolic eigenvalues
and the pair $\lambda_{1,2}$ obtained in equation~\eref{pp1-0_0-eigenvalues}.

\begin{figure}[tbh]
  \centering
  \includegraphics[width=0.45\textwidth]{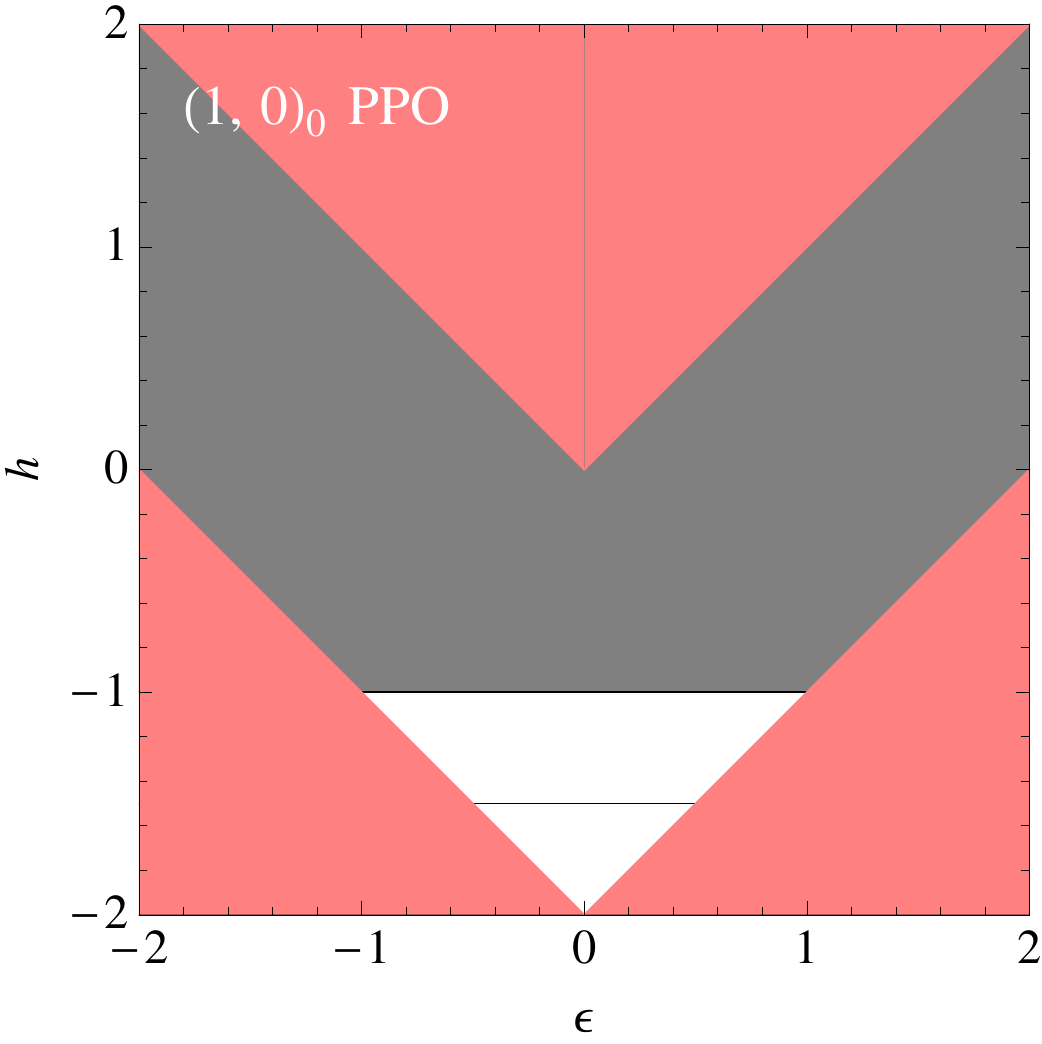}
  \caption{Results of the linear stability analysis of the $(1,0)_0$
    PPOs plotted in the parameter space, with the dynamical parameter
    $\epsilon$ on the horizontal axes, and geometrical parameter $h$ on the
    vertical axis. The light red areas represent the regions outside the range 
    of allowed parameter values. The grey areas correspond to regions where the
    eigenvalues \eref{pp1-0_0-eigenvalues} are hyperbolic, and 
    white where they are elliptic. The line which appears in the elliptic
    region is a line of parabolic eigenvalues.}
  \label{fig.pp1-0_0_stab}
\end{figure}

This result is summarised in  \fref{fig.pp1-0_0_stab}, where the
parameter space of the $(1,0)_0$ PPOs is colored, when the orbit exists, in
white where the eigenvalues \eref{pp1-0_0-eigenvalues} are elliptic and
gray where they are hyperbolic, and in light red where the orbits do not
exist. Similar conventions will be used throughout the paper.

The conclusion we infer from this calculation is that the parameter region
$-2 < h < -1$ corresponds to a regime with mixed phase space, where the
period-2 orbits in the plane of the billiard give rise to an elliptic
island. Though each of these orbits is only marginally stable, it belongs
to a continuous family of similar orbits, which is moreover complete,
meaning that it extends from one corner of the billiard to the opposite 
one. The family of orbits therefore corresponds to a plane of elliptic
stability. 

We note that the $(1,0)_0$ PPOs are peculiar in the sense that their
nonlinear stability analysis yields non-trivial corrections only to cubic
order in the perturbations $\delta\theta$, $\delta\xi$ and $\delta
w$. Turning to the class of $(1,n)_0$ PPOs, we will see that, not
surprisingly, the $(1,0)_0$ PPOs are not the only planar periodic orbits 
which give rise to a stable regime. In fact, a perturbation of the $(1,0)_0$ PPOs
in the plane of the billiard will typically be closer to a family of
$(1,n)_0$ PPOs with $n$ large. Interestingly, such periodic orbits may
exhibit a mechanism for non-linear stability quadratic in the
perturbations $\delta\theta$, $\delta\xi$, $\delta w$, absent for the
$(1,0)_0$ family. 

\subsection{$(1,1)_0$- planar periodic orbits 
\label{sec.pp1-1_0}}

\subsubsection{Existence \label{sec.pp1-1_0Exs}}

The orbit with unit speed at angle $\pi/4$ with respect to the cylinder axes
is the period-4 orbit which cycles through the phase points
\begin{eqnarray}
  \begin{array}{l}
    \vect{q}_0 = \left(0, -2 -h, \epsilon\right)\\ 
    \vect{p}_0 = \frac{1}{\sqrt 2}
    \left(0,1,1\right)   
  \end{array}
  &\mapsto
  \begin{array}{l}
    \vect{q}_1 = \left(0,- \epsilon,2 + h\right)\\
    \vect{p}_1 = \frac{1}{\sqrt2} \left(0,1,-1\right) 
  \end{array}
  \mapsto
  \begin{array}{l}
    \vect{q}_2 = \left(0,2 + h,-\epsilon\right)\\ 
    \vect{p}_2 = \frac{1}{\sqrt 2}
    \left(0,-1,-1\right) 
  \end{array}
  \nonumber\\
  &\mapsto
  \begin{array}{l}
    \vect{q}_3 = \left(0,\epsilon,- 2 - h\right)\\ 
    \vect{p}_3 = \frac{1}{\sqrt2}
    \left(0,-1,1\right) 
  \end{array}
  \mapsto
  \begin{array}{l}
    \vect{q}_4 = \vect{q_0} \\
    \vect{p}_4 = \vect{p_0} 
  \end{array},
  \label{pp1-1_0-ic}
\end{eqnarray}
where $-(2+h) < \epsilon < 2+h$ if $-2 < h \leq 0$ and $h - 2 < \epsilon <
2 - h$ if $0 \leq h < 2$.

\subsubsection{Linear stability \label{sec.pp1-1_0Sta}}

Going through the analysis described in \sref{sec.pp1-0_0Sta}, we
obtain a linear mapping describing the evolution of the perturbation vector
along the periodic orbit, which, in Birkhoff coordinates, is identical
to equation \eref{pp1-n_0-linearmap} with matrix elements
\numparts
\begin{eqnarray}
  m_{\theta\theta}&=&1 - 4 \epsilon + 8 [h (2 + h) - \epsilon^2]
  [(1 + h)^2 - \epsilon - \epsilon^2],
  \label{pp1-1_0-linearmap11}
  \\
  m_{\theta\xi}&=&-4 [2 h (h+2)-2 \epsilon ^2+1] [h
  (h+3)-\epsilon ^2+2],
  \label{pp1-1_0-linearmap12}
  \\
  m_{\xi\theta}&=&-4 [2 h (h+2)-2 \epsilon ^2+1]
  [h^2+h-\epsilon ^2]
  \label{pp1-1_0-linearmap21}
  \\
  m_{\xi\xi}&=&8 [h (h+2)-\epsilon ^2]
  [(h+1)^2-\epsilon ^2+\epsilon ]+4\epsilon +1,
  \label{pp1-1_0-linearmap22}
  \\
  m_{zw}&=& 8 \sqrt{2}(2+h).
  \label{pp1-1_0-mz}
\end{eqnarray}
\endnumparts

\begin{figure}[thb]
  \centering 
  \includegraphics[width=0.5\textwidth]{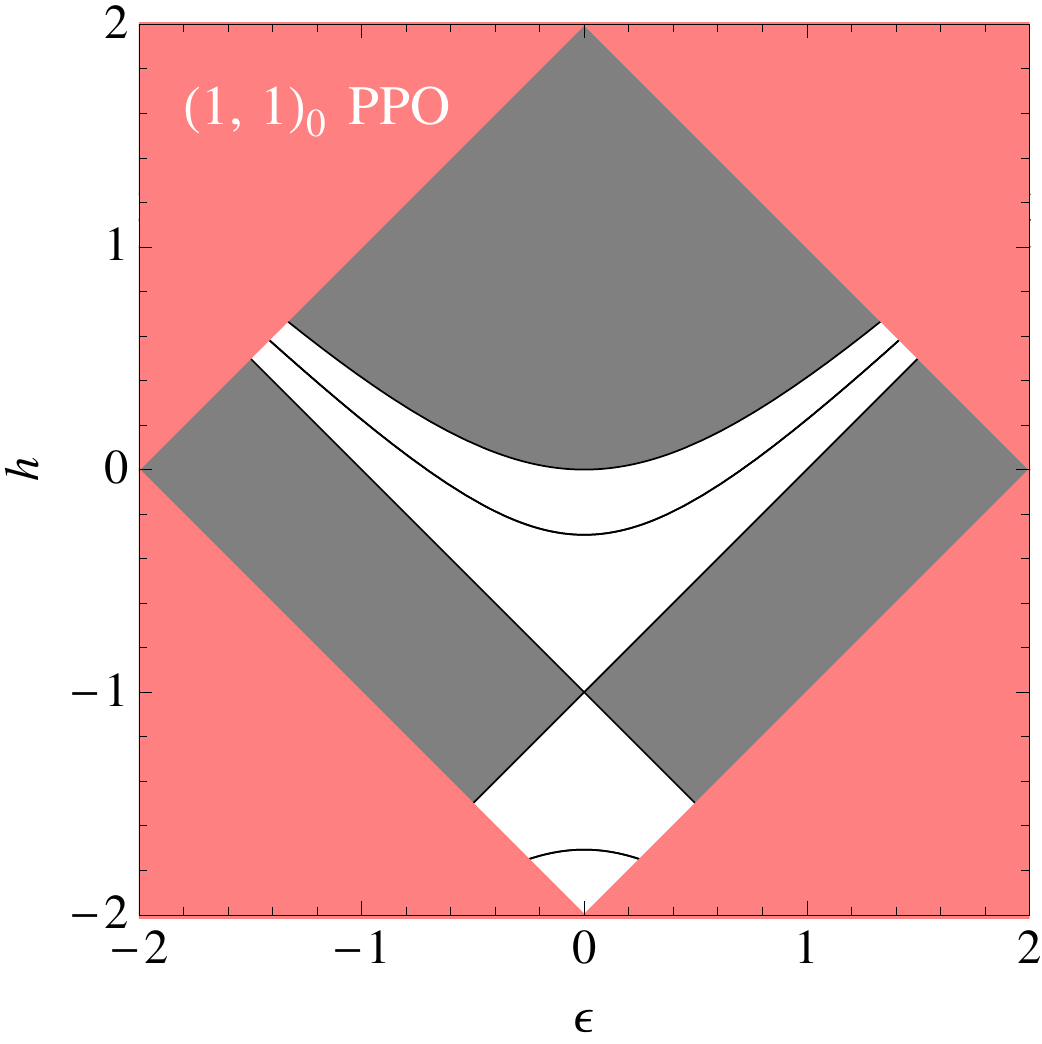}
  \caption{Results of the linear stability analysis of the $(1,1)_0$
    PPOs plotted in the parameter space. The conventions are identical to
    \fref{fig.pp1-0_0_stab}.}
  \label{fig.pp1-1_0_stab}
\end{figure}

The two non-trivial eigenvalues are
\begin{eqnarray}
  \lambda_{1,2} &=& 1 + 8 [-2 h (h+2)\epsilon ^2+h (h+1)^2 (h+2)+\epsilon
  ^4 - \epsilon^2] \nonumber\\
  &&\pm4 \sqrt{(h-\epsilon +1) (h+\epsilon +1)
    [2 h (h+2)-2 \epsilon ^2+1]^2 [h(h+2)-\epsilon^2]}.
  \label{pp1-1_0-eigenvalues}
\end{eqnarray}
The regions of elliptic and hyperbolic eigenvalues are displayed in 
\fref{fig.pp1-1_0_stab}. There are two separate elliptic regions, bounded by
$h = -1 \pm \epsilon$ and $h = -1 \pm \sqrt{1 + \epsilon^2}$. For $h = -1$
and $2/3\leq h <2$, there are no elliptic regions along the $\epsilon$ axis
and, where they exist, the $(1,1)_0$ PPOs are everywhere hyperbolic for
these values of the geometric parameter.

The transitions from elliptic to hyperbolic regimes are understood as
follows. The segment of orbit between, say, $\vect{q}_1$ and $\vect{q}_2$ 
lies in the transverse plane $z - y = 2 + h -\epsilon$, which cuts the
cylinders along the $y$- and $z$-axes in the form of two partial ellipses,
glued one against the other. The semi-minor axis of these ellipses is unity
(the radius of the cylinders) and the semi-major axis $\sqrt{2}$. The
segment of orbit that joins $\vect{q}_1$ to $\vect{q}_2$ goes along the
major axes of these two ellipses. The orbit is thus stable only so long as
the distance traveled along the major axes is less than the radius
$\sqrt{2}/4$ of the disk inscribed into the ellipses at $\vect{q}_1$ and
$\vect{q}_2$. This condition yields the critical values
$\epsilon = \pm (1+h)$, which separates the hyperbolic and elliptic
regions in \fref{fig.pp1-1_0_stab}.

Our analysis shows that the family of $(1,1)_0$ PPOs under consideration is
linearly stable and complete only for $-2<h \leq -3/2$. For larger values
of $h$, though some orbits display elliptic behavior, they  do not
correspond to linearly stable oscillations in the vicinity of the periodic
orbit. The reason, according to the linear stability analysis, is that a
trajectory initially close to an elliptic PPO will oscillate in
the planes transverse to the cylinder's axes, and simultaneously move along
the $z$-axis due to perturbations along the $w$-axis, which remain
unchanged under iterations of the map \eref{pp1-n_0-linearmap}, whose
coefficient $m_{zw}$, given by equation \eref{pp1-1_0-mz}, depends
solely on $h$. In other words, the orbit moves in the parameter space of the 
$(1,1)_0$ PPOs along the $\epsilon$-axis due to the action of the
perturbation along the vertical velocity axis.
In the absence of nonlinear effects acting on $\delta w$, the trajectory
would eventually have to cross over from the elliptic to the hyperbolic
region and therefore lose stability as the parameter $\epsilon$ translates
away from the interval of elliptic eigenvalues.

\subsubsection{Nonlinear stability \label{sec.pp1-1_0NLSta}}

Linear stability would have us conclude that periodic orbit families whose
non-trivial eigenvalues display bifurcations from elliptic to hyperbolic
regimes must correspond to unstable oscillations, even when the reference
periodic orbit we initially perturb away from is marginally stable. This,
however, is in general incorrect. Going beyond linear stability, we find
that the perturbations 
along the $w$ and $z$ coordinates are nonlinear functions of the
oscillations that take place in the $\theta$--$\xi$ plane. At  second
order in the stability analysis, the $\theta$--$\xi$ oscillations give rise
to quadratic terms which act on the perturbations along the parabolic
directions according to
\numparts
\begin{eqnarray}
  \delta w&\mapsto& \delta w + Q_w(\delta \theta, \delta\xi),
  \label{pp1-1_0-quadraticmapw}
  \\
  \delta z&\mapsto& \delta z + m_{z w} \delta w 
  + Q_z(\delta \theta, \delta\xi),
  \label{pp1-1_0-quadraticmapz}
\end{eqnarray}
\endnumparts
where $Q_w(\delta \theta, \delta\xi)$ and $Q_z(\delta \theta, \delta\xi)$
are quadratic forms in their variables, which we write as
\numparts
\begin{eqnarray}
  Q_w(\delta \theta, \delta\xi) &=& 
  \Big(\delta \theta\quad\delta\xi\Big)
  \left(
    \begin{array}{cc}
      a_w & \frac{1}{2}b_w\\
      \frac{1}{2}b_w & c_w
    \end{array}
  \right)
  \left(
    \begin{array}{c}
      \delta\theta \\
      \delta\xi
    \end{array}
  \right),
  \label{pp1-1_0-quadraticformw}
  \\
  Q_z(\delta \theta, \delta\xi) &=& 
  \Big(\delta \theta\quad\delta\xi\Big)
  \left(
    \begin{array}{cc}
      a_z & \frac{1}{2}b_z\\
      \frac{1}{2}b_z & c_z
    \end{array}
  \right)
  \left(
    \begin{array}{c}
      \delta\theta \\
      \delta\xi
    \end{array}
  \right).
  \label{pp1-1_0-quadraticformz}
\end{eqnarray}
\endnumparts
The coefficients $a_w$, $b_w$, $c_w$, $a_z$, $b_z$ and $c_z$ are 
functions of the two parameters $h$ and $\epsilon$.

Now, if $\delta w$ is large enough to start with, then the quadratic terms
in \eref{pp1-1_0-quadraticmapw}--\eref{pp1-1_0-quadraticmapz} will do
little to affect the motion along 
the $w$ and $z$ axes. Thus the problem we need to address is the following:
assuming\footnote{Identifying the initial position
  along the $z$ axis with the dynamical parameter $\epsilon$, we can always
  set $\delta z = 0$.} that $\delta w$ is small with respect to quadratic
terms in $\delta\theta$ and $\delta\xi$, we must find out whether the
quadratic forms $Q_w$ and $Q_z$ are or are not able to induce oscillations
of $\delta w$ and $\delta z$, i.e.\ $\epsilon$. Such oscillations would
prevent the contribution  $m_{zw}\delta w$ in \eref{pp1-n_0-linearmap}
from inducing a translation in the parameter $\epsilon$ (along the $z$
axis) and would thus keep the trajectory oscillating within the elliptic
region. In this case the corresponding family of $(1,1)_0$ PPOs would
remain stable as a result of the nonlinear oscillations.

Fortunately, the solution to this problem is simpler than it may seem, for,
since the coefficient $m_{zw}$ in \eref{pp1-1_0-mz} is positive,
a mechanism inducing oscillations of $w$ around $\epsilon = 0$ (or, for that
matter, around any fixed value $\epsilon = \epsilon_0$) exists whenever $Q_w$
has the opposite sign from $\epsilon$ (or $\epsilon-\epsilon_0$ if
$\epsilon_0\neq0$), i.e.\ whenever $Q_w$ is positive 
where $\epsilon$ is negative and $Q_w$ is negative where $\epsilon$ is
positive. These oscillations in $\epsilon$ will be further amplified by
$Q_z$, constructively so provided $Q_z$ also takes the opposite sign from
$\epsilon$. If, on the other hand, $Q_w$ is positive where $\epsilon$ is
positive and is negative where $\epsilon$ is negative, then, whether or not
$Q_z$ keeps the opposite sign from $\epsilon$, $\delta w$ will keep
increasing or decreasing, steadily if $Q_w$ has the sign of $\epsilon$, or
on average if it has the opposite sign from $\epsilon$, until the trajectory
reaches the limits of the elliptic region and thus destabilises, either
because of a transition to a hyperbolic regime, or because the
limits of the range of allowed $\epsilon$ values is reached.

\begin{figure}[thb]
  \centering 
  (a)
  \includegraphics[width=0.45\textwidth]{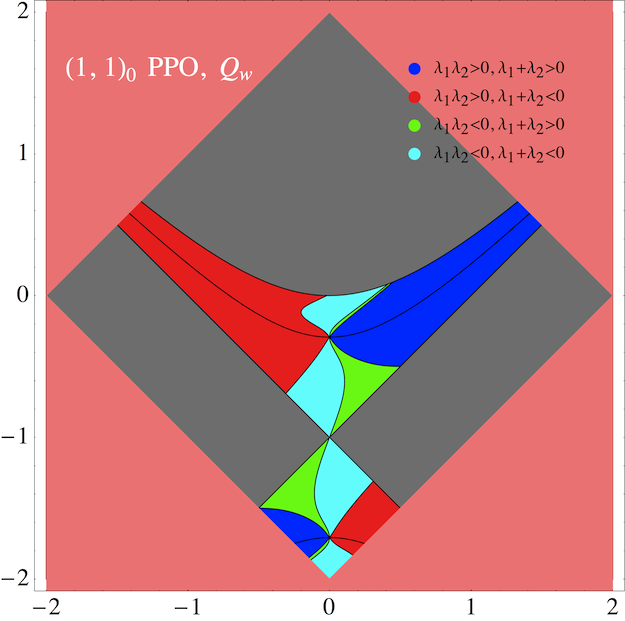}
  (b)
  \includegraphics[width=0.45\textwidth]{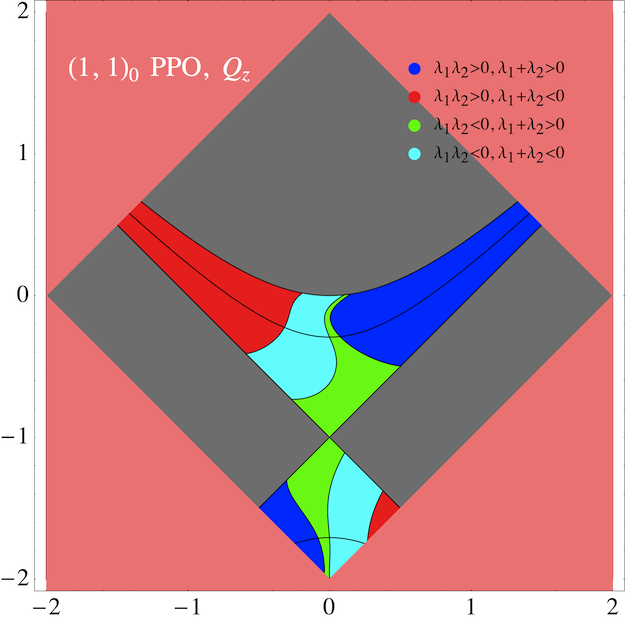}
  \caption{Results of the nonlinear stability analysis of the $(1,1)_0$
    PPOs plotted in the parameter space. The conventions are similar to
    \fref{fig.pp1-1_0_stab}, with the elliptic regions filled
    according to the signs of the determinant and trace of the quadratic
    forms (a) $Q_w$ and (b) $Q_z$:
    in blue where both eigenvalues are positive ($\lambda_1 \lambda_2 >0$,
    $\lambda_1 + \lambda_2 > 0$), red where both are negative ($\lambda_1
    \lambda_2 >0$, $\lambda_1 + \lambda_2 < 0$), green where one is more
    positive than the other is negative ($\lambda_1 \lambda_2 <0$,
    $\lambda_1 + \lambda_2 > 0$), and cyan where one is more negative than
    the other is positive ($\lambda_1 \lambda_2 <0$, $\lambda_1 + \lambda_2
    < 0)$.}
  \label{fig.pp1-1_0_nlstab}
\end{figure}

The quadratic forms
\eref{pp1-1_0-quadraticformw}--\eref{pp1-1_0-quadraticformz} are positive-
or negative-definite whenever the eigenvalues of their defining matrices
have the same signs. (Since these matrices are symmetric and have real
entries, their eigenvalues are always real.) Thus, irrespective of the
oscillatory motion of the $\theta$--$\xi$ variables, $Q_w$ (resp. $Q_z$) is
positive when its two eigenvalues are positive and negative when they are
negative. When the two eigenvalues are of opposite signs, on the other
hand, we must bear in mind that the oscillations of the $\theta$--$\xi$
variables keep going at their own pace, and they will probe the two
eigendirections of $Q_w$ as they do so. Thus, as $Q_w$
(resp. $Q_z$) oscillates, it will be expected to take, on average, the 
sign of the eigenvalue that has the largest magnitude.

\Fref{fig.pp1-1_0_nlstab} shows the eigenvalue spectrum of the matrices
of the quadratic forms
\eref{pp1-1_0-quadraticformw}--\eref{pp1-1_0-quadraticformz}, superimposed
on the 
elliptic regions found from the linear stability analysis in 
\fref{fig.pp1-1_0_stab}. The coefficients of these quadratic forms are given
explicitly in \ref{app.1-1_0-PPO}, equations
\eref{pp1-1_0-quadraticformaw}--\eref{pp1-1_0-quadraticformcw} and
\eref{pp1-1_0-quadraticformaz}--\eref{pp1-1_0-quadraticformcz}. For
convenience, we analyze them by using a representation in four different
colors, in terms of the sign and 
magnitude of their eigenvalues, which are denoted in both cases by
$\lambda_1$ and $\lambda_2$. We observe that the elliptic regions below and
above the lines at $h = -1$ in both cases display clearly different
patterns. Whereas the upper elliptic regions are everywhere unstable under
the quadratic perturbations, the lower elliptic regions are stable
around $\epsilon = 0$.

\begin{figure}[phtb]
  \centering
  (a)
  \includegraphics[width = .25\textwidth]{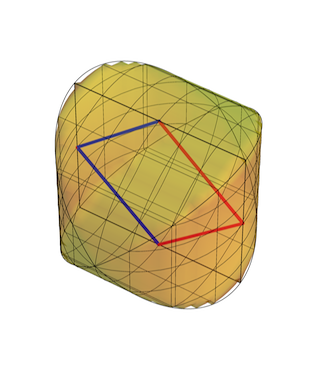}
  \hfill
  (b)
  \includegraphics[width = .3\textwidth]{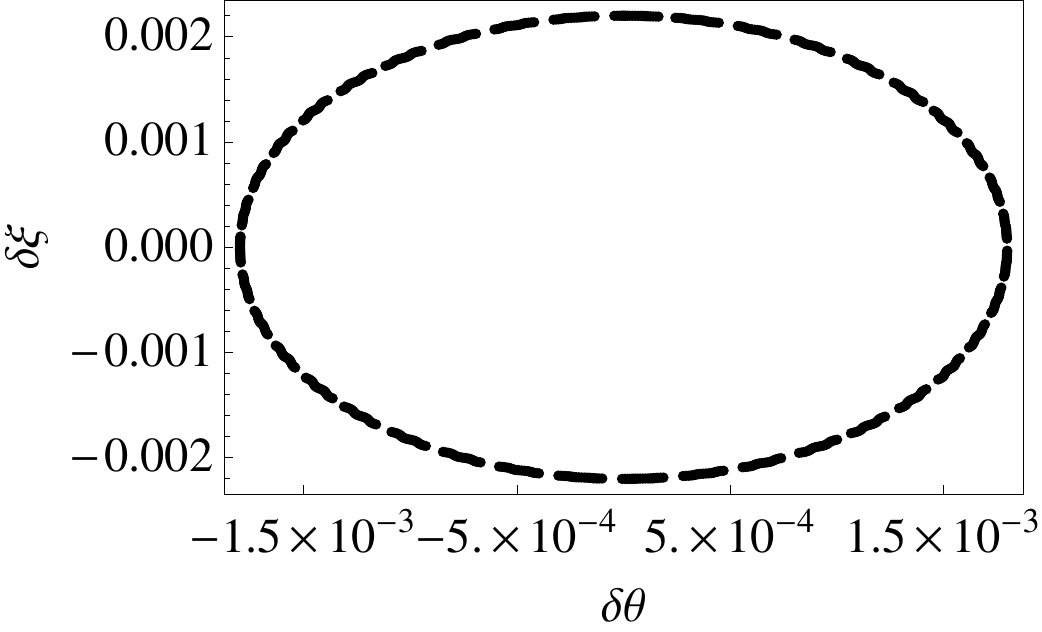}
  \hfill
  (c)
  \includegraphics[width = .3\textwidth]{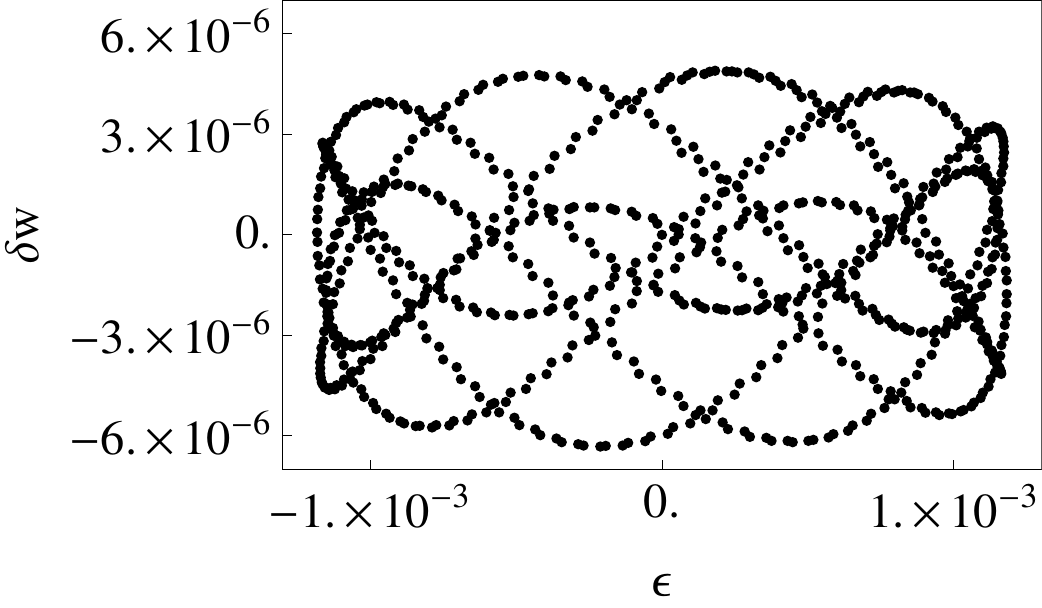}
  \caption{Nonlinearly stable oscillations of the $(1,1)_0$ PPOs measured
    at $h = -1.2$. (a) Actual trajectory; (b) Linear oscillations measured
    in the $\theta$--$\xi$ plane; (c) Nonlinear oscillations measured in
    the $w$--$z$ plane (here we identify the $z$ coordinate with the
    dynamical parameter $\epsilon$). Notice the order of magnitude of the
    $\epsilon$ oscillations as opposed to those of $\delta w$.}
  \label{fig.pp1-1_0_trajs}
\end{figure}

\begin{figure}[tbh]
  \centering
  (a)
  \includegraphics[width = .45\textwidth]{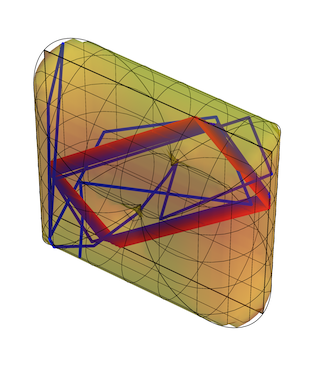}
  \hfill
  (b)
  \includegraphics[width = .45\textwidth]{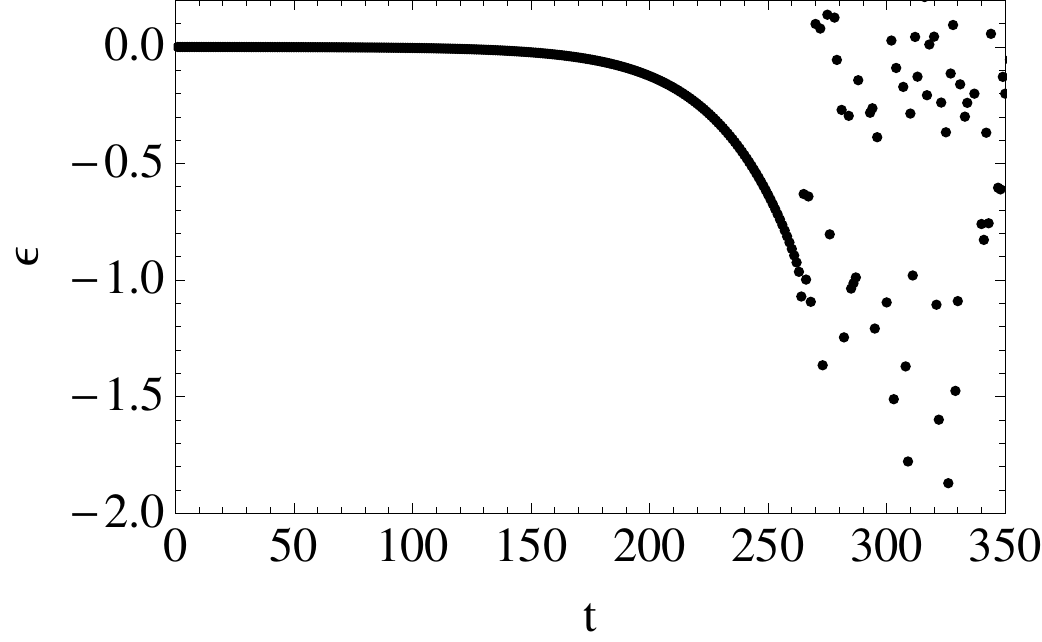}
  \hfill
  (c)
  \includegraphics[width = .45\textwidth]{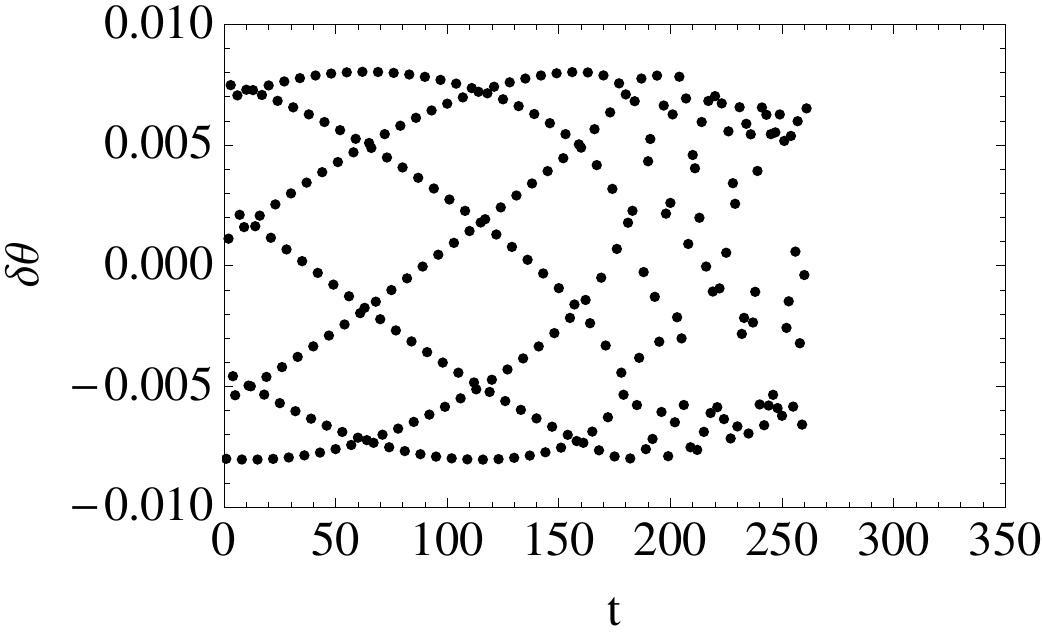}
  (d)
  \includegraphics[width = .45\textwidth]{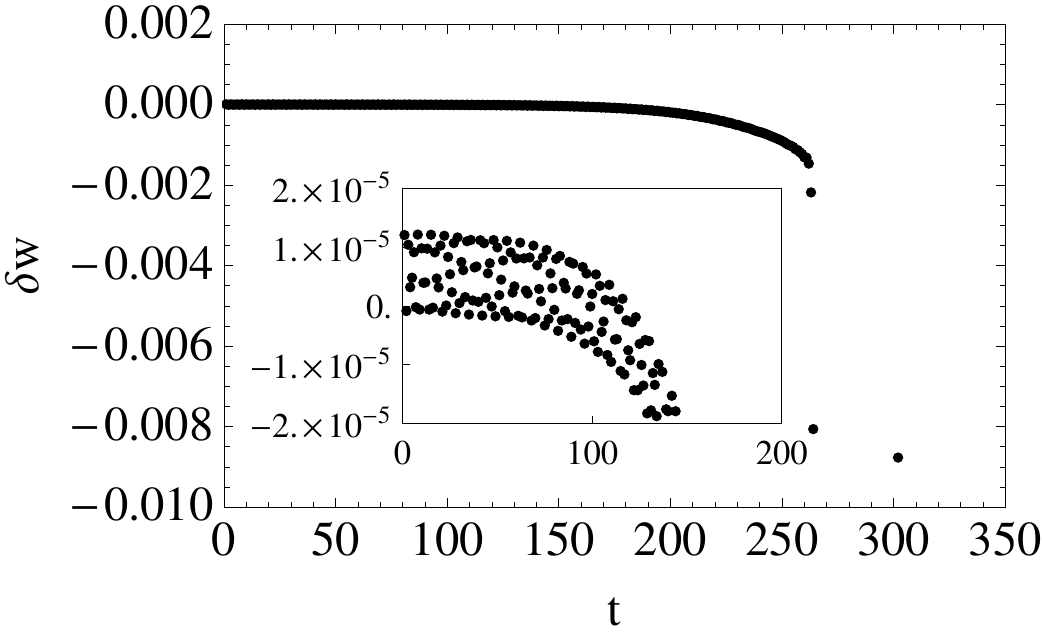}
  \caption{Nonlinearly unstable oscillations of the $(1,1)_0$ PPOs measured
    at $h = -0.1$. (a) The iterations of the actual trajectory
    remain stable over a short time only.
    (b) The motion along the $z$ axis presents a drift as the dynamical
    parameter $\epsilon$ decreases to the negative values and away from the
    elliptic regime. (c) The oscillations of $\theta$ (as well as $\xi$,
    not shown) 
    are initially elliptic, but soon hit the hyperbolic regime as
    $\epsilon$ crosses the bifurcation point. (d) The velocity along the
    cylinder's axes displays small oscillations, but becomes more negative
    under the influence of the oscillations in the $\theta$--$\xi$ plane.}
  \label{fig.pp1-1_0_traju}
\end{figure}

The separation between nonlinearly stable and unstable elliptic regions at
$h=-1$ is supported by numerical results such as those shown in figures
\ref{fig.pp1-1_0_trajs} and \ref{fig.pp1-1_0_traju}.

\subsection{$(1,1)_k$- planar periodic orbits 
\label{sec.pp1-1_k}}

\subsubsection{Existence \label{sec.pp1-1_kExs}}

Orbits of the plane with unit ratio between the $y$ and $z$ velocity
components are peculiar in that they remain periodic even when, for $h>0$, 
they hit the inner walls of the cylinders, i.e.\ at $y, z = \pm
h$. This is generally not so of $(1,n)$ PPOs, which typically cease to exist
after a collision with the inner walls of the cylinders. 

\begin{figure}[htb]
  \centering 
  (a)
  \includegraphics[width=0.29\textwidth]{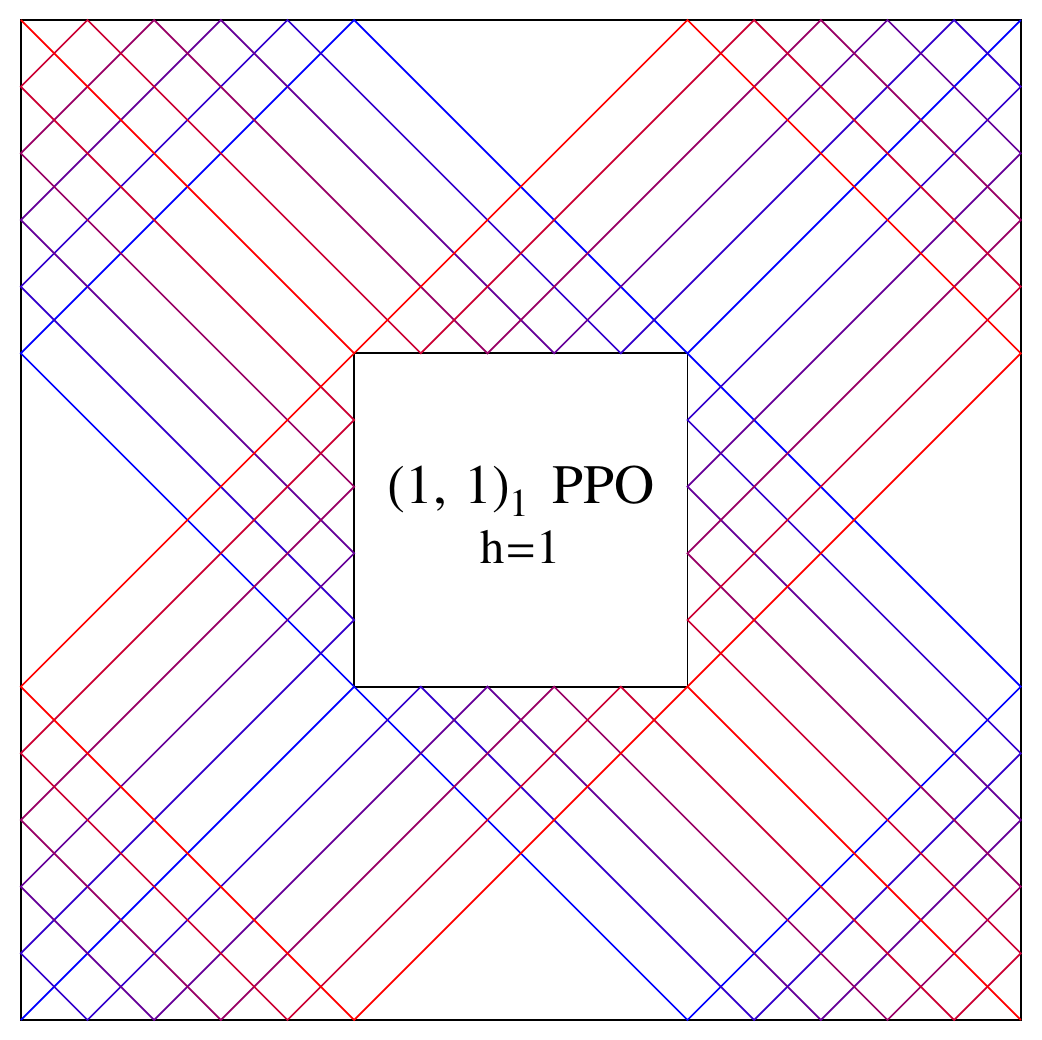}
  \hfill
  (b)
  \includegraphics[width=0.29\textwidth]{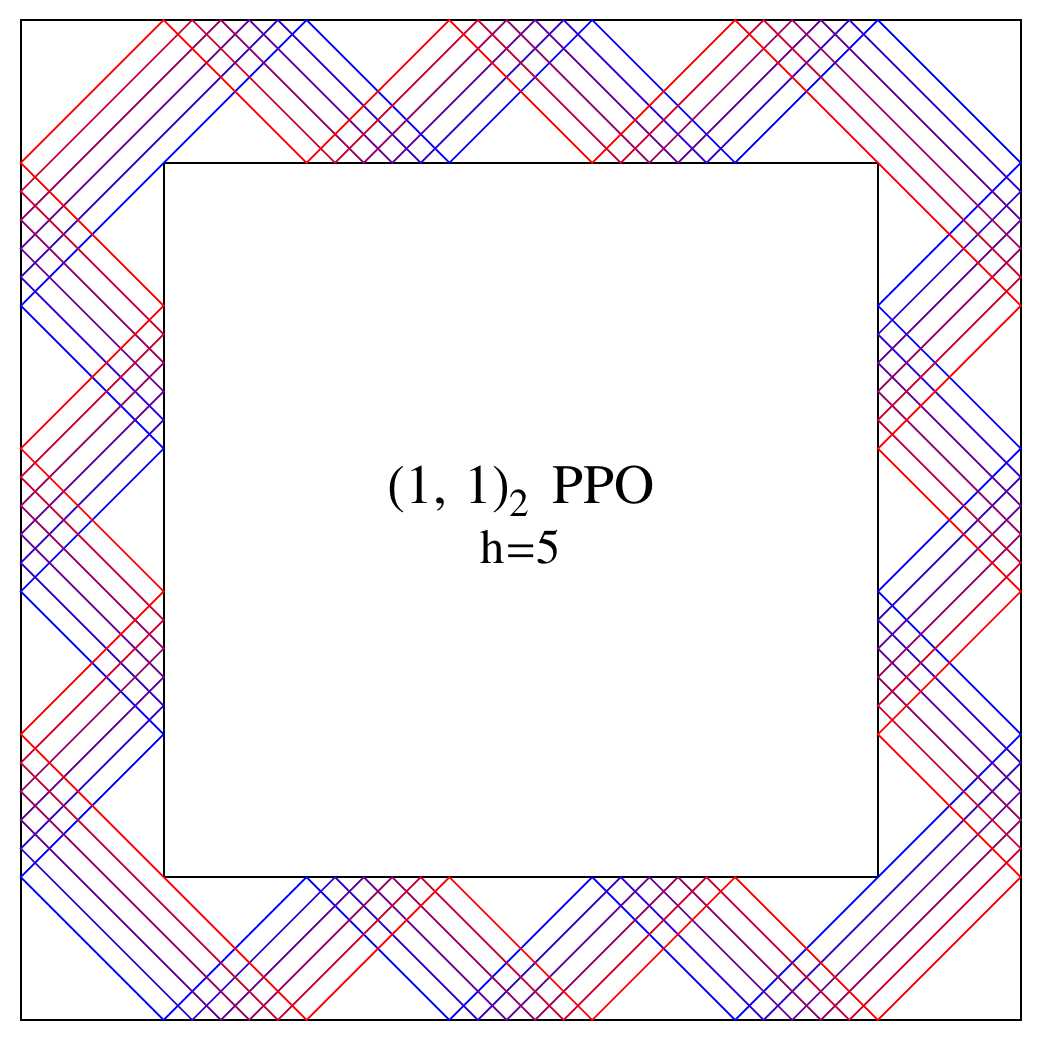}
  \hfill
  (c)
  \includegraphics[width=0.29\textwidth]{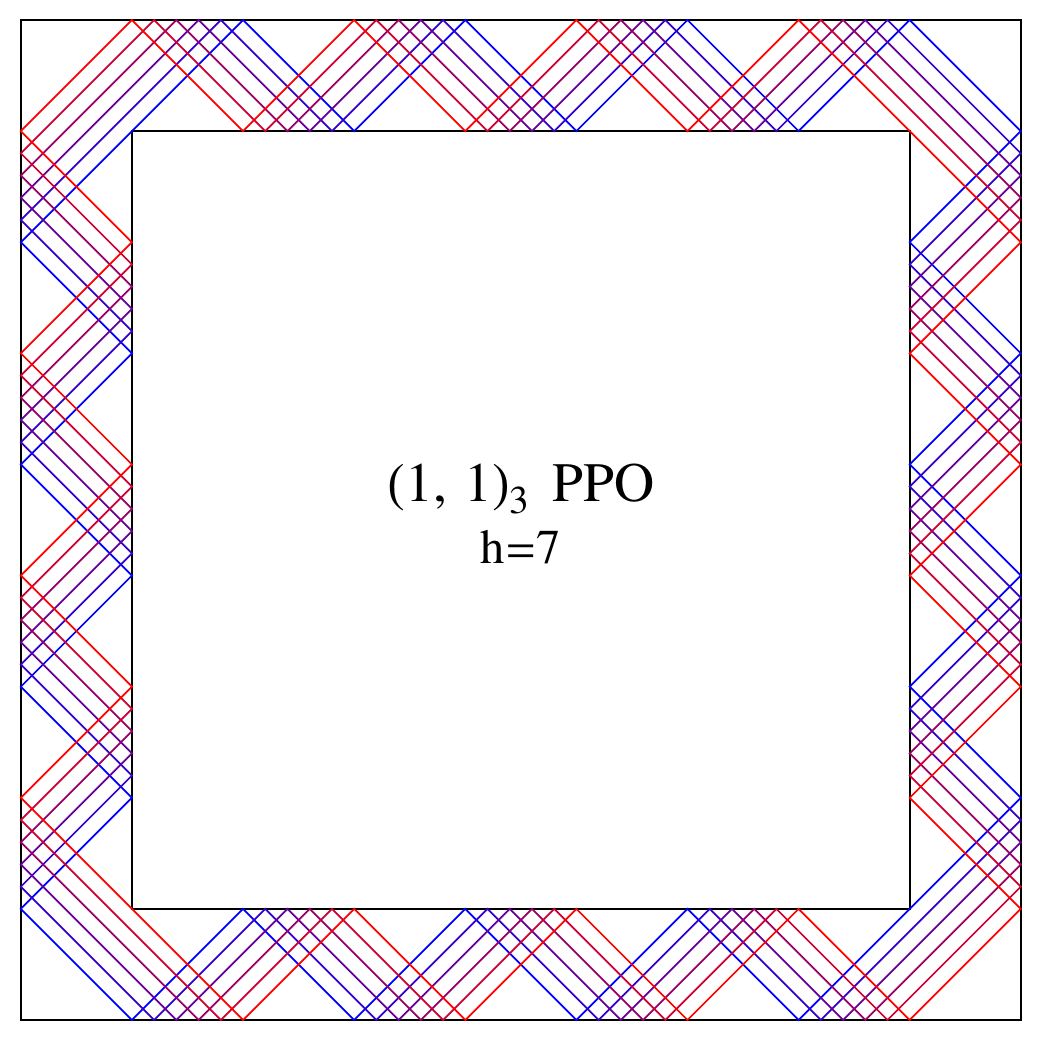}
  \caption{Families of $(1,1)_k$ PPOs shown in the plane of the billiard,
    (a) $k=1$, (b) $k=2$, (c) $k=3$.
    }
  \label{fig.pp1-1_k-families}
\end{figure}

We define the
$(1,1)_k$ planar periodic orbits to be those among such 
orbits which make exactly $k$ collisions with the inner walls of each
cylinder over a periodic cycle. Examples are shown in 
\fref{fig.pp1-1_k-families}. 

We identify these orbits by the initial coordinates
\begin{equation}
  \left(
    \begin{array}{c}
      x_0\\
      y_0\\
      z_0
    \end{array}
  \right)
  =
  \left(
    \begin{array}{c}
      0\\
      -2 - h\\
      \epsilon
    \end{array}
  \right), \quad k\,\mathrm{even}, \qquad
  =
  \left(
    \begin{array}{c}
      0\\
      - h\\
      \epsilon
    \end{array}
  \right), \quad k\,\mathrm{odd},
  \label{pp1-1_k-icpos}
\end{equation}
with the corresponding velocity 
\begin{equation}
  \left(
    \begin{array}{c}
      u_0\\
      v_0\\
      w_0
    \end{array}
  \right)
  =
  \left(
    \begin{array}{c}
      0\\
      -1/\sqrt{2}\\
      1/\sqrt{2}
    \end{array}
  \right), \quad k\,\mathrm{even}, \qquad
  =
  \left(
    \begin{array}{c}
      0\\
      1/\sqrt{2}\\
      1/\sqrt{2}
    \end{array}
  \right), \quad k\,\mathrm{odd}.
  \label{pp1-1_k-icvel}
\end{equation}

The values of the geometrical parameter, $h$, for which these orbits can be
observed are restricted to the interval $h_\mathrm{min} < h \leq
h_\mathrm{max}$, whose bounds correspond respectively to the value of $h$
for which the orbit hits the corners at the intersection of outer cylinders
walls (in which case the segments of the orbits have constant lengths), and
to the value of $h$ for which the longest segment grazes the corner at the 
intersection of the inner cylinder walls:
\numparts
\begin{eqnarray}
  h_\mathrm{min} &=& 2(k-1),\\    
  h_\mathrm{max} &=& 2(k+1). 
\end{eqnarray}
\endnumparts

Given a value of $h$ in this interval, the dynamical parameter, $\epsilon$,
which characterises the orbits of this family, is itself 
restricted to the interval $\epsilon_\mathrm{min} \leq \epsilon <
\epsilon_\mathrm{max}$, where 
\numparts
\begin{eqnarray}
  \epsilon_\mathrm{min} &=& \mathrm{max}[h - 2(k+1),
  2(k-1) - h],\\  
  \epsilon_\mathrm{max} &=& \mathrm{min}[2(k+1) - h,
  h - 2(k-1)].   
\end{eqnarray}
\endnumparts

Interestingly, the $(1,1)_{k+1}$
orbit at $h = h_\mathrm{min}$ is identical to the $(1,1)_{k-1}$
orbit at $h = h_\mathrm{max}$, except for eight segments which connect the
inner to the outer corners back to back.

\subsubsection{Stability \label{sec.pp1-1_kSta}}

The stability analysis of the $(1,1)_k$ PPOs proceeds along the lines of
sections \ref{sec.pp1-1_0Sta} and \ref{sec.pp1-1_0NLSta}. The linear map is
of the form \eref{pp1-n_0-linearmap} and the quadratic terms similar to
equations \eref{pp1-1_0-quadraticmapw}--\eref{pp1-1_0-quadraticmapz} and 
\eref{pp1-1_0-quadraticformw}--\eref{pp1-1_0-quadraticformz}. The
coefficients are, however, different and their expressions get more
complicated as $k$ increases. 

\begin{figure}[phtb]
  \centering
  (a)
  \includegraphics[width=0.4\textwidth]{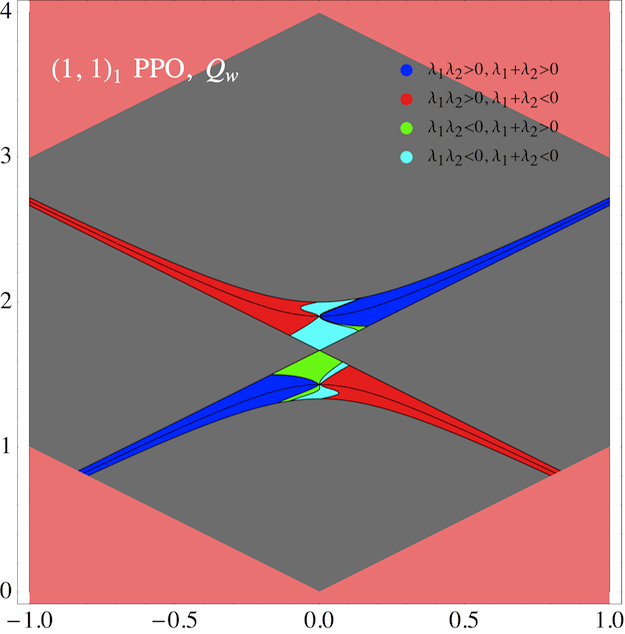}
  \hfill
  (b)
  \includegraphics[width=0.4\textwidth]{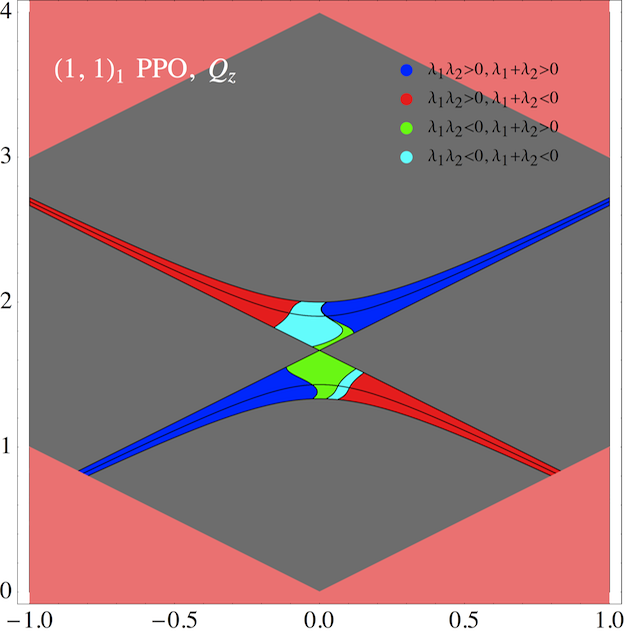}
  \\
  (c)
  \includegraphics[width=0.4\textwidth]{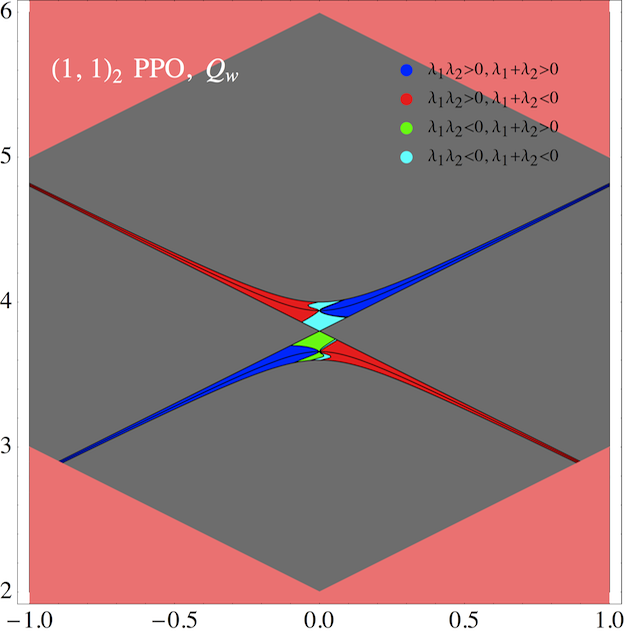}
  \hfill
  (d)
  \includegraphics[width=0.4\textwidth]{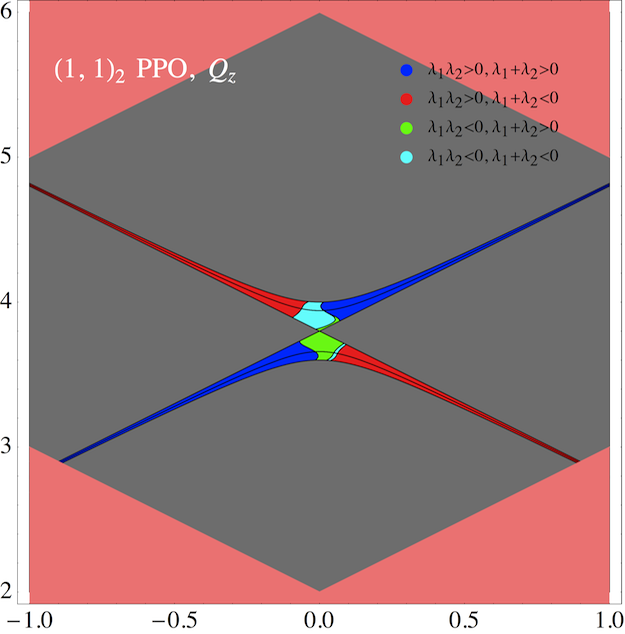}
  \\
  (e)
  \includegraphics[width=0.4\textwidth]{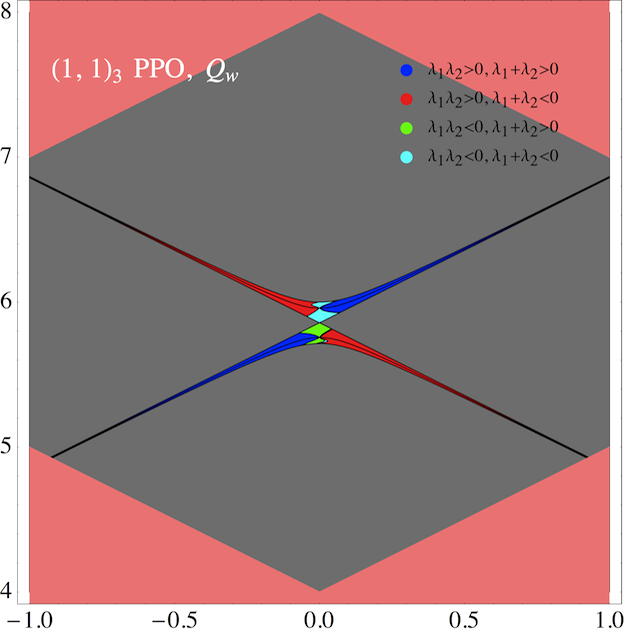}
  \hfill
  (f)
  \includegraphics[width=0.4\textwidth]{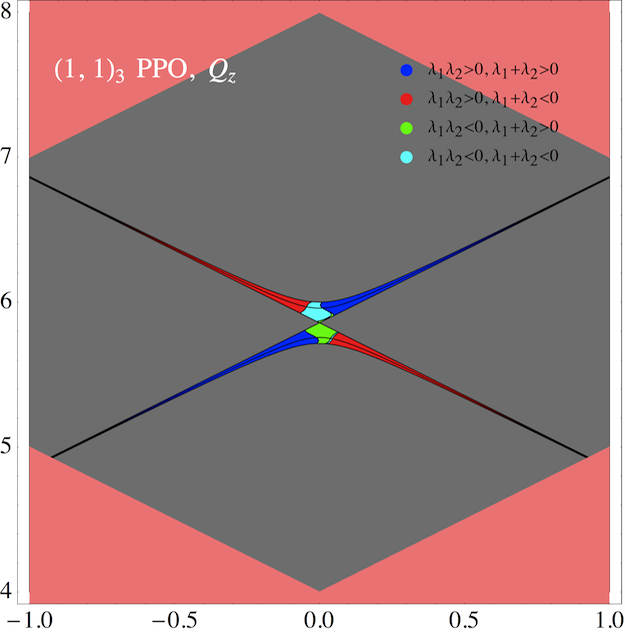}
  \caption{Results of the combined linear and nonlinear stability analysis
    of the $(1,1)_k$ PPOs, for k = 1,2,3. The conventions are the same as in
    \fref{fig.pp1-1_0_nlstab}.}
  \label{fig.pp1-1_k_nlstab}
\end{figure}

The results of this analysis are shown in \fref{fig.pp1-1_k_nlstab}. The
patterns are very similar to those displayed 
in \fref{fig.pp1-1_0_nlstab} for the $(1,1)_0$ PPOs,
with the conclusion that elliptic regions recur 
for every value of $k$ at geometrical parameter values of $h$ near $h =
2k$. To be more precise, the elliptic regions come in the shape of two
tongues, elongated along the lines $h =  2 k -1/(2k+1) \pm \epsilon$, which
are symmetric reflections of each other along the line $h = 2 k
-1/(2k+1)$. Moreover, the lower tongues are nonlinearly stable around
$\epsilon = 0$ in the interval $2 k - 2/(2k+1) \leq h <  2 k
-1/(2k+1)$ and the upper tongues unstable. 

\subsection{$(m,n)_0$- planar periodic orbits 
  \label{sec.pp1-n_0}}

\subsubsection{Existence \label{sec.pp1-n_0Exs}}

The $(m=1,n)_0$ planar periodic orbits can be identified by the phase point
with position coordinates $\qq_0$ identical to that of the $(1,1)_0$
orbits, equation \eref{pp1-1_0-ic}, and velocity
\begin{equation}
  \pp_0 = \frac{1}{\sqrt{1+n^2}}
  \left(0,1,n\right).
  \label{pp1-n_0-icvel}
\end{equation}

These orbits exist for geometric parameter values between the bounds
\numparts
\begin{eqnarray}
  h_\mathrm{min} &=& -2,\\    
  h_\mathrm{max} &=& \frac{2}{n}.
\end{eqnarray}
\endnumparts

Given a value of $h$ in this interval, the dynamical parameter, $\epsilon$,
is restricted, for $n$ odd, to the interval $\epsilon_\mathrm{min} \leq
\epsilon < \epsilon_\mathrm{max}$, where 
\numparts
\begin{eqnarray}
  \epsilon_\mathrm{min} &=& \mathrm{max}(n h - 2, - 2 - h),\\  
  \epsilon_\mathrm{max} &=& \mathrm{min}(2 - n h, h + 2),  
\end{eqnarray}
\endnumparts
and, for $n$ even and $h<0$, to the interval $\epsilon_\mathrm{min} \leq
\epsilon < \epsilon_\mathrm{max}$, where
\numparts
\begin{eqnarray}
  \epsilon_\mathrm{min} &=& -(2 + h), \\
  \epsilon_\mathrm{max} &=& 2 + h,
\end{eqnarray}
\endnumparts
or, for $h\geq0$, to the two intervals $\epsilon_\mathrm{min} \leq
|\epsilon| < \epsilon_\mathrm{max}$, where
\numparts
\begin{eqnarray}
  \epsilon_\mathrm{min} &=& (2n + 1)h,\\
  \epsilon_\mathrm{max} &=& 2 + h.
\end{eqnarray}
\endnumparts

\begin{figure}[phtb]
  \centering 
  (a)
  \includegraphics[width=0.29\textwidth]{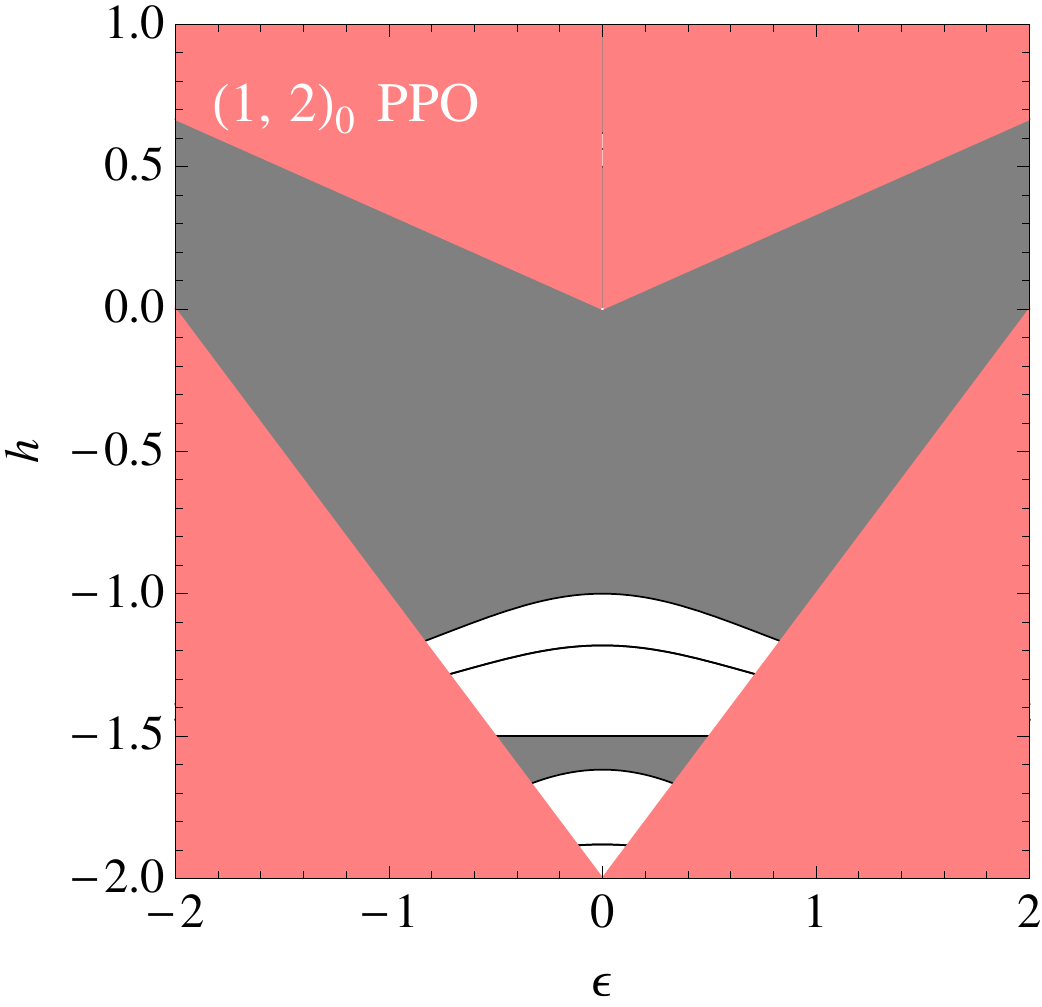}
  \hfill
  (b)
  \includegraphics[width=0.29\textwidth]{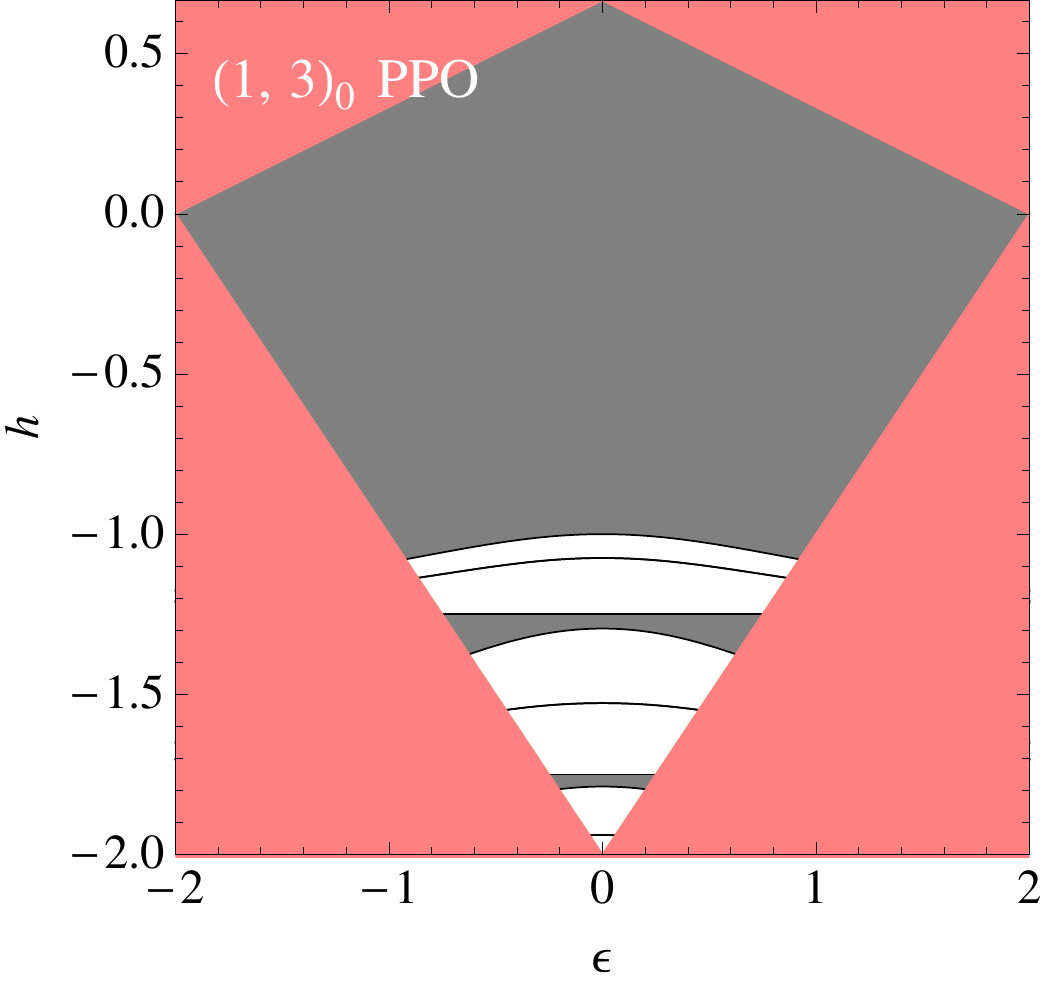}
  \hfill
  (c)
  \includegraphics[width=0.29\textwidth]{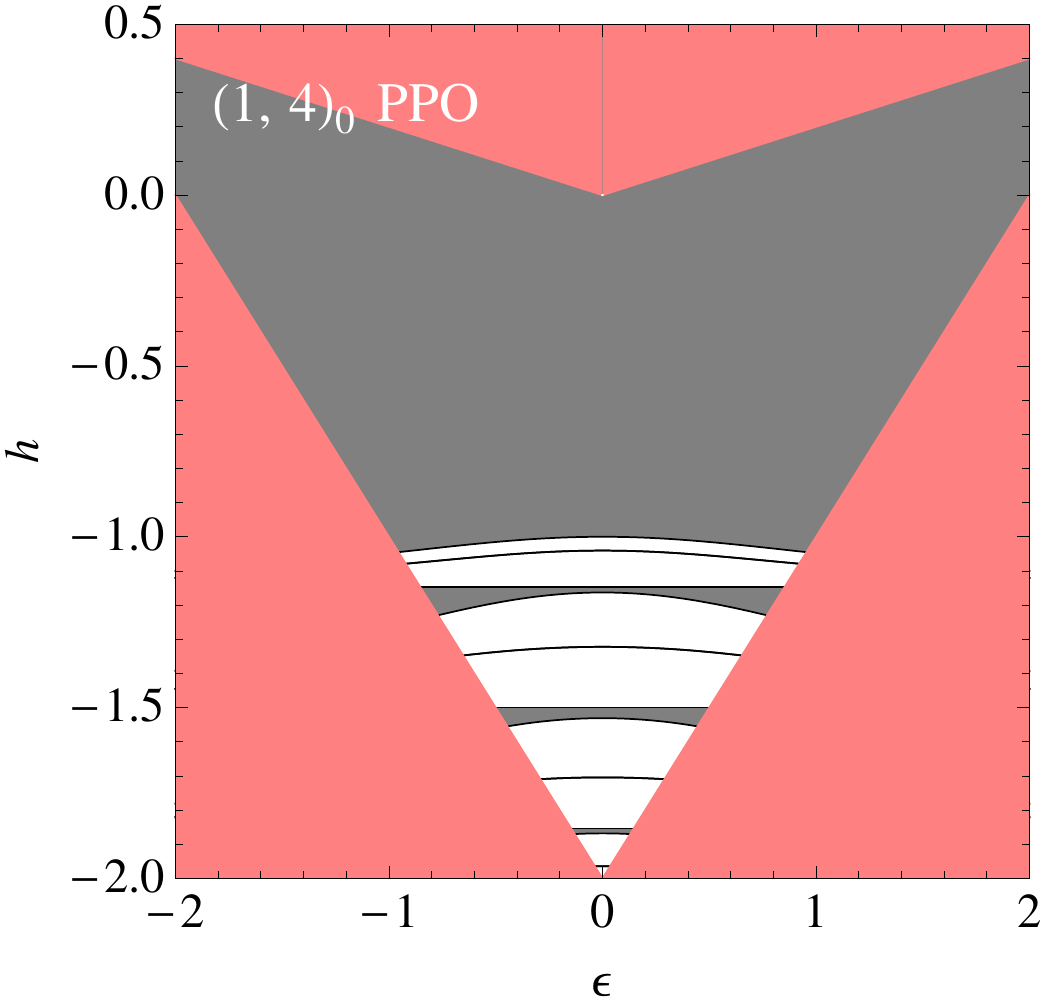}
  \\
  (d)
  \includegraphics[width=0.29\textwidth]{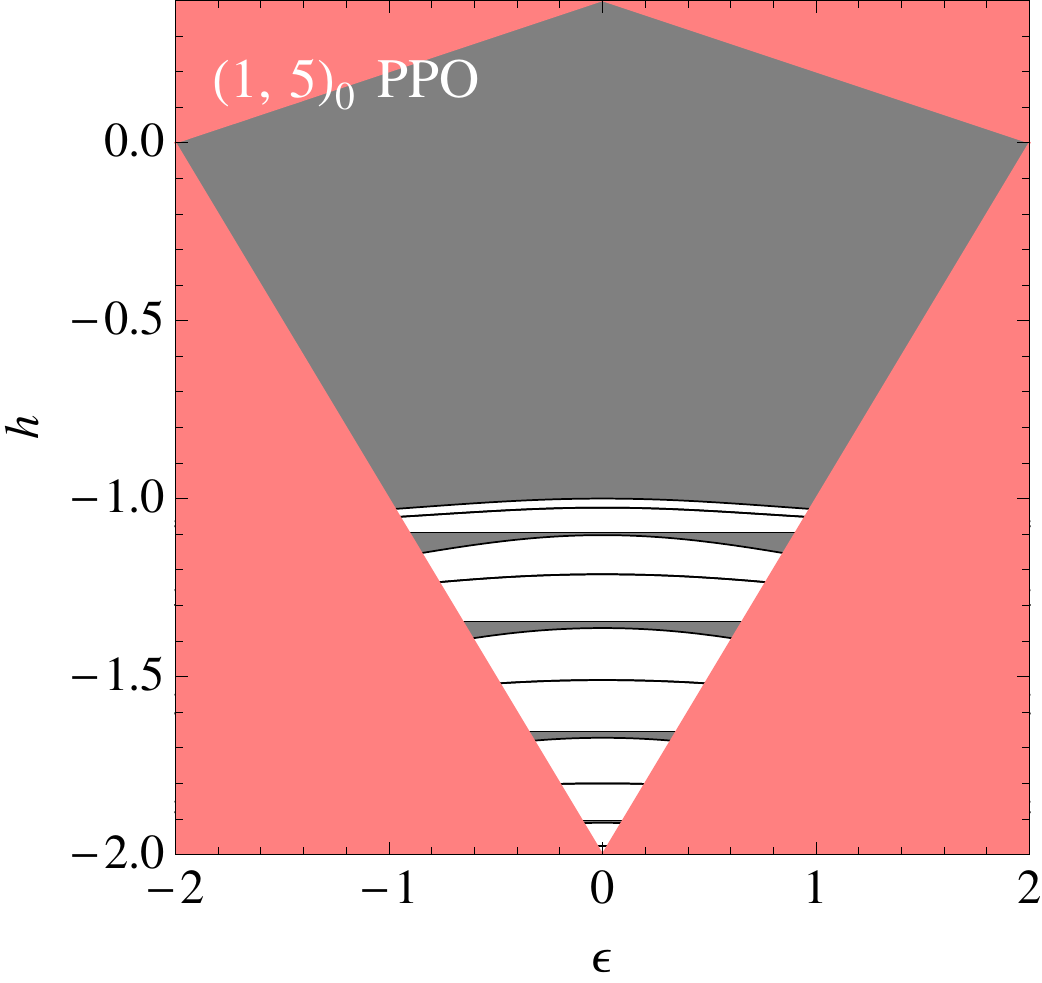}
  \hfill
  (e)
  \includegraphics[width=0.29\textwidth]{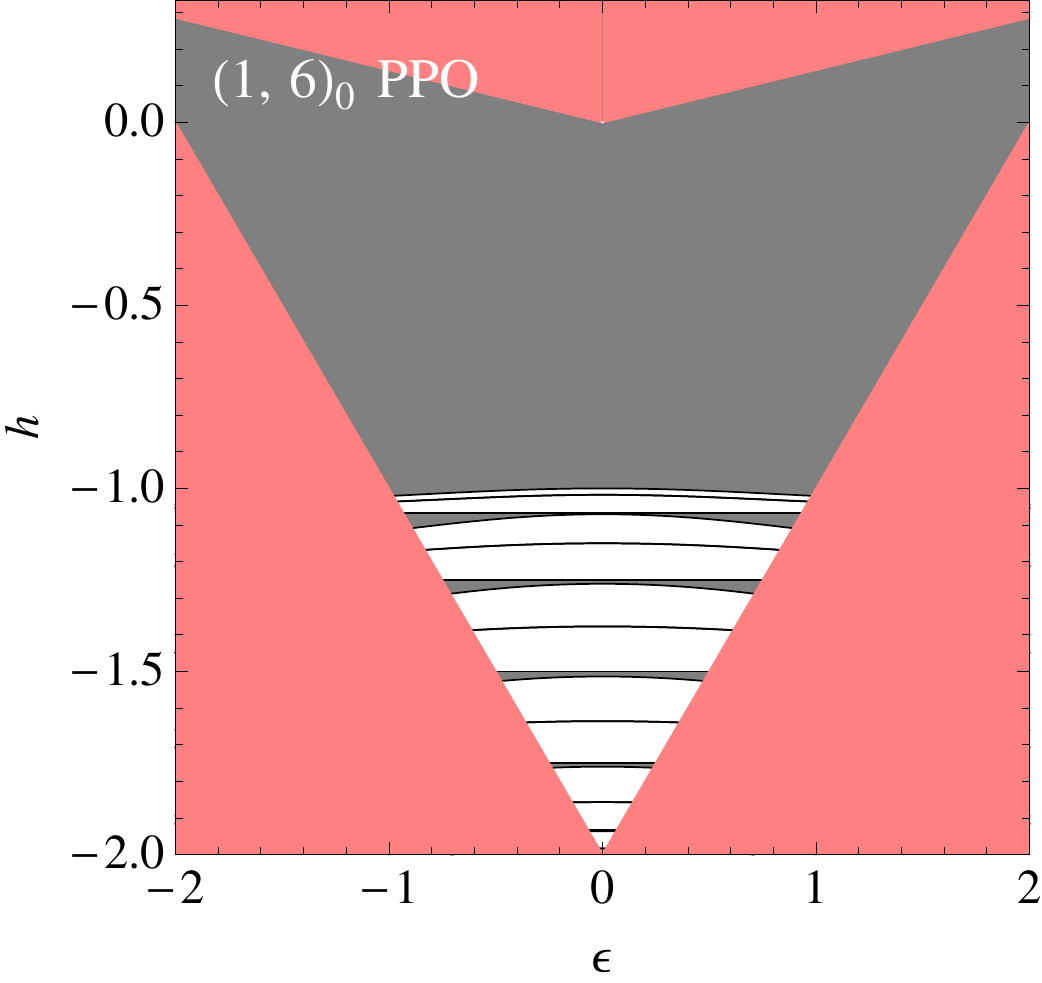}
  \hfill
  (f)
  \includegraphics[width=0.29\textwidth]{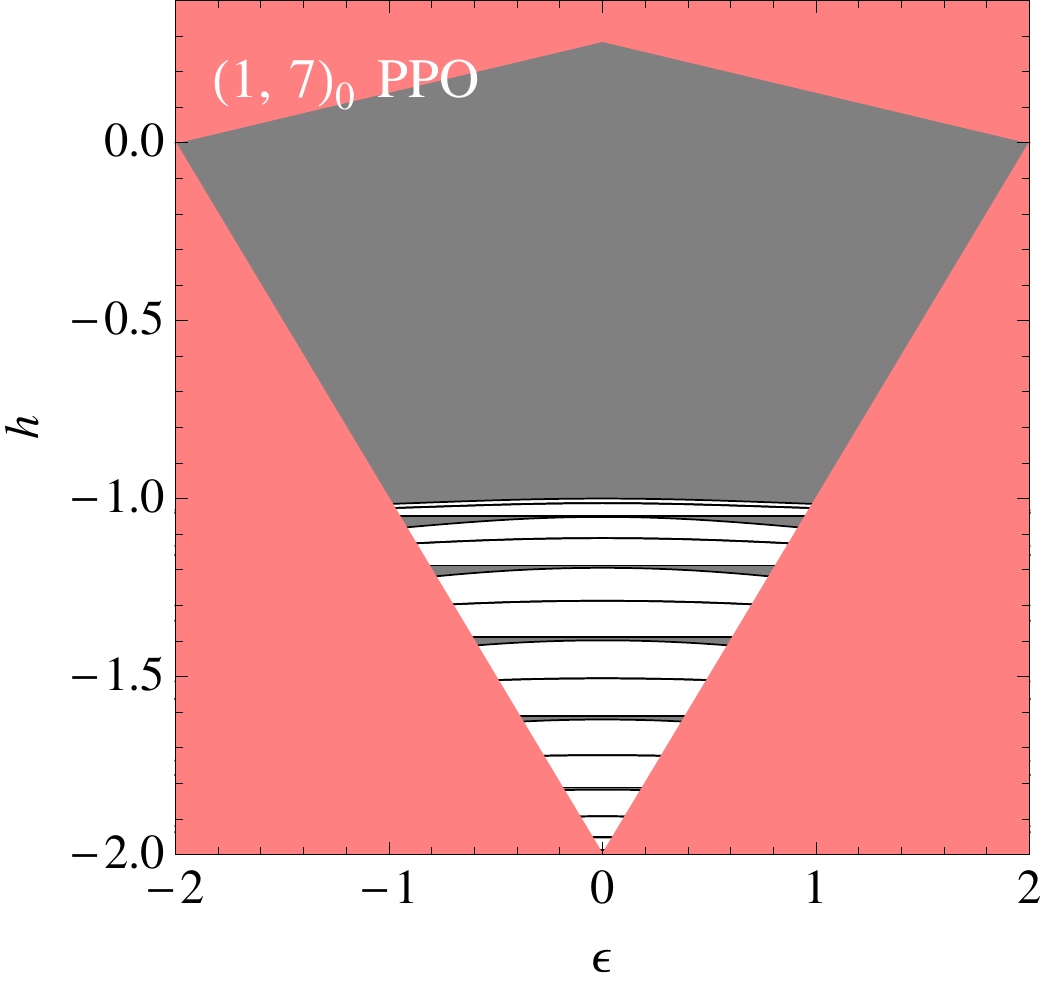}
  \caption{Results of the linear stability analysis of the $(1,n)_0$
    PPOs plotted in the parameter space. The conventions are identical to
   \fref{fig.pp1-0_0_stab}.}
  \label{fig.pp1-n_0_stab}
\end{figure}

\subsubsection{Stability \label{sec.pp1-n_0Sta}}

The linear stability analysis of the $(1,n)_0$ PPOs can be carried out
along the lines of the analysis of the $(1,0)_0$ PPOs described in 
\sref{sec.pp1-0_0Sta}. The result yields a linear map of the form
\eref{pp1-n_0-linearmap}, with coefficients, given, for the
$n=2$ case in \ref{app.1-2_0-PPO}, equations
\eref{pp1-2_0-linearmap11}--\eref{pp1-2_0-linearmap43}. The results are 
summarised in \fref{fig.pp1-n_0_stab}.

Regions of the parameter space corresponding to elliptic eigenvalues of the
linear form \eref{pp1-n_0-linearmap} are restricted to the $h < -1$
area. They are divided into $n$ strips, each extending over the whole
interval $-2-h <\epsilon < 2 + h$. These strips are all bounded from below
by a straight line and crossed in the middle by a line of parabolic
eigenvalues. Their upper bounds are rather complicated functions of the
parameters. 

We note that, as $n$ increases, the elliptic areas fill out the whole area
$-2 < h < -1$, $-2 - h < \epsilon < 2 + h$, resembling the pattern
observed for the $(1,0)_0$ PPOs shown in \fref{fig.pp1-0_0_stab}. This
is not surprising, since, as $n$ increases, the $(1,n)_0$ PPOs can be
viewed as perturbations in the plane of the billiard of the  the $(1,0)_0$
PPOs. 

\begin{figure}[phtb]
  \centering 
  (a)
  \includegraphics[width=0.4\textwidth]{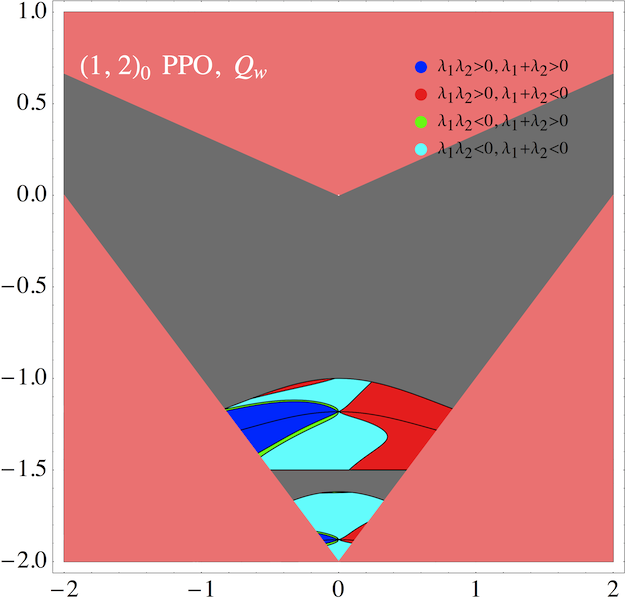}
  \hfill
  (b)
  \includegraphics[width=0.4\textwidth]{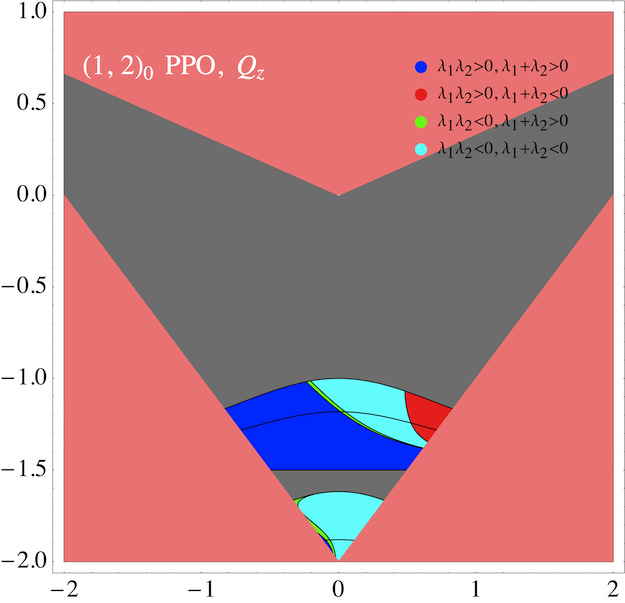}
  \\
  (c)
  \includegraphics[width=0.4\textwidth]{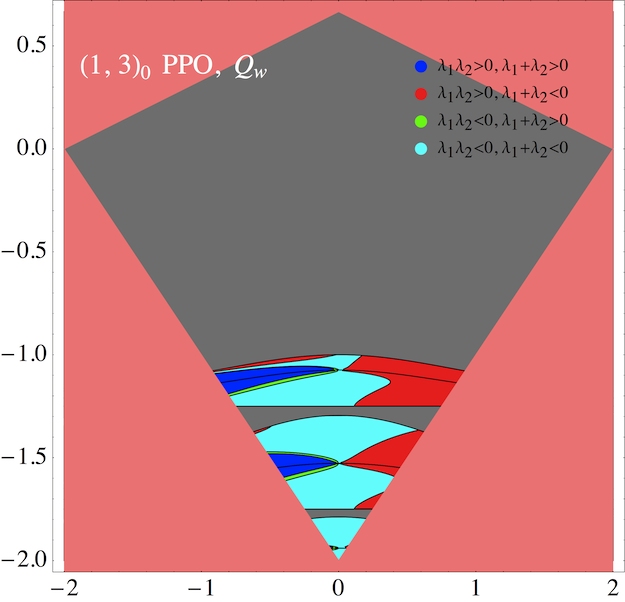}
  \hfill
  (d)
  \includegraphics[width=0.4\textwidth]{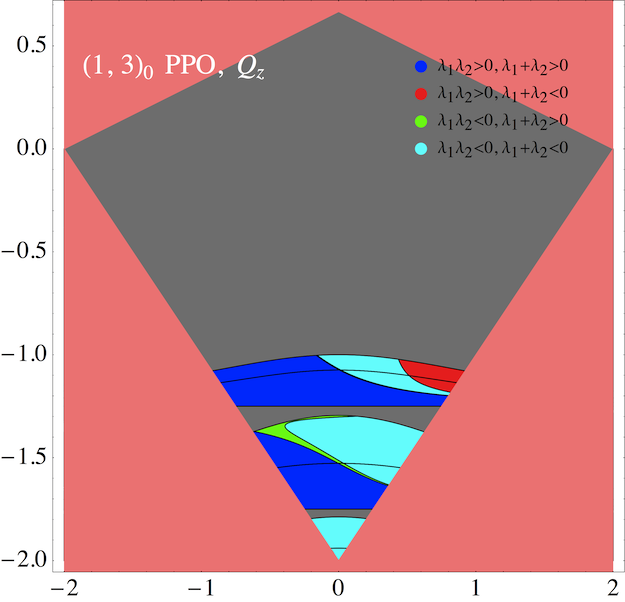}
  \\
  (e)
  \includegraphics[width=0.4\textwidth]{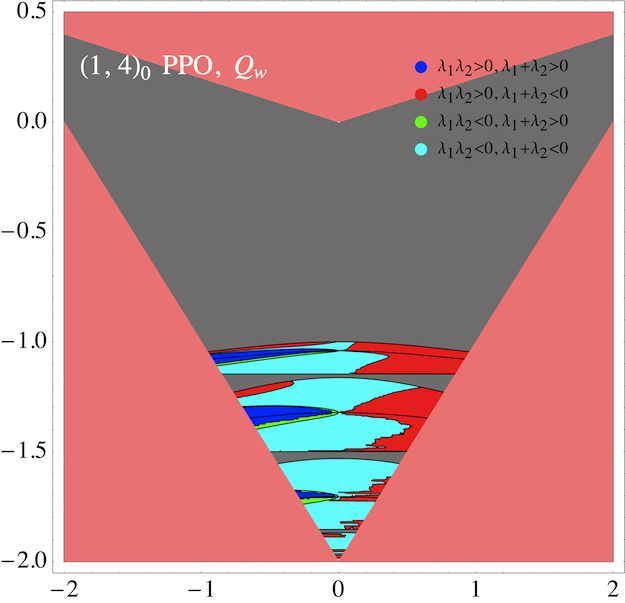}
  \hfill
  (f)
  \includegraphics[width=0.4\textwidth]{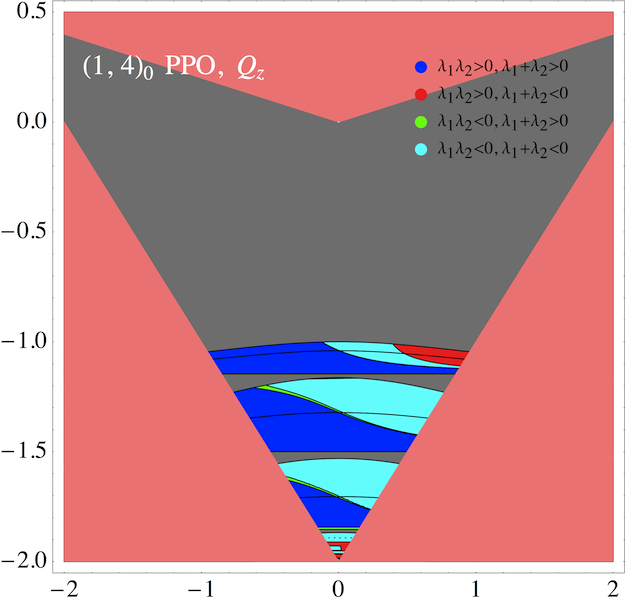}
  \caption{Results of the nonlinear stability analysis of the $(1,n)_0$
    PPOs, for n = 2,3,4, combined with those of the linear analysis shown
    in \fref{fig.pp1-n_0_nlstab}. The conventions are the same as in
    \fref{fig.pp1-1_0_nlstab}.}
  \label{fig.pp1-n_0_nlstab}
\end{figure}

The interesting feature of the $(1,n)_0$ PPOs is that, as opposed to the
$(1,0)_0$ PPOs, their nonlinear stability analysis is not trivial. As with
the linear stability analysis, we proceed in a way similar to 
\sref{sec.pp1-1_0NLSta} to obtain quadratic corrections to the linear
stability map \eref{pp1-0_0-linearmap}, similar to equations
\eref{pp1-1_0-quadraticmapw}--\eref{pp1-1_0-quadraticmapz} and
\eref{pp1-1_0-quadraticformw}--\eref{pp1-1_0-quadraticformz}. We provide in 
\sref{app.1-2_0-PPO} the explicit coefficients of the quadratic forms
associated to the $(1,2)_0$ PPOs, see equations 
\eref{pp1-2_0-quadraticformaw}--\eref{pp1-2_0-quadraticformcw} and
\eref{pp1-2_0-quadraticformaz}--\eref{pp1-2_0-quadraticformcz}. The
expressions coefficients of the quadratic forms associated to the $(1,n)_0$
PPOs, $n\geq 3$, will not be given explicitly.

The results of the nonlinear stability analysis are displayed in 
\fref{fig.pp1-n_0_nlstab} and suggest that every elliptic strip may give
rise to nonlinearly stable oscillations, at least over some range of values
of the geometric parameter $h$. It is also interesting to note that the
coefficients of the quadratic forms are all identically zero at the $n$
values of $h$ which correspond to the straight-line lower bounds of the
tongues of elliptic stability of these orbits.

\begin{figure}[htb]
  \centering 
  (a)
  \includegraphics[width=0.29\textwidth]{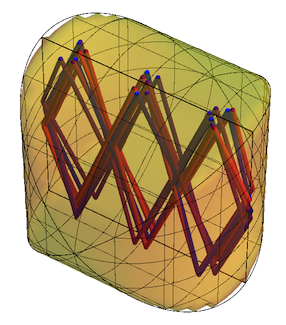}
  \hfill
  (b)
  \includegraphics[width=0.29\textwidth]{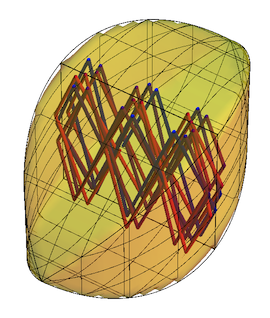}
  \hfill
  (c)
  \includegraphics[width=0.29\textwidth]{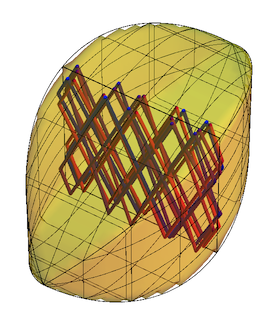}
  \caption{Examples of stable oscillations close to $(m,n)$-periodic orbits
    observed (a) $h = -1.1$ for a $(1,3)_0$ PPO and at $h = -1.5$ for (b)
    a $(1,3)$ PPO and (c) a $(2,5)_0$ PPO.} 
  \label{fig.pp1-n_traj}
\end{figure}
Examples of such stable oscillations are shown from actual trajectories in
\fref{fig.pp1-n_traj}, including the example of nonlinear
oscillations around a $(2,5)_0$ PPO. This orbit, as well as other $(m,n)_0$
PPOs, display features similar to those described above for the $(1,n)_0$
PPOs, and will we not go into further details.

\section{Helical orbits\label{sec.HPO}}

We now proceed to analyse orbits of helical shape, which wind around each
cylinder  in the squared billiard in turn. 

By symmetry, helical orbits of
any pitch may occur along the surfaces of infinitely long, straight
cylinders. When these orbits cross the intersection between two cylinders,
however, they typically lose focus and go astray as they hit a corner
between the two cylindrical surfaces. Nonetheless, there are isolated
points where the intersection of the surfaces of the two cylinders is
smooth, and  a helical orbit will thus be transmitted unscathed if it
passes through these specific points. One may further 
expect these orbits to generate stable oscillations in their immediate
vicinity. Indeed, so long as the orbits remain close enough to the surface
of the billiard, with a short separation between successive collisions, no
defocusing can take place. We emphasise that this
phenomenon is specific to higher-dimensional billiards with curved
surfaces. The problem of assessing the stability of such orbits is again
a fully nonlinear one which requires a suitable treatment.

In the case under consideration, where the cylinders' axes are in the $x=0$
plane and intersect at right angles, the points of smooth intersections
are the points $x = \pm 1$,  where the curvature with respect to the other
$y$ or $z$ coordinate vanishes. Thus a helical orbit which crosses over
from one cylinder to the other at $x = \pm 1$, $y = \pm(1 + h)$ and $z=y$,
or $z = -y$, is a smooth periodic orbit of the squared cylindrical
stadium. (The conditions must be verified at every intersection.) 
These are rather stringent conditions and impose restrictive symmetries on
the orbits. 

\begin{figure}[phbt]
  \centering
  (a) 
   \includegraphics[width=0.29\textwidth]{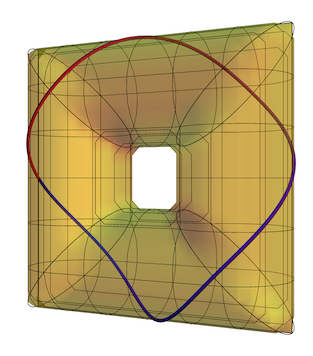}
  \hfill
  (b) 
   \includegraphics[width=0.29\textwidth]{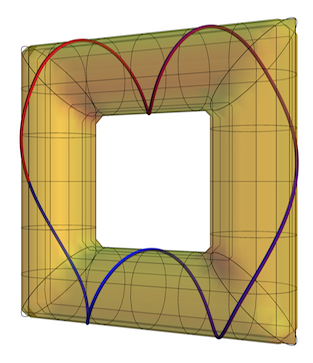}
  \hfill
  (c) 
   \includegraphics[width=0.29\textwidth]{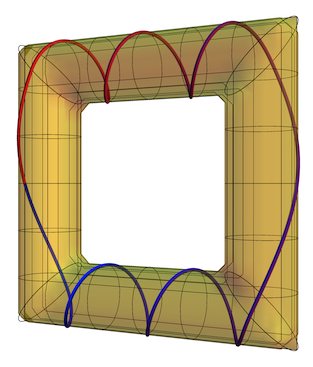}
  \\
  (d)
   \includegraphics[width=0.29\textwidth]{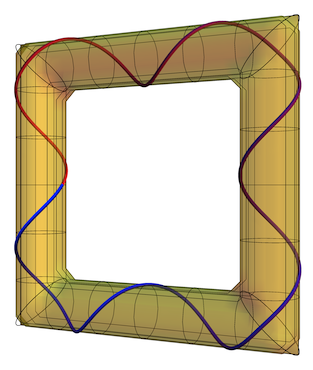}
  \hfill
  (e)
   \includegraphics[width=0.29\textwidth]{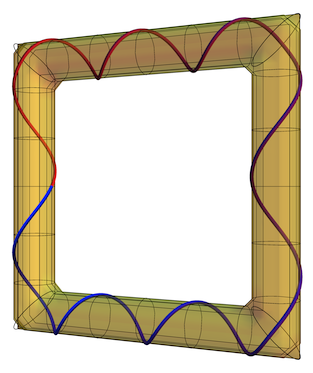}
  \hfill
  (f)
   \includegraphics[width=0.29\textwidth]{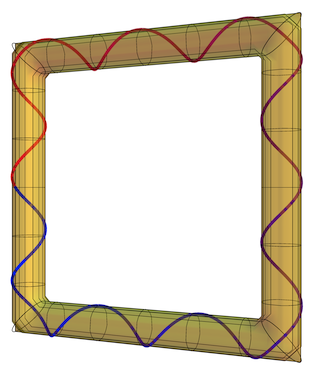}
  \caption{Periodic orbits approximating exact helical periodic orbits of
    the squared cylindrical stadia. They correspond respectively to: (a) a
    $(1,1)_{101}$-periodic orbit approximating the corresponding helix near
    $h = \pi/2 - 1$; (b) a $(1,3)_{101}$-HPO near $h =
    \sqrt{3}\pi/2 - 1$; (c) a $(1,5)_{101}$-HPO near
    $h = \sqrt{5}\pi/2 - 1$; (d) a $(3,3)_{101}$-HPO near $h = 3\pi/2 - 1$;
    (e) a $(3,5)_{101}$-HPO near
    $h = \sqrt{15}\pi/2 - 1$; and (f) a $(5,5)_{101}$-HPO near $h = 5\pi/2
    - 1$. The exact values of $h$ where the actual 
    $(n_1, n_2)_{k=101}$ periodic orbits can be found are provided in the 
    text. Regardless of the actual period, they occur in an interval of
    values of $h$ for symmetric cases such as (a), (d) and (f), but only at
    isolated values of $h$ for asymmetric cases such as (b), (c) and
    (e).} 
  \label{fig.helices}
\end{figure}

We characterise these helical periodic orbits in terms of the number of
times that they whirl around specific axes. Let $n_1$ and $n_2$ denote the
number of half periods the helices make about the cylinders along the $y$
and $z$ axes, respectively. By symmetry, $n_1$ and $n_2$ must be odd
integers, and it is easily found that the $(n_1,n_2)$-helix exists only
at $h = \sqrt{n_1 n_2} \pi/2 -1$, with initial condition in the $z=0$ plane
at $x=0$, $y = h$ if $(n_1-1)/2$ is even, $y = h + 2$ if odd, and with initial
velocity
components $\pm \sqrt{n_1/(n_1 + n_2)}$ along the $x$-axis and $\pm
\sqrt{n_2/(n_1 + n_2)}$ along the $z$-axis. Technically, these orbits are
of infinite period and are therefore impossible to realise. However, they give
rise to an infinite sequence of periodic orbits of finite periods which
approximate them. We will denote these approximate helices by
$(n_1,n_2)_k$-helical periodic orbits (HPO), meaning that 
each half period (of which there are $2(n_1 + n_2)$ along the orbit) is
made out of $k$ segments. Examples of such long periodic orbits are shown in
\fref{fig.helices}; see also related media
\footnote{\url{http://www.ulb.ac.be/~tgilbert/research/}
}.

An important point is that, irrespective of their stability,
finite-period approximations to helical periodic orbits can be moved around
in some neighborhood, so long as they pass through the intersection between
two cylinders away from their boundaries and do not hit a sharp 
angle. Finite-period approximations to a helical periodic orbit thus
belong to one-parameter continuous families of such orbits, indexed by the
dynamical parameter $\epsilon$ which we define below.

We now turn to the properties of these orbits and analyze their stability
for specific classes.

\subsection{$(1,1)_k$ helical periodic orbits
  \label{sec.1-1_k}}

\subsubsection{Existence\label{sec.1-1_kExs}}

Finite-length $(1,1)_k$ $(k\in\mathbb{N}$, $k \geq 2$) helical periodic
orbits which approximate the $(1,1)$ helix cross the plane intersecting
pairs of cylinders perpendicularly\footnote{This property is shared with other
  symmetrical $(n,n)_k$ HPOs.}. Their velocity at the intersections thus
has the 
form $(0, \frac{1}{\sqrt{2}}, \frac{1}{\sqrt{2}})$. The complete orbits can be
identified as going through the point with Cartesian
coordinates\footnote{Although we list only one, there are in total four
  such similar orbits, which are easily obtained from each other by
  symmetry. This property is shared by all helical periodic orbits of any
  length.}:
\begin{equation}
  \left(
    \begin{array}{c}
      x_0\\
      y_0\\
      z_0
    \end{array}
  \right)
  =
  \left(
    \begin{array}{c}
      \sin\frac{\pi}{2k}\\
      -1 - h -\cos\frac{\pi}{2k}\\
      \sin\frac{\pi}{2k} + \epsilon
    \end{array}
  \right), \quad k\,\mathrm{even}, \qquad
  =
  \left(
    \begin{array}{c}
      0\\
      -2 - h\\
      \epsilon
    \end{array}
  \right), \quad k\,\mathrm{odd},
  \label{hp1-1_k-icpos}
\end{equation}
with the corresponding velocity 
\begin{equation}
  \left(
    \begin{array}{c}
      u_0\\
      v_0\\
      w_0
    \end{array}
  \right)
  =
  \left(
    \begin{array}{c}
      \frac{1}{\sqrt{2}}\cos\frac{\pi}{k}\\
      \frac{1}{\sqrt{2}}\sin\frac{\pi}{k}\\
       \frac{1}{\sqrt{2}}
    \end{array}
  \right), \quad k\,\mathrm{even}, \qquad
  =
  \left(
    \begin{array}{c}
      \frac{1}{\sqrt{2}}\cos\frac{\pi}{2 k}\\
      \frac{1}{\sqrt{2}}\sin\frac{\pi}{2 k}\\
       \frac{1}{\sqrt{2}}
    \end{array}
  \right), \quad k\,\mathrm{odd}.
  \label{hp1-1_k-icvel}
\end{equation}

The values of the geometrical parameter, $h$, for which these orbits can be
observed are
restricted to the interval $h_\mathrm{min} \leq h < h_\mathrm{max}$, whose
bounds correspond, respectively, to the value of $h$ for which the lengths of
the segments of the orbit going through the intersections of the cylinders
shrink exactly to zero, and to the value of $h$ for which these segments
have their maximal length, grazing the surface of the cylinders at the
intersection, 
\numparts
\begin{eqnarray}
  h_\mathrm{min} &=& (k-2) \sin \frac{\pi}{2 k} -1,
  \label{hp1-1_k-hmin}
  \\
  h_\mathrm{max} &=& (k+2) \sin \frac{\pi}{2 k} -1.
  \label{hp1-1_k-hmax}
\end{eqnarray}
\endnumparts
Note that, for $k$ large, the width of this interval shrinks to zero
like $1/k$, with $\lim_k h_\mathrm{max} - h_\mathrm{min} \simeq 2\pi/k$. The
limiting value $\lim_k h_\mathrm{max} = \lim_k h_\mathrm{min}  = \pi/2 -1$
is the only value of the geometrical parameter for which the $(1,1)$
helix exists.

\begin{figure}[phbt]
  \centering
  (a) 
   \includegraphics[width=0.29\textwidth]{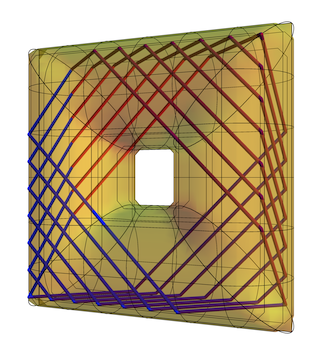}
  \hfill
  (b) 
   \includegraphics[width=0.29\textwidth]{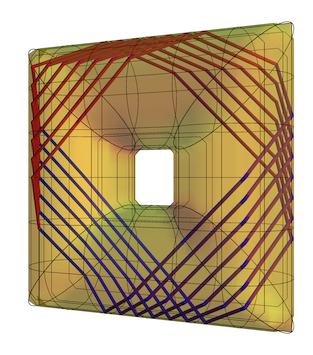}
  \hfill
  (c) 
   \includegraphics[width=0.29\textwidth]{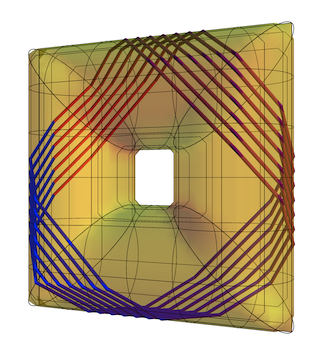}
  \\
  (d)
   \includegraphics[width=0.29\textwidth]{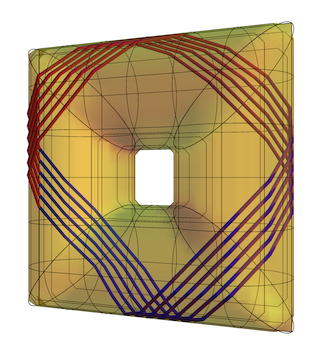}
  \hfill
  (e)
   \includegraphics[width=0.29\textwidth]{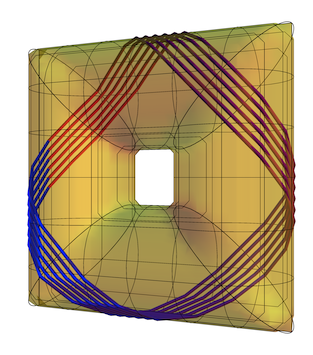}
  \hfill
  (f)
   \includegraphics[width=0.29\textwidth]{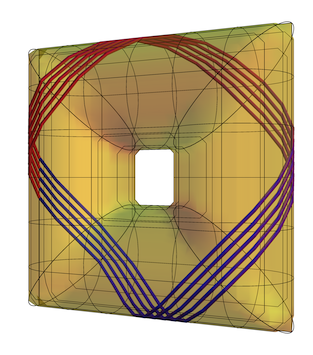}
  \\
  (g)
   \includegraphics[width=0.29\textwidth]{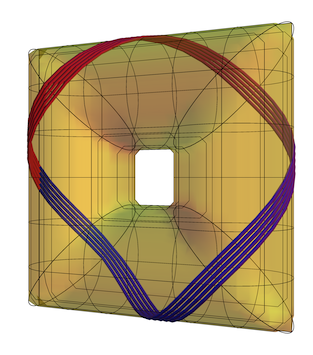}
  \hfill
  (h)
   \includegraphics[width=0.29\textwidth]{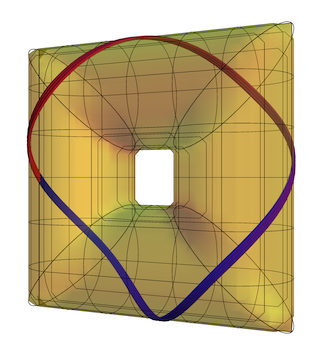}
  \hfill
  (i)
   \includegraphics[width=0.29\textwidth]{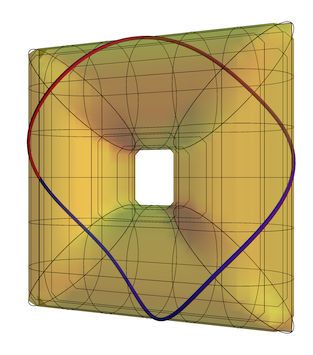}
  \caption{Families of $(1,1)_k$ HPOs found at $h = 0.5$. As opposed to
    the exact helix, which is unique, its finite-length approximations are
    parameterised by a continuous parameter in the 
    interval \eref{hp1-1_k-emin}--\eref{hp1-1_k-emax}. We show them for
    different values of 
    the parameter taken across the interval, for (a) $k=2$, (b) $k=3$,
    (c) $k=4$, (d) $k=5$, (e) $k=6$, (f) $k=7$, (g) $k=11$, (h) $k=23$, (i)
    $k=41$.}
  \label{fig.hp1-1_k-family}
\end{figure}

Given a value of $h$ in this interval, the
dynamical parameter, $\epsilon$, which characterises the orbits of this family,
is itself
restricted to the interval 
$\epsilon_\mathrm{min} \leq \epsilon < \epsilon_\mathrm{max}$, where
\numparts
\begin{eqnarray}
  \epsilon_\mathrm{min} &=& \mathrm{max}\left[
  - 1 - h + (k-2) \sin \frac{\pi}{2 k},
  1 + h - (k+2) \sin \frac{\pi}{2 k}\right],  
  \label{hp1-1_k-emin}
  \\
  \epsilon_\mathrm{max}  &=& \mathrm{min}\left[
  1 + h - (k-2) \sin \frac{\pi}{2 k},
  -1 - h + (k+2) \sin \frac{\pi}{2 k}\right].
  \label{hp1-1_k-emax}
\end{eqnarray}
\endnumparts
As with $h$, these bounds correspond to the values of $\epsilon$ for which
the periodic orbit either hits a corner at the intersection between two 
cylinders, or, the other way around, collides with the surface of the
cylinders in the middle of its longest segment connecting the orbit between
two transverse cylinders. Examples are displayed in \fref{fig.hp1-1_k-family}. 

As opposed to the parameter $h$ which fixes the geometry of the 
billiard, the parameter $\epsilon$ is a dynamical one: it changes as
perturbations of the periodic orbit are introduced. Just as was the case
with the planar periodic orbits, displacements of the helical periodic
orbit along the $z$ axes as the trajectory 
returns periodically close to its initial condition \eref{hp1-1_k-icpos}
can be interpreted as changes in the value of the parameter $\epsilon$; 
the trajectory comes back closer to another periodic orbit than the one it
was perturbed away from. In particular the trajectory moves away from this
family of orbits as soon as the value of $\epsilon$ exceeds either of its
two bounds, $\epsilon_\mathrm{min}$  or $\epsilon_\mathrm{max}$, since it
ceases to exist beyond them.

\subsubsection{Stability \label{sec.1-1_kSta}}

\paragraph{Birkhoff coordinates.}
To analyze the stability of the $(1,1)_k$ orbits, we change coordinates to
the Birkhoff coordinates \eref{pp1-0_0-theta}--\eref{pp1-0_0-w}, here given
by 
\numparts
\begin{eqnarray}
  \theta_0 &=& \arctan\frac{y_0 + 1 + h}{x_0} = 
  \left\{
    \begin{array}{l@{\quad}l}
      - \frac{\pi}{2},&k\,\mathrm{odd},\\
      - \frac{\pi}{2} + \frac{\pi}{2k},&k\,\mathrm{even},
    \end{array}
  \right.
  \label{hp1-1_k-theta}\\
  \xi_0 &=& \sin \left(\arctan\frac{v_0}{u_0} - \theta_0\right)
  = \sin\frac{(k-1)\pi}{2k},
  \label{hp1-1_k-xi}\\
  z_0 &=& 
  \left\{
    \begin{array}{l@{\quad}l}
      \sin\frac{\pi}{2k} + \epsilon,&k\,\mathrm{odd},\\
      \epsilon,&k\,\mathrm{even},
    \end{array}
  \right.
  \label{hp1-1_k-z}\\
  w_0 &=& \frac{1}{\sqrt 2}.
  \label{hp1-1_k-w}
\end{eqnarray}
\endnumparts

\paragraph{Linear stability analysis.}
Introducing perturbations $\delta\theta$, $\delta\xi$, $\delta z$ and
$\delta w$, the linear stability analysis yields the symplectic map
\begin{equation}
  \left(
    \begin{array}{c}
      \delta\theta\\
      \delta\xi\\
      \delta w\\
      \delta z
    \end{array}
  \right)
  \mapsto
  \left(
    \begin{array}{cccc}
      m_{\theta\theta}&m_{\theta\xi}&0&0\\
      m_{\xi\theta}&m_{\xi\xi}&0&0\\
      0&0&1&0\\
      m_{z\theta}&m_{z\xi}&m_{zw}&1
    \end{array}
  \right)
  \left(
    \begin{array}{c}
      \delta\theta\\
      \delta\xi\\
      \delta w\\
      \delta z
    \end{array}
  \right).
  \label{hp1-1_k-linearmap}
\end{equation}
This form is similar to that of the planar periodic orbits,
\eref{pp1-n_0-linearmap}, but with extra non-zero coefficients
$m_{z\theta}$ and $m_{z\xi}$. Its eigenvalues remain similar: there are two
unit eigenvalues associated to the motion in the $z$--$w$ plane, and two
nontrivial eigenvalues, $\eta_1$ and $\eta_2$, associated to the motion in
the $\theta$--$\xi$ plane. These two eigenvalues can be either 
hyperbolic, when they are real, $\eta_1 = \eta_2^{-1}$, or elliptic, when
they have a non-zero imaginary part, $\eta_1 = \eta_2^*$. The matrix
elements $m_{\alpha\beta}$ are generally complicated functions of the two
parameters $h$ and $\epsilon$.

\begin{figure}[tbh]
  \centering
   \includegraphics[width=0.45\textwidth]{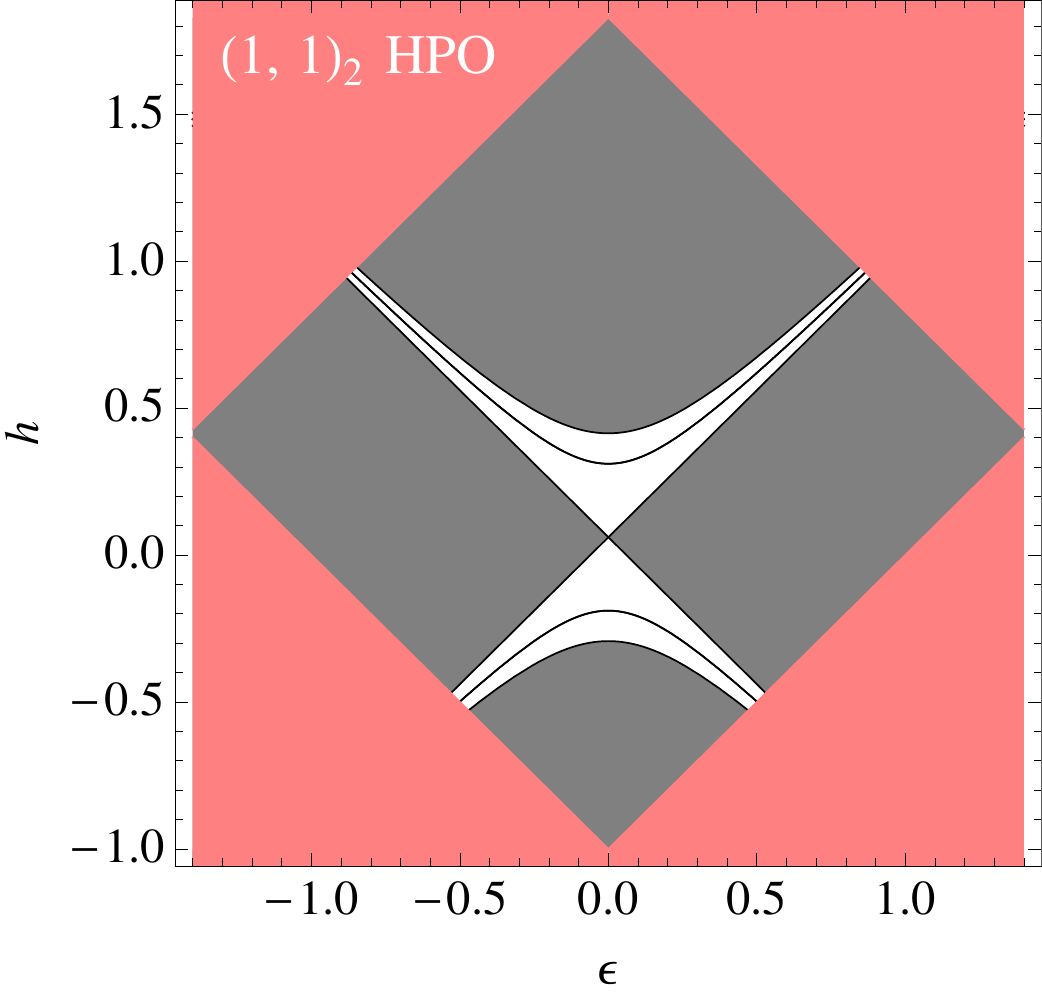}
  \hfill
   \includegraphics[width=0.45\textwidth]{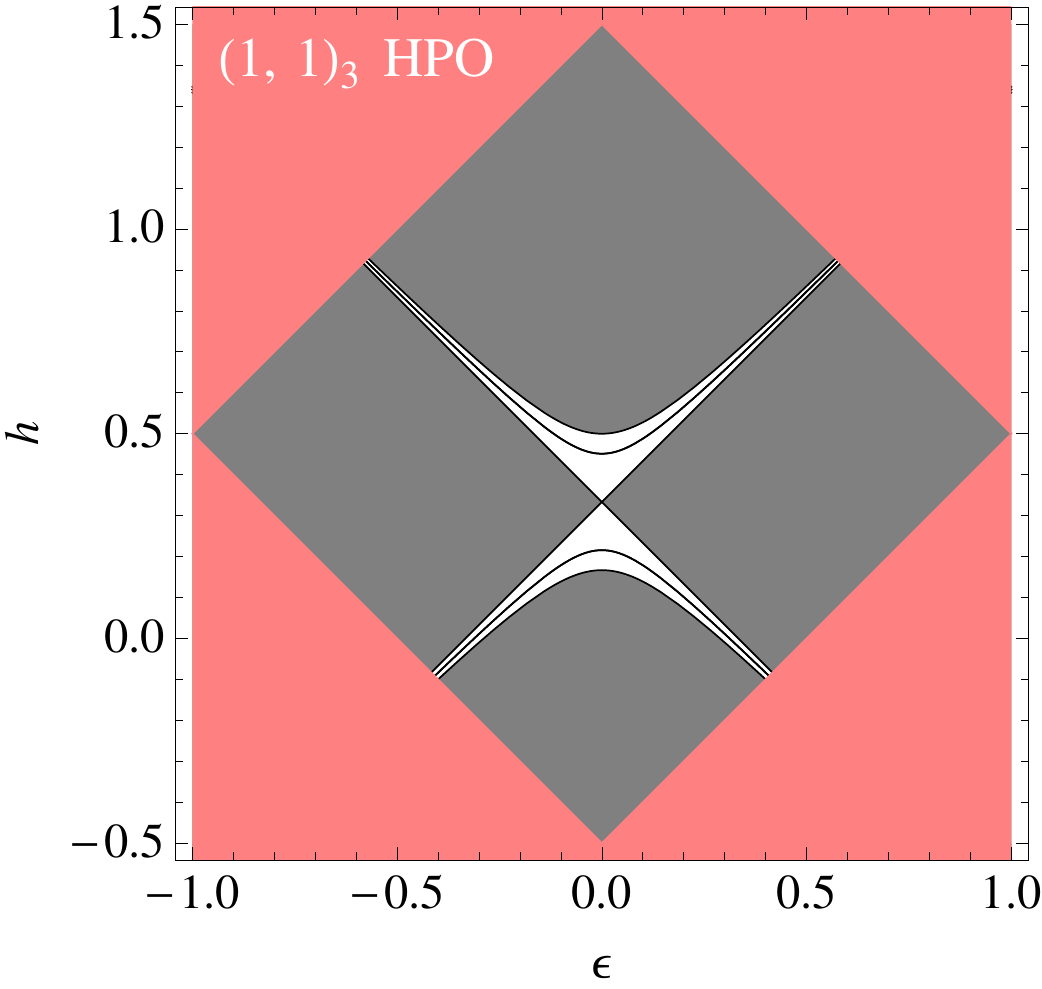}
  \caption{Results of the linear stability analysis of the $(1,1)_2$ and
    $(1,1)_3$ HPOs. The light red areas represent the regions outside the range
    $[\epsilon_\mathrm{min}, \epsilon_\mathrm{max}]$, defined in equation
    \eref{hp1-1_k-emin}--\eref{hp1-1_k-emax}. The grey areas correspond to
    regions where the 
    eigenvalues of the map \eref{hp1-1_k-linearmap} are hyperbolic, and 
    white where they are elliptic. The lines which appear in the elliptic
    regions are lines of parabolic eigenvalues.}
  \label{fig.hp1-1_stab}
\end{figure}

The cases $k=2$ and $k=3$ are the simplest and yield results shown in
\fref{fig.hp1-1_stab}, exhibiting features rather similar to the $(1,1)_k$
PPOs, which are common to all $(1,1)_k$ HPOs. (Explicit values of the
coefficients $m_{\alpha\beta}$ in the linear form \eref{hp1-1_k-linearmap}
for these two cases are provided in \ref{app.HPO}.) Namely,
there are regions in the $\epsilon$--$h$ parameter plane where $\eta_1$ and
$\eta_2$ are elliptic, which are in the shape of two symmetric tongues. 
For $k=2$ the symmetry line corresponds to $h = 3\sqrt{2}/4-1$, and to $h =
1/3$ for $k=3$. This value keeps growing for larger values of $k$ towards
$h = \pi/2-1$. 

Here, as with the $(1,1)_k$ PPOs, we see that the coefficient $m_{zw}$ in
the linear form \eref{hp1-1_k-linearmap} depends only on $h$ -- see 
equations~\eref{hp1-1_2-mzw}--\eref{hp1-1_3-mzw} in \sref{app.HPO}. Repeating the
arguments given in \sref{sec.pp1-1_0Sta}, we would conclude that all these
periodic orbits are linearly unstable because of the existence of a
bifurcation from elliptic to hyperbolic regimes in the range of dynamical
parameter values the orbit would visit as it translates further away from
its initial position under the action of $\delta w$. 

\paragraph{Nonlinear stability analysis.}
There are, however, again nonlinear effects which may affect this scenario, due to
terms which are quadratic in the small perturbations. The picture here is
actually slightly simpler than with the planar periodic orbits. Indeed,
since the $\theta$, $\xi$ and $z$ variables pick up nontrivial
perturbations at the order of linear contributions, it is enough to perform
the quadratic expansion for the $w$ axis alone. This yields the quadratic
map
\begin{equation}
  \delta w\mapsto \delta w + Q_w(\delta \theta, \delta\xi),
  \label{hp1-1_k-quadraticmap}
\end{equation}
where $Q_w(\delta \theta, \delta\xi)$ is a quadratic form, given by
\begin{equation}
  Q_w(\delta \theta, \delta\xi) = 
  \Big(\delta \theta\quad\delta\xi\Big)
  \left(
    \begin{array}{cc}
      a_w & \frac{1}{2}b_w\\
      \frac{1}{2}b_w & c_w
    \end{array}
  \right)
  \left(
    \begin{array}{c}
      \delta\theta \\
      \delta\xi
    \end{array}
  \right),
  \label{hp1-1_k-quadraticform}
\end{equation}
similar to \eref{pp1-1_0-quadraticformw}. The coefficients $a_w$, $b_w$ and
$c_w$ are here again functions of the two parameters $h$ and $\epsilon$.

\begin{figure}[tbh]
  \centering
   \includegraphics[width=0.45\textwidth]{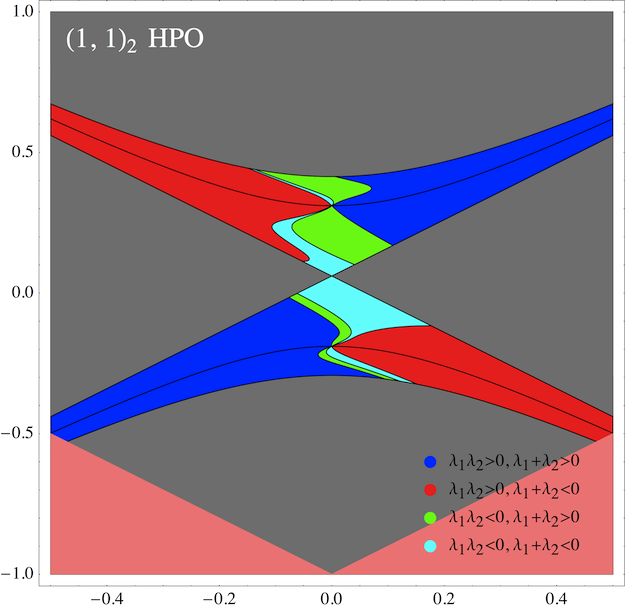}
  \hfill
   \includegraphics[width=0.45\textwidth]{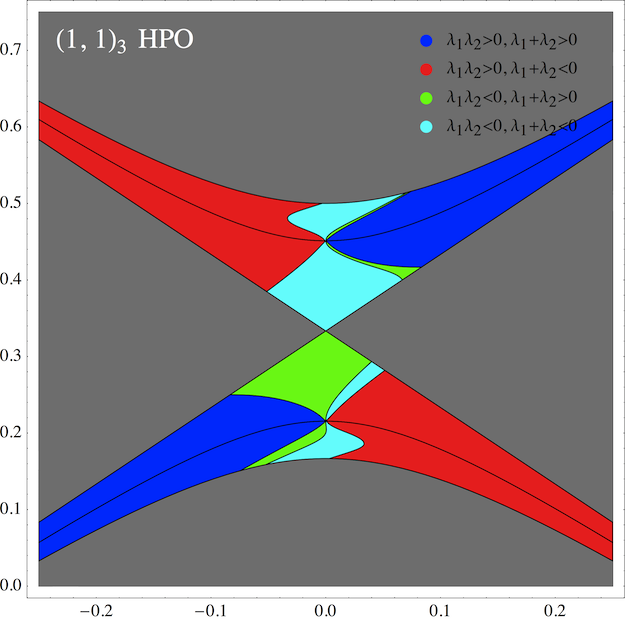}
  \caption{Nonlinear stability analysis of the $(1,1)_2$ and $(1,1)_3$
    HPOs in terms of the eigenvalue spectrum of the quadratic form
    \eref{hp1-1_k-quadraticform}. The conventions are the same as in 
    \fref{fig.pp1-1_0_nlstab}.}
  \label{fig.hp1-1_nlstab}
\end{figure}

Recapitulating the arguments given in \sref{sec.pp1-1_0NLSta}, the sign of
the quadratic form \eref{hp1-1_k-quadraticform} is determined by that of
the eigenvalues of its matrix. When the eigenvalues are of the same sign,
$Q_w$ will be positive when they are positive and negative when they are
negative. If, on the other hand, the eigenvalues are of opposite signs,
$Q_w$ will oscillate, but, on average, will take the sign of the eigenvalue
that  has the largest magnitude. Stable nonlinear oscillations can take
place in the parameter regions where the eigenvalue spectrum of $Q_w$
changes sign so as to confine the small oscillations to an interval of the
dynamical parameter where the eigenvalues of the linear form
\eref{hp1-1_k-linearmap} are elliptic.

\Fref{fig.hp1-1_nlstab} shows, for the $k=2$ and $k=3$ HPO
families, the eigenvalue spectrum of the matrix of the quadratic form
\eref{hp1-1_k-quadraticform}, superimposed on the elliptic regions found
from the linear stability analysis in \fref{fig.hp1-1_stab}. The
coefficients of these quadratic forms are given in \ref{app.HPO}. The color
convention is the same as the one used in \sref{sec.pp1-1_0NLSta}, in terms
of the sign and magnitude of their eigenvalues, which are denoted by
$\lambda_1$ and $\lambda_2$. We observe that the elliptic regions below and
above the lines at $h = 3\sqrt{2}/4-1$ for $k=2$ and $h = 1/3$ for $k=3$
display clearly different 
patterns. Whereas the upper elliptic regions are everywhere unstable under
the quadratic perturbations, the lower elliptic regions contain stable
regions around $\epsilon = 0$ and $h$ above  $h = -1 + 1 /\sqrt 2$ for
$k=2$ and $h = 1/6$ for $k=3$ (these bounds correspond to the lowest point
of intersection between the elliptic and hyperbolic regions at $\epsilon =
0$). Though the lower bound we read from this analysis is sharp, the upper
bound appears perhaps less so. More precise results might be obtained from
a more detailed analysis of the eigenspectrum of $Q_w$, but this will not
concern us here. Numerical results, such as shown below, support the claim
that the upper bound is actually sharp too. The main result here is the
existence of a region of parameter space where the helical periodic orbits
under consideration are nonlinearly stable. 

\begin{figure}[phtb]
  \centering
   \includegraphics[width = .45\textwidth]{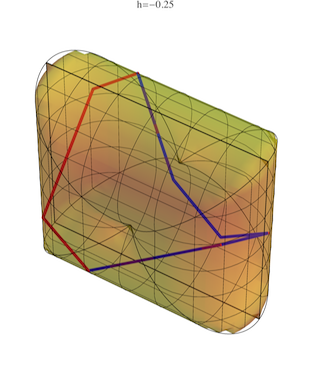}
  \hfill
   \includegraphics[width = .45\textwidth]{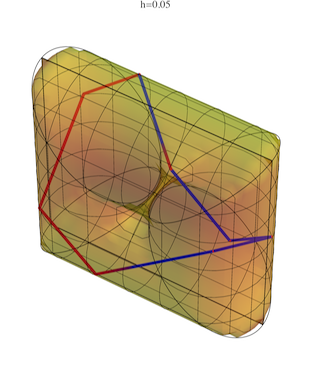}
   \includegraphics[width = .45\textwidth]{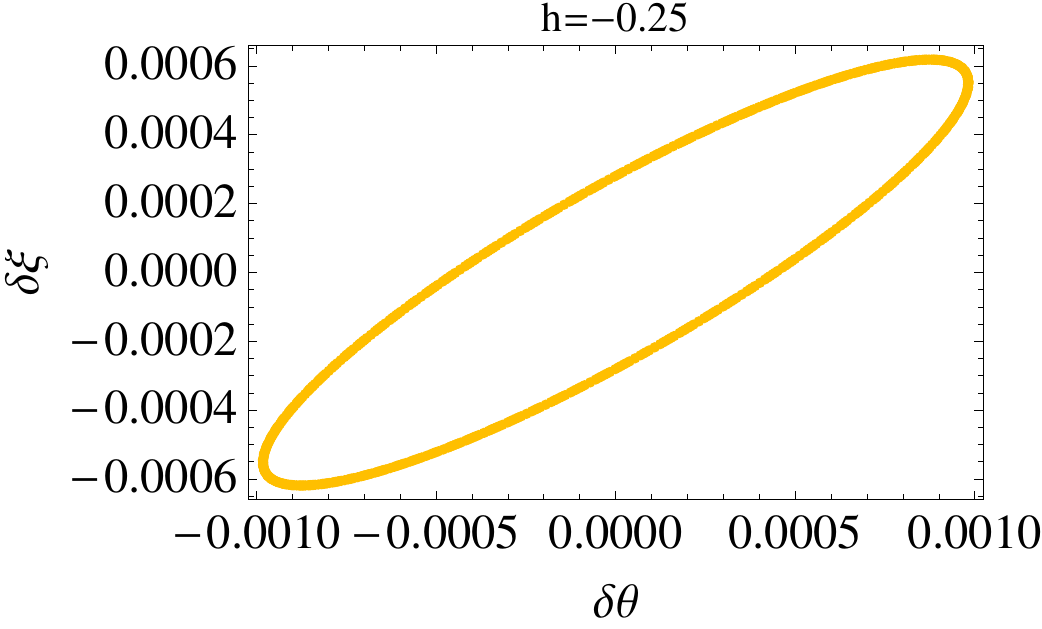}
  \hfill
   \includegraphics[width = .45\textwidth]{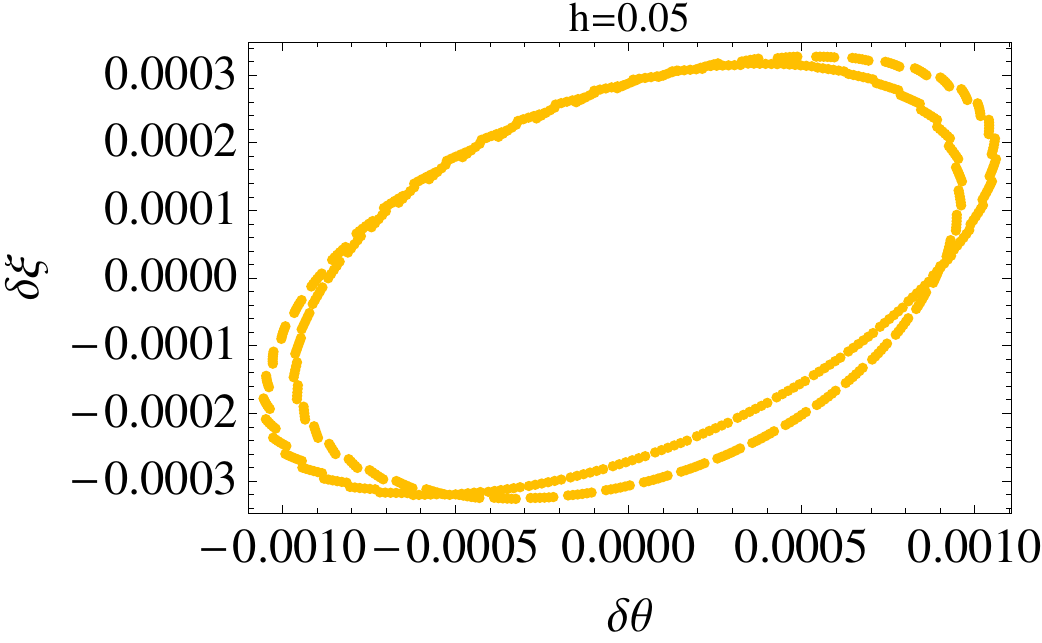}
   \includegraphics[width = .45\textwidth]{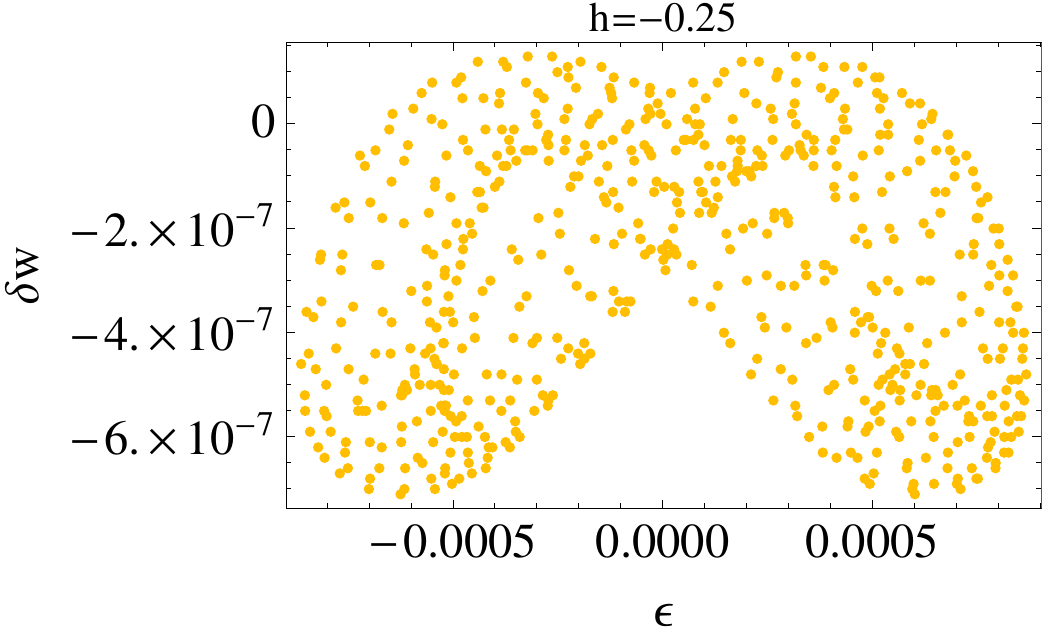}
  \hfill
   \includegraphics[width = .45\textwidth]{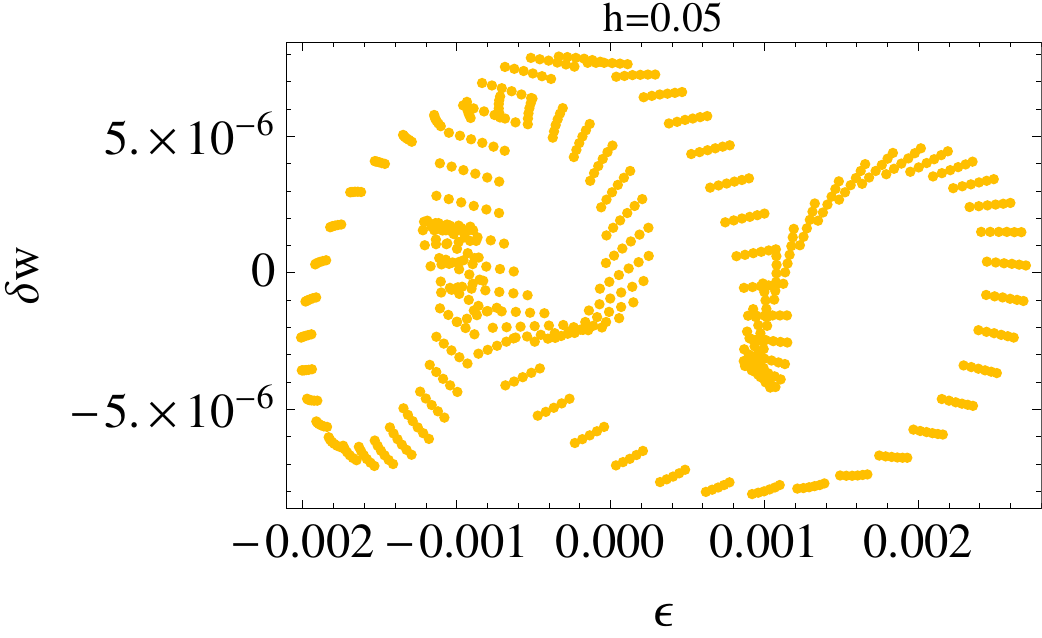}
  \caption{Nonlinearly stable oscillations of the $(1,1)_2$ HPOs measured
    from actual trajectories at $h = -0.25$ (left column) and $h = +0.05$
    (right column). The oscillations measured in the $\theta$--$\xi$
    (resp. $w$--$z$) plane are shown on the second (resp. third)
    line. Notice the order of magnitude of the $epsilon$ oscillations as
    opposed to those of $\delta w$.}
  \label{fig.hp1-1_2_traj}
\end{figure}

The stability can be checked numerically by perturbing actual trajectories
about these periodic orbits, as shown in \fref{fig.hp1-1_2_traj}.

\begin{figure}[phtb]
  \centering
   \includegraphics[width=0.45\textwidth]{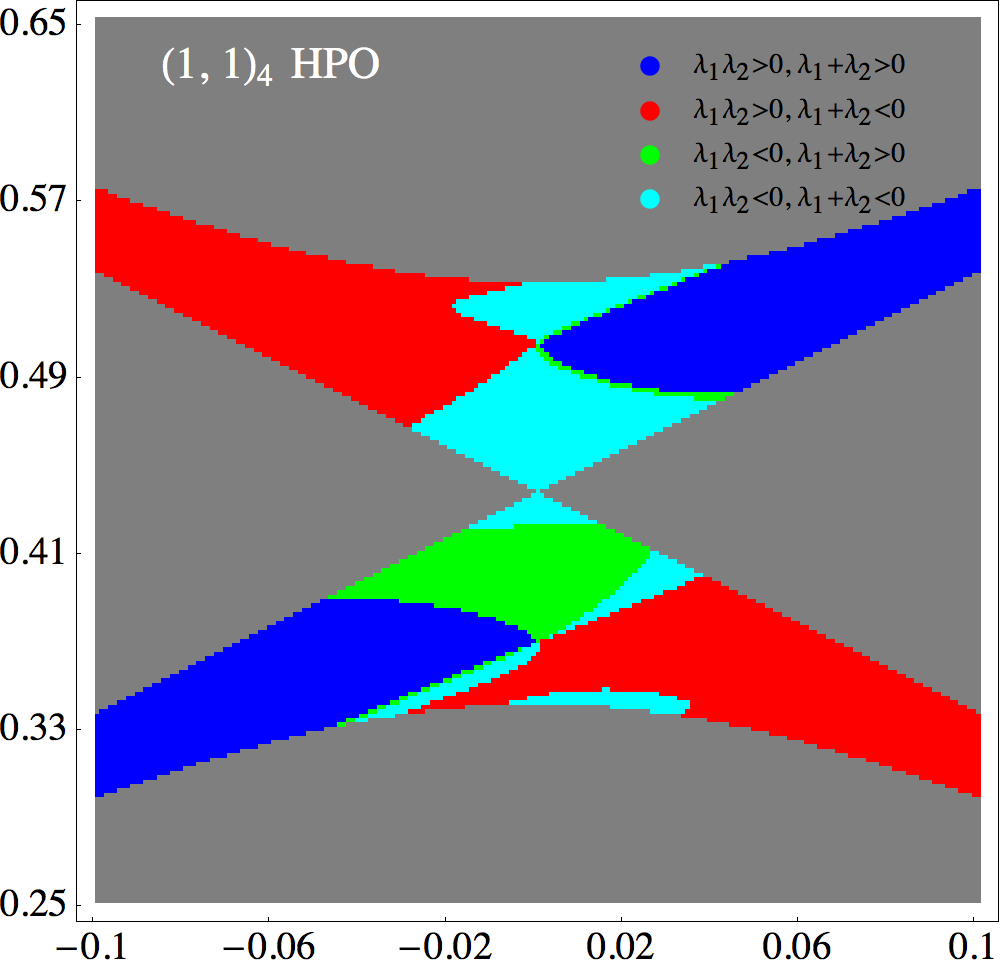}
  \hfill
   \includegraphics[width=0.45\textwidth]{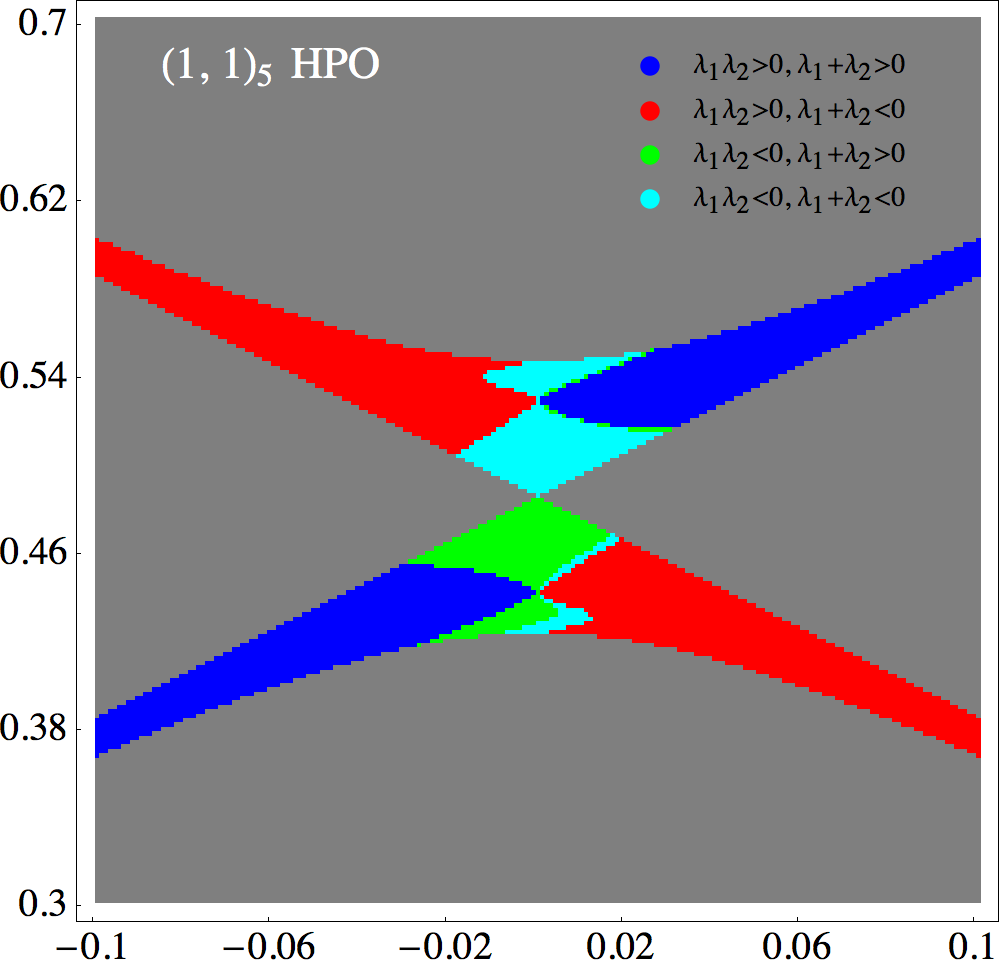}
   \includegraphics[width=0.45\textwidth]{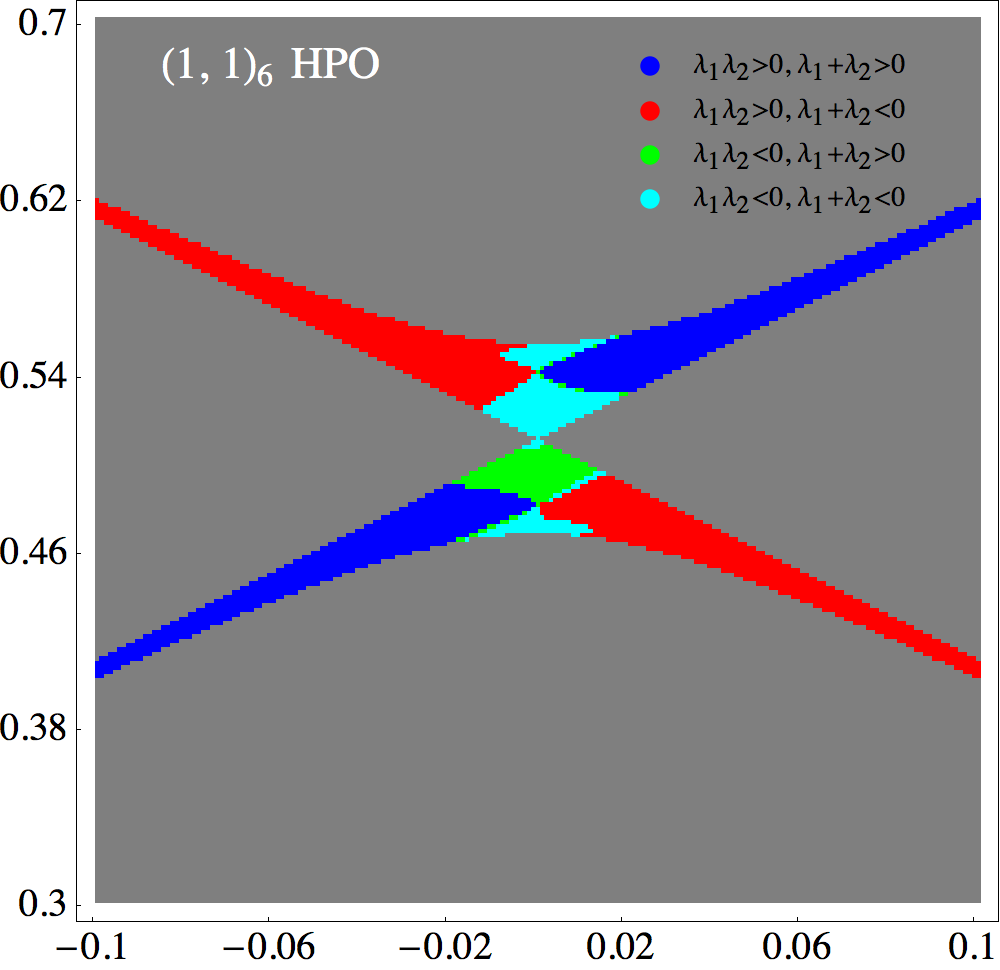}
  \hfill
  \includegraphics[width=0.45\textwidth]{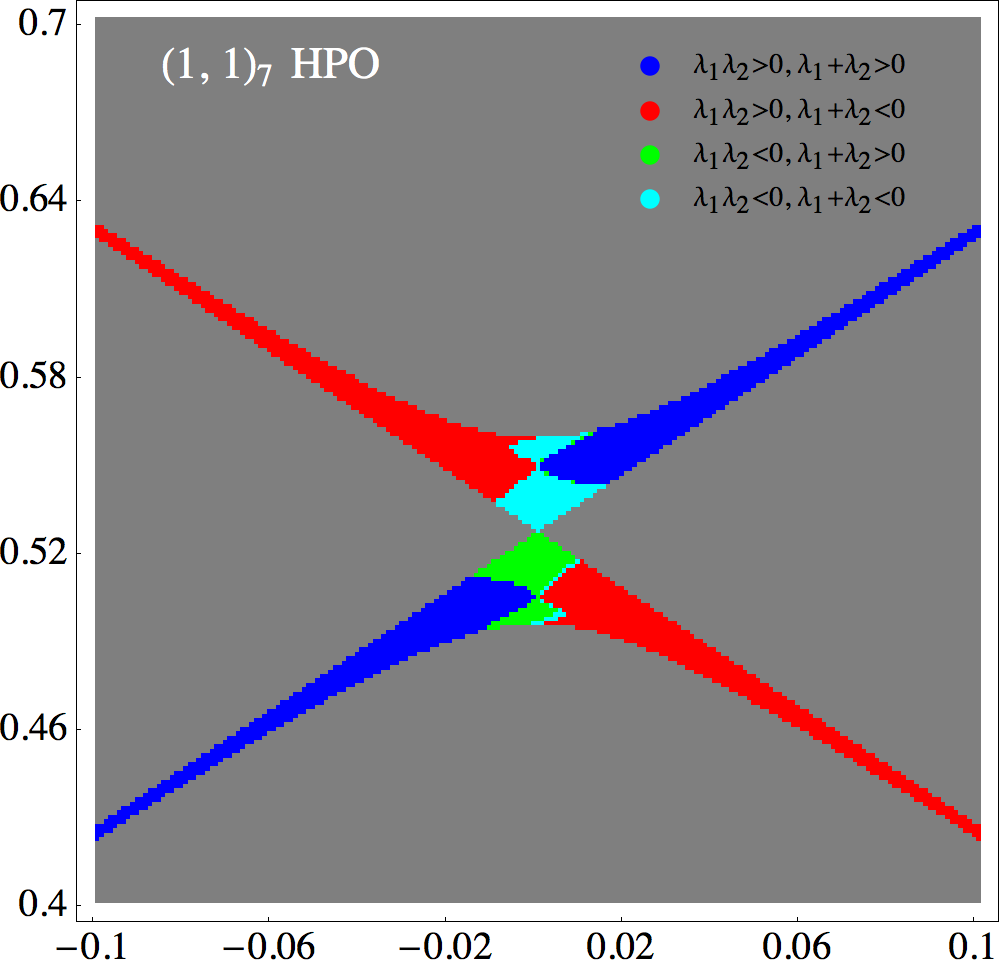}
  \includegraphics[width=0.45\textwidth]{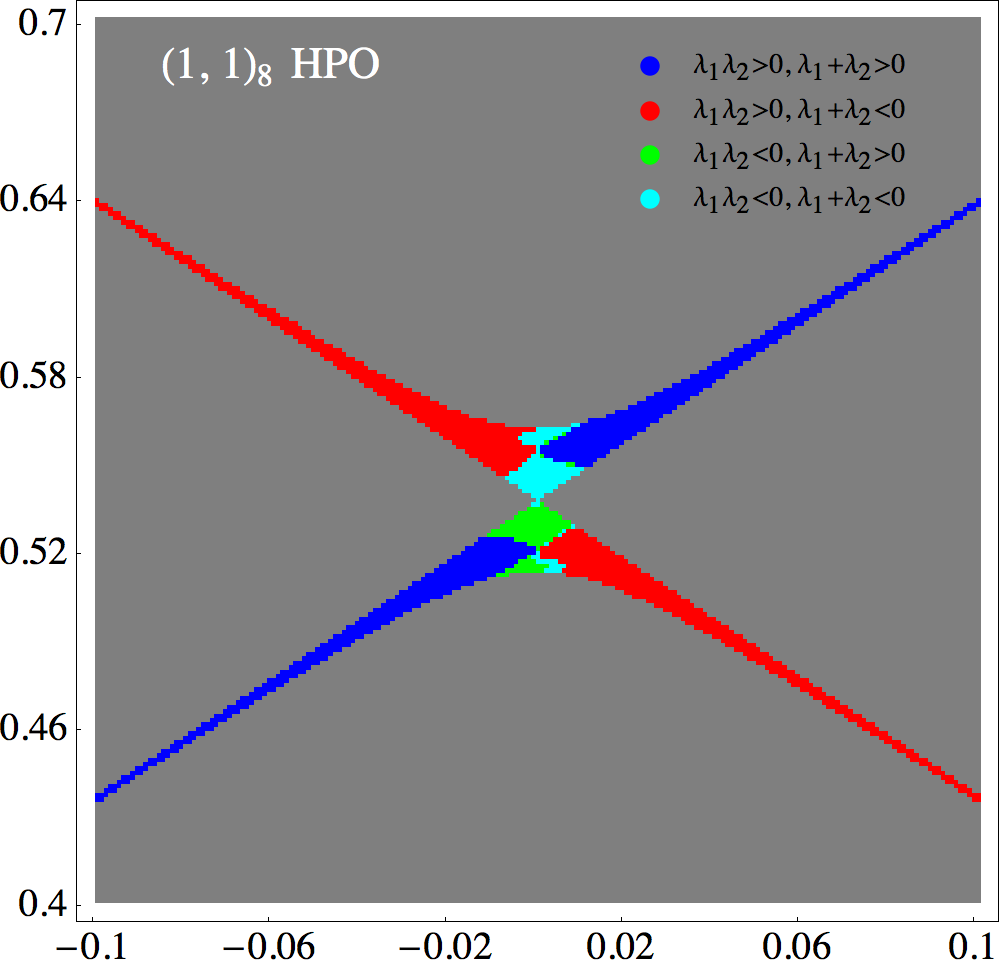}
  \hfill
  \includegraphics[width=0.45\textwidth]{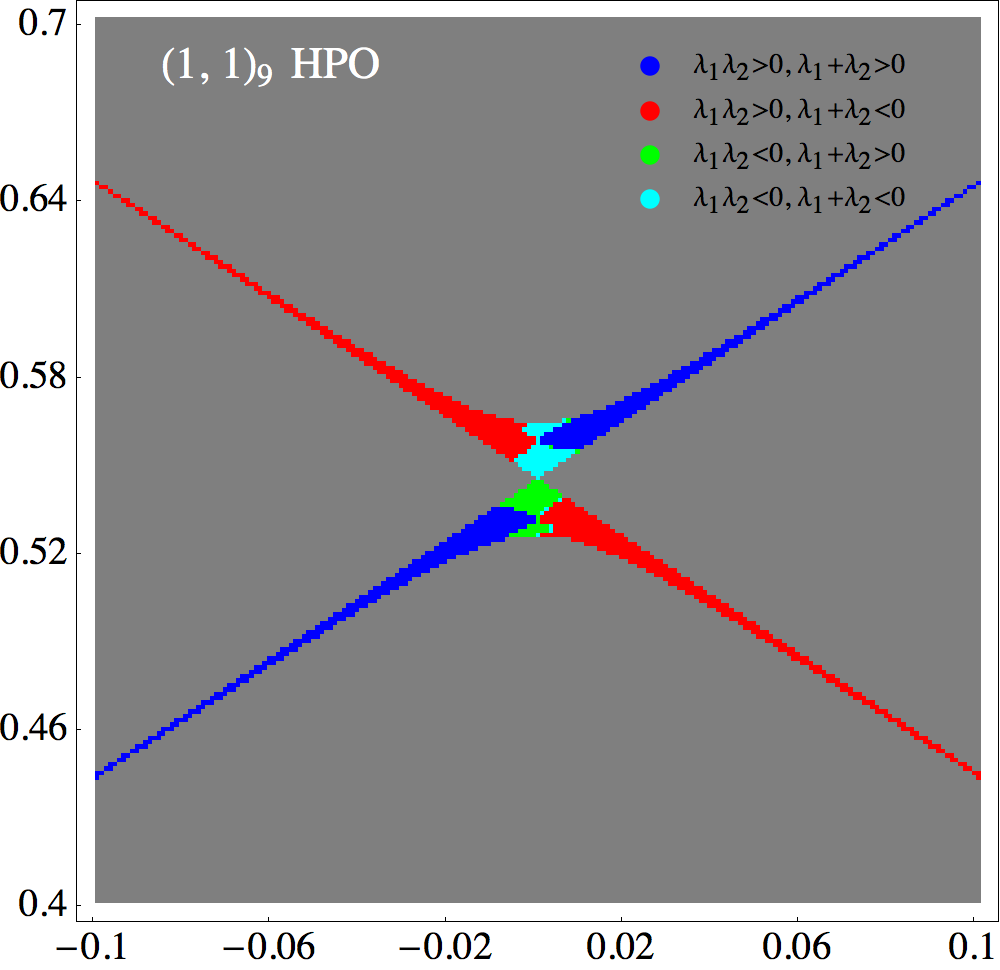}
  \caption{Numerical results of the nonlinear stability analysis of the
    $(1,1)_k$ HPOs, for k = 4,\dots,9. The conventions are the same as in
    \fref{fig.pp1-1_0_nlstab}.}
  \label{fig.hp1-1_k_nlstab}
\end{figure}

Similar results, obtained for larger values of $k$, are shown in 
\fref{fig.hp1-1_k_nlstab}, there resorting to a numerical computation
of the quadratic form $Q_w$ on a grid of points of the parameter
space. That is to say, we fix numerical values of the two parameters $h$
and $\epsilon$, 
identify the corresponding periodic orbit, and perform a perturbation
analysis around that orbit, which yields numerical values of the
coefficients $m_{\alpha \beta}$ in equation \eref{hp1-1_k-linearmap} and
$a_w$, $b_w$ and $c_w$ in equation \eref{hp1-1_k-quadraticform}. As $k$
gets larger, the regions of stability get thinner, but they persist and can
be observed. For example, we were able to check the existence
of nonlinearly stable periodic orbits for the family of $(1,1)_{101}$ HPO
at $h=0.5707$, which belong to class of orbits displayed in
\fref{fig.helices}(a). The same orbit is, however, found to be unstable
at $h = 0.5708$. 

\subsection{$(n,n)_k$  Helical Periodic Orbits
  \label{sec.n-n_k}}

\subsubsection{Existence\label{sec.n-n_kExs}}

\begin{figure}[tbh]
  \centering
  (a) 
   \includegraphics[width=0.29\textwidth]{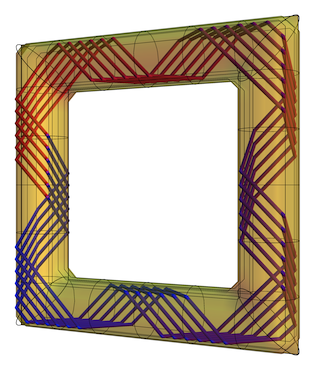}
  \hfill
  (b) 
   \includegraphics[width=0.29\textwidth]{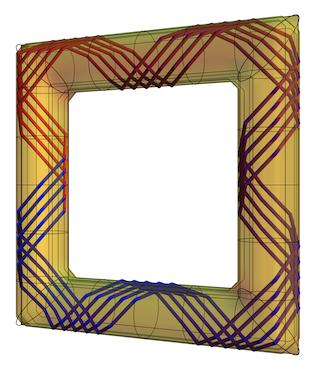}
  \hfill
  (c) 
   \includegraphics[width=0.29\textwidth]{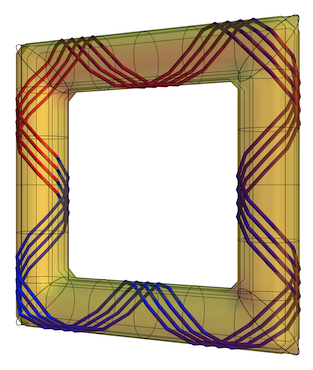}
  \caption{Families of $(3,3)_k$ HPOs found at $h = 3.5$, with parameter
    values across the interval \eref{hpn-n_k-emin}--\eref{hpn-n_k-emax}:
    (a) $k=2$, (b) $k=3$, (c) $k=4$.}
  \label{fig.hp3-3_k-family}
\end{figure}

\begin{figure}[tbh]
  \centering
  (a) 
   \includegraphics[width=0.29\textwidth]{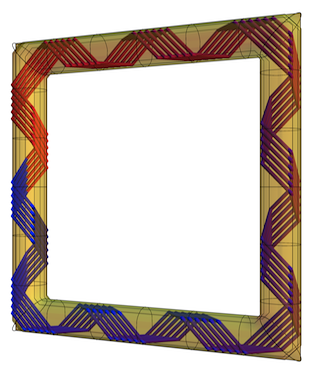}
  \hfill
  (b) 
   \includegraphics[width=0.29\textwidth]{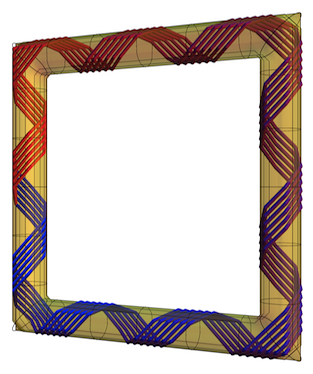}
  \hfill
  (c) 
   \includegraphics[width=0.29\textwidth]{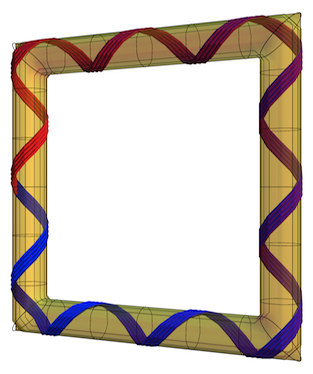}
  \caption{Families of $(5,5)_k$ HPOs found at $h = 6.5$, with parameter
    values across the interval \eref{hpn-n_k-emin}--\eref{hpn-n_k-emax}:
    (a) $k=2$, (b) $k=3$, (c) $k=5$.}
  \label{fig.hp5-5_k-family}
\end{figure}

We can identify, for arbitrary odd integer $n$, any finite length $(n,n)_k$
HPO approximating the $(n,n)$ helices, with the same initial conditions
\eref{hp1-1_k-icpos}--\eref{hp1-1_k-icvel}. These are basically the same
orbits, only the number of collisions along the periodic orbit changes
from $2k$ to $2nk$ and the values of the parameters
\eref{hp1-1_k-hmin}--\eref{hp1-1_k-hmax} and
\eref{hp1-1_k-emin}--\eref{hp1-1_k-emax} are accordingly changed to 
\numparts
\begin{eqnarray}
  h_\mathrm{min} &=& (nk-2) \sin \frac{\pi}{2 k} -1,
  \label{hpn-n_k-hmin}
  \\
  h_\mathrm{max} &=& (nk+2) \sin \frac{\pi}{2 k} -1.
  \label{hpn-n_k-hmax}
\end{eqnarray}
\endnumparts
\numparts
\begin{eqnarray}
  \epsilon_\mathrm{min} &=& \mathrm{max}[
  -1 - h + (nk-2) \sin \frac{\pi}{2 k},
  1 + h - (nk+2) \sin \frac{\pi}{2 k}],
  \label{hpn-n_k-emin}
  \\
  \epsilon_\mathrm{max}  &=& \mathrm{min}[
  1 + h - (nk-2) \sin \frac{\pi}{2 k},
  -1 - h + (nk+2) \sin \frac{\pi}{2 k}].
  \label{hpn-n_k-emax}
\end{eqnarray}
\endnumparts
Some examples of $(3,3)_k$ and $(5,5)_k$ HPOs are shown in figures
\ref{fig.hp3-3_k-family}  and \ref{fig.hp5-5_k-family}, respectively.

\subsubsection{Stability \label{sec.n-n_kSta}}

\begin{figure}[tbh]
  \centering
   \includegraphics[width=0.6\textwidth]{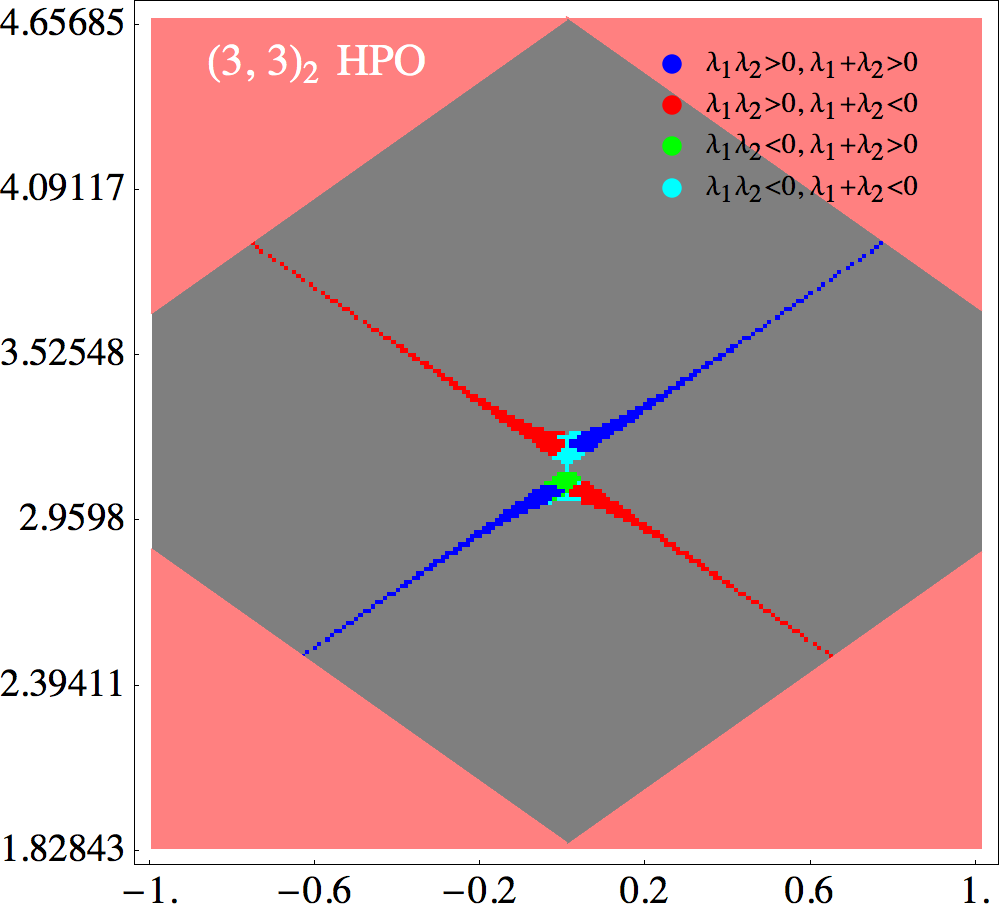}
  \caption{Nonlinear stability analysis of the $(3,3)_2$ HPO.
    The conventions are the same as in \fref{fig.pp1-1_0_nlstab}.}
  \label{fig.hp3-3_2-nlstab}
\end{figure}

The problem of determining the stability of $(n,n)_k$ HPOs transposes
verbatim from \sref{sec.1-1_kSta}. The computation gets more
challenging as the lengths of the orbits increase, but similar results can
be obtained. This is supported by the results shown in
\fref{fig.hp3-3_2-nlstab}, obtained for the $(3,3)_2$ HPO.  

\subsection{$(n_1,n_2)_k$ Helical Periodic Orbits
  \label{sec.n1-n2_k}}

\subsubsection{Existence.\label{sec.n1-n2_kExs}}

Given odd integers $n_1$ and $n_2$, $n_1 \neq n_2$, the $(n_1,n_2)_k$ HPOs
which approximate the $(n_1, n_2)$ helical periodic orbit can be identified
starting with initial positions \eref{hp1-1_k-icpos} and velocity components
adapted according to the ratio between the numbers of half periods to be
completed along the two axes $y$ and $z$:
\begin{equation}
  \left(
    \begin{array}{c}
      u_0\\
      v_0\\
      w_0
    \end{array}
  \right)
  =
  \left(
    \begin{array}{c}
      \sqrt{\frac{n_1}{n_1 + n_2}}\cos\frac{\pi}{k}\\
      \sqrt{\frac{n_1}{n_1 + n_2}}\sin\frac{\pi}{k}\\
      \sqrt{\frac{n_1}{n_1 + n_2}}
    \end{array}
  \right), \quad k\,\mathrm{even}, \qquad
  =
  \left(
    \begin{array}{c}
      \sqrt{\frac{n_1}{n_1 + n_2}}\cos\frac{\pi}{2 k}\\
      \sqrt{\frac{n_1}{n_1 + n_2}}\sin\frac{\pi}{2 k}\\
      \sqrt{\frac{n_2}{n_1 + n_2}}
    \end{array}
  \right), \quad k\,\mathrm{odd}.
  \label{hpn1-n2_k-icvel}
\end{equation}

Because of the asymmetry between the $y$ and $z$ motions, there is here,
and for any given $k$, a unique value of the geometric parameter at which
these orbits are realised,
\begin{equation}
  h = \sqrt{n_1 n_2} k \sin\frac{\pi}{2k} - 1.
  \label{hpn1-n2_k-hvalues}
\end{equation}
The corresponding interval of values of the dynamical parameter, $\epsilon$, has
the
bounds 
\numparts
\begin{eqnarray}
  \epsilon_\mathrm{max} = 
  \left\{
    \begin{array}{l@{\quad}l}
      2\sqrt{n_2/n_1}\sin[\pi/(2k)],&k\,\mathrm{even},\\
      (1+\sqrt{n_2/n_1})\sin[\pi/(2k)],&k\,\mathrm{odd},
    \end{array}
  \right.
  \label{hpn1-n2_k-emax}
  \\
  \epsilon_\mathrm{min} = 
  \left\{
    \begin{array}{l@{\quad}l}
      -2\sin[\pi/(2k)],&k\,\mathrm{even},\\
      -(1+\sqrt{n_2/n_1})\sin[\pi/(2k)],&k\,\mathrm{odd}.
    \end{array}
  \right.
  \label{hpn1-n2_k-emin}
\end{eqnarray}
\endnumparts

\begin{figure}[tbh]
  \centering
  (a) 
   \includegraphics[width=0.29\textwidth]{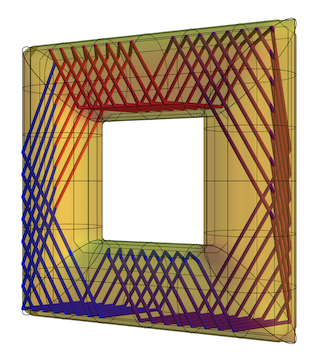}
  \hfill
  (b) 
   \includegraphics[width=0.29\textwidth]{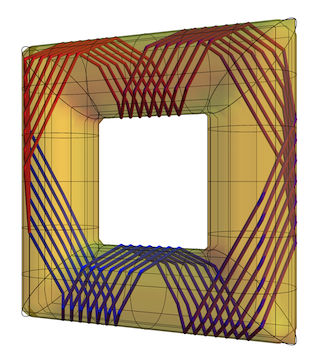}
  \hfill
  (c) 
   \includegraphics[width=0.29\textwidth]{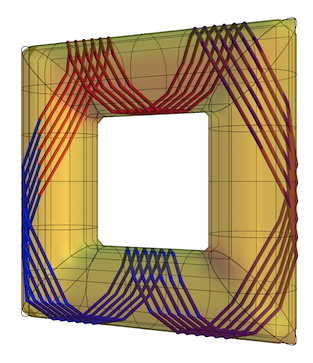}
  \\
  (d) 
   \includegraphics[width=0.29\textwidth]{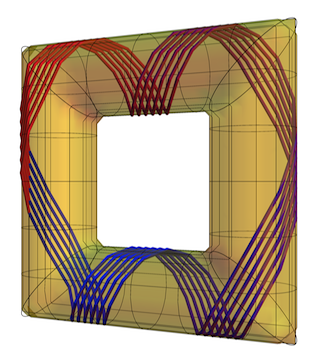}
  \hfill
  (e) 
   \includegraphics[width=0.29\textwidth]{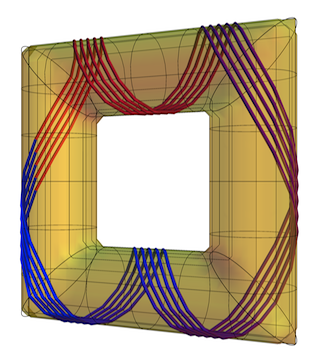}
  \hfill
  (f) 
   \includegraphics[width=0.29\textwidth]{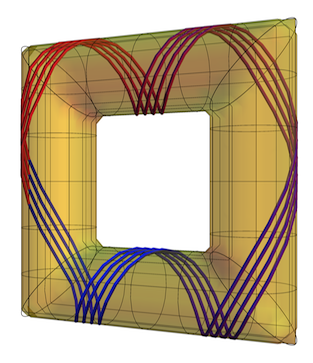}
  \caption{Families of $(1,3)_k$ HPOs found at the values of the geometric
    parameter \eref{hpn1-n2_k-hvalues}, with dynamical parameter
    values across the interval
    \eref{hpn1-n2_k-emin}--\eref{hpn1-n2_k-emax}: (a) $k=2$, (b) 
    $k=3$, (c) $k=4$, (d) $k=5$, (e) $k=6$, (f) $k=7$.}
  \label{fig.hp1-3_k-family}
\end{figure}

\begin{figure}[tbh]
  \centering
  (a) 
   \includegraphics[width=0.29\textwidth]{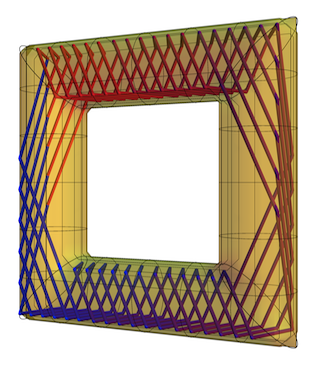}
  \hfill
  (b) 
   \includegraphics[width=0.29\textwidth]{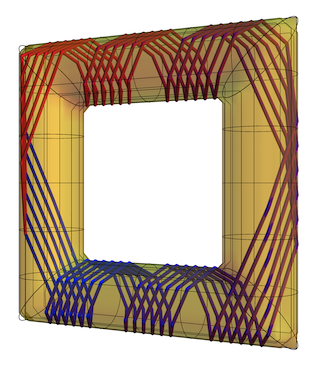}
  \hfill
  (c) 
  \includegraphics[width=0.29\textwidth]{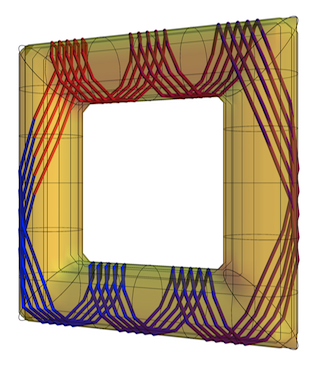}
  \caption{Families of $(1,5)_k$ HPOs found at the values of the geometric
    parameter \eref{hpn1-n2_k-hvalues}, with dynamical parameter
    values across the interval \eref{hpn1-n2_k-emin}--\eref{hpn1-n2_k-emax}: 
    (a) $k=2$, (b) $k=3$, (c) $k=4$.}
  \label{fig.hp1-5_k-family}
\end{figure}

\begin{figure}[tbh]
  \centering
  (a) 
  \includegraphics[width=0.29\textwidth]{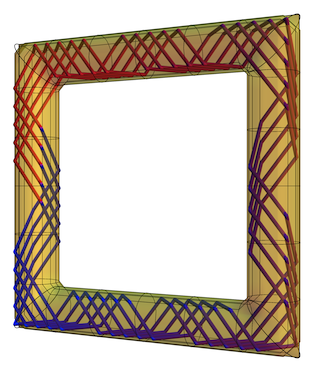}
  \hfill
  (b) 
  \includegraphics[width=0.29\textwidth]{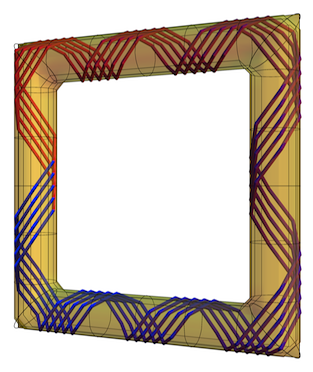}
  \hfill
  (c) 
  \includegraphics[width=0.29\textwidth]{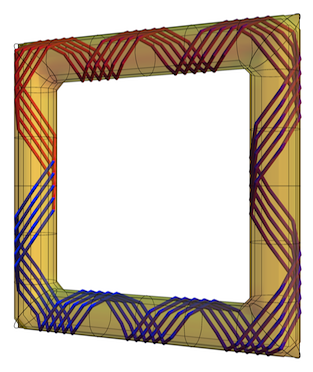}
  \caption{Families of $(3,5)_k$ HPOs found at the values of the geometric
    parameter \eref{hpn1-n2_k-hvalues}, with dynamical parameter
    values across the interval
    \eref{hpn1-n2_k-emin}--\eref{hpn1-n2_k-emax}: (a) $k=2$, (b) 
    $k=3$, (c) $k=4$.}
  \label{fig.hp3-5_k-family}
\end{figure}

We will focus our attention on three groups of such orbits, namely the
$(1,3)_k$, the $(1,5)_k$, and the $(3,5)_k$ HPOs. Specific
examples of families of these classes of $(n_1,n_2)_k$ orbits are shown in
figures \ref{fig.hp1-3_k-family}, \ref{fig.hp1-5_k-family} and
\ref{fig.hp3-5_k-family}, respectively. 

\subsubsection{Stability \label{sec.n1-n2_kSta}}

The nonlinear stability analysis proceeds along the lines described in
\sref{sec.1-1_kSta}. However, since each $(n_1,n_2)_k$ HPO exists
only for a singular value of the geometric parameter, we can consider the
problem of identifying the elliptic regime of the map
\eref{hp1-1_k-linearmap} and the quadratic form
\eref{hp1-1_k-quadraticmap}--\eref{hp1-1_k-quadraticform} as a function of
the dynamical parameter $\epsilon$ alone. 

\begin{figure}[tbh]
  \centering
  (a) 
  \includegraphics[width=0.45\textwidth]{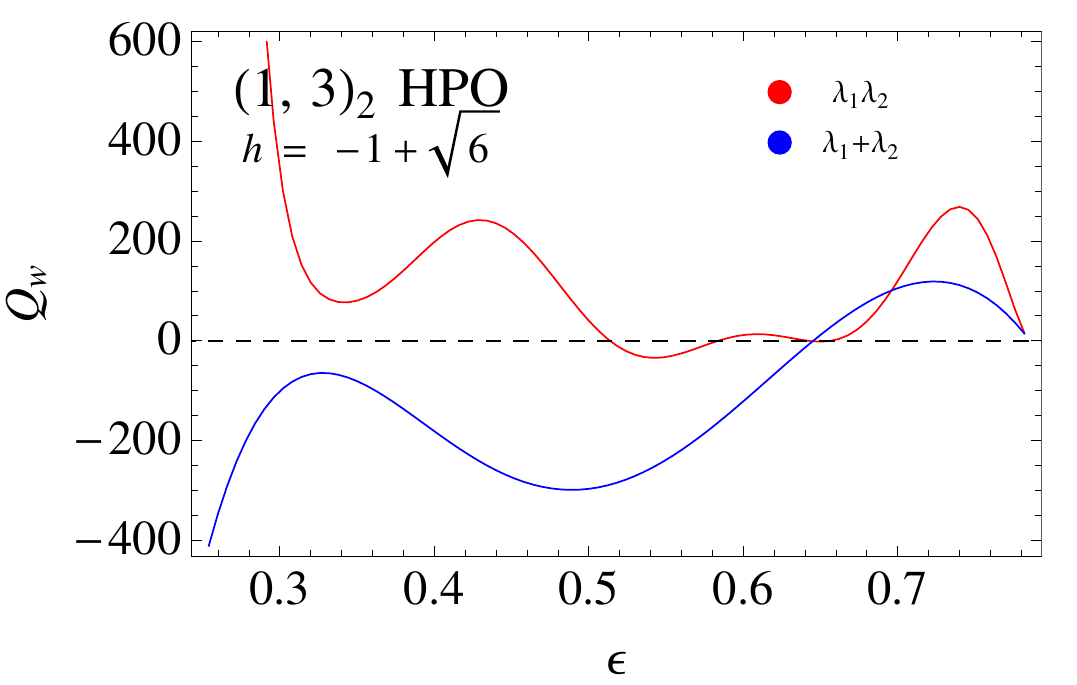}
  \hfill
  (b) 
  \includegraphics[width=0.45\textwidth]{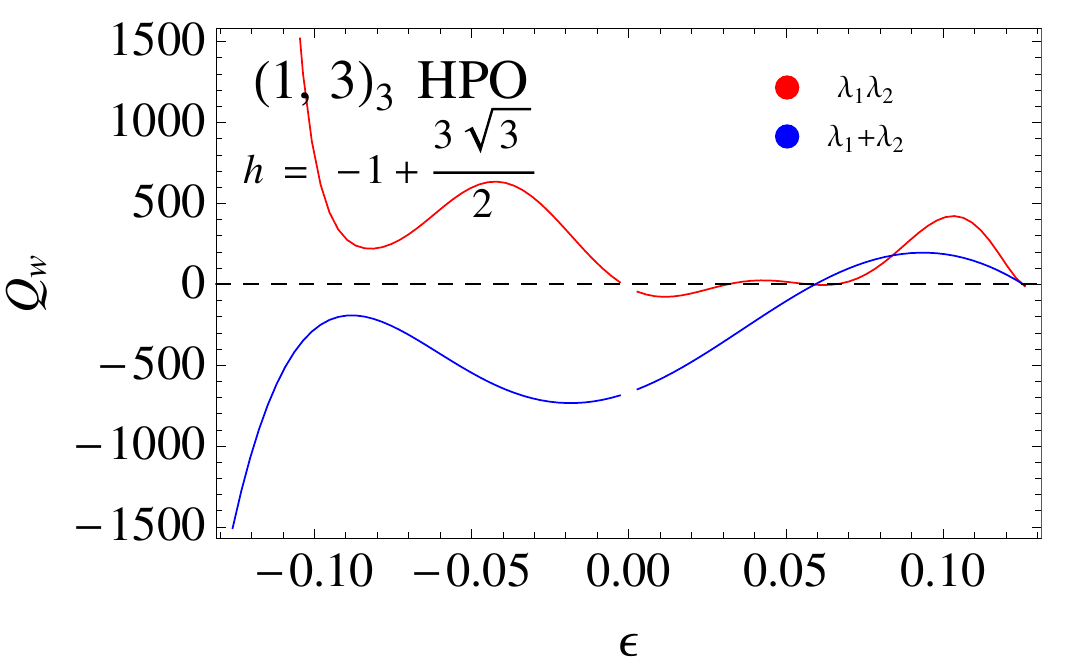}
  \\
  (c) 
  \includegraphics[width=0.45\textwidth]{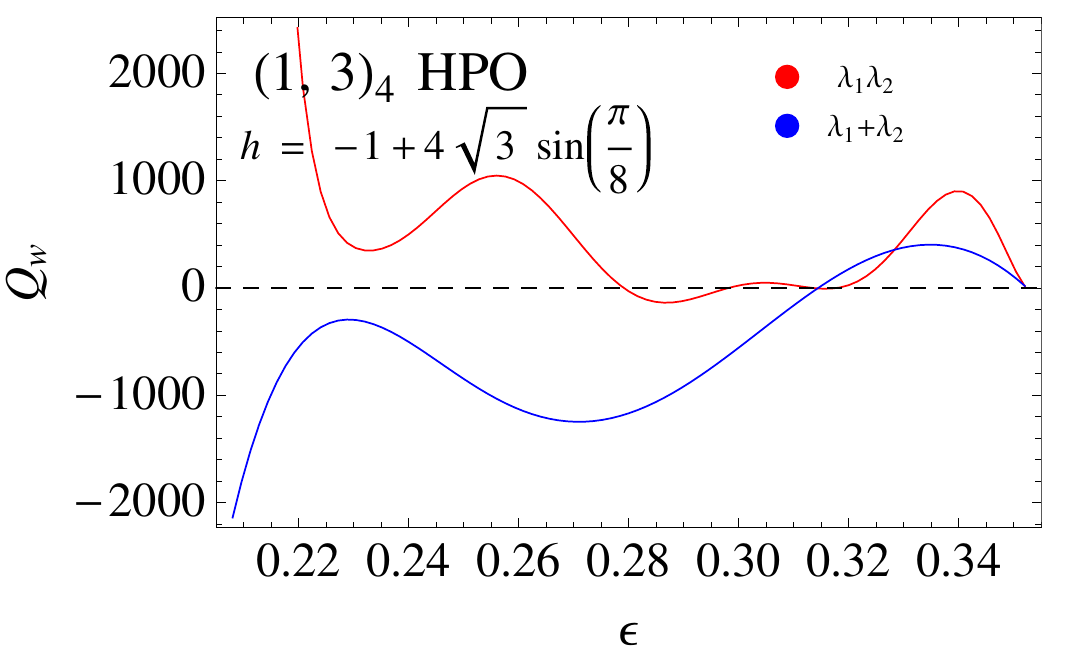}
  \hfill
  (d) 
  \includegraphics[width=0.45\textwidth]{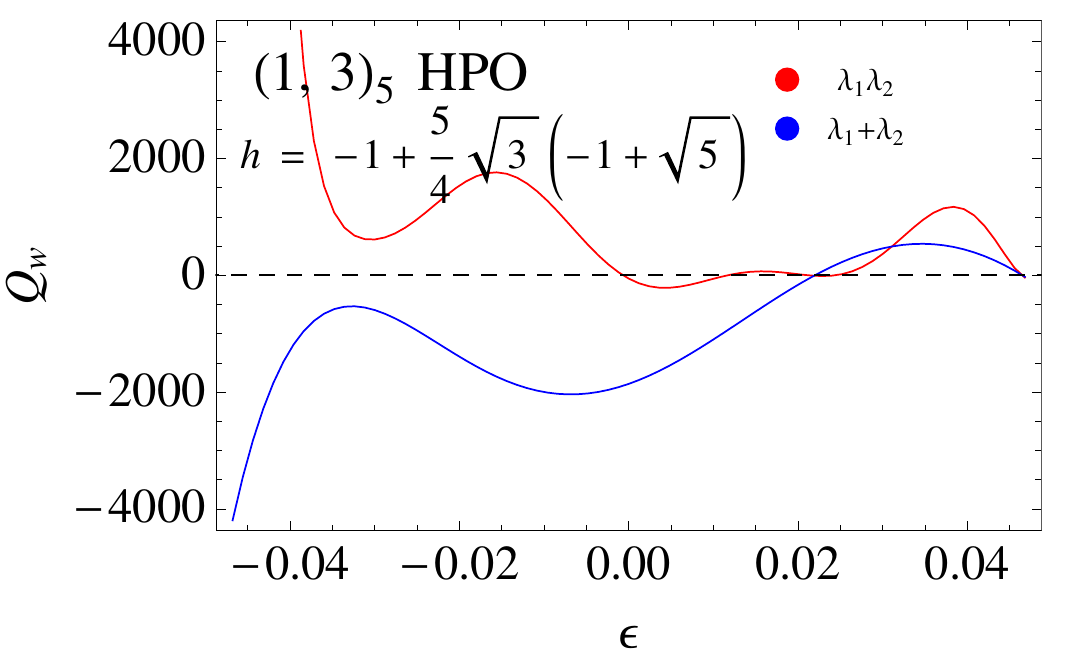}
  \\
  (e) 
  \includegraphics[width=0.45\textwidth]{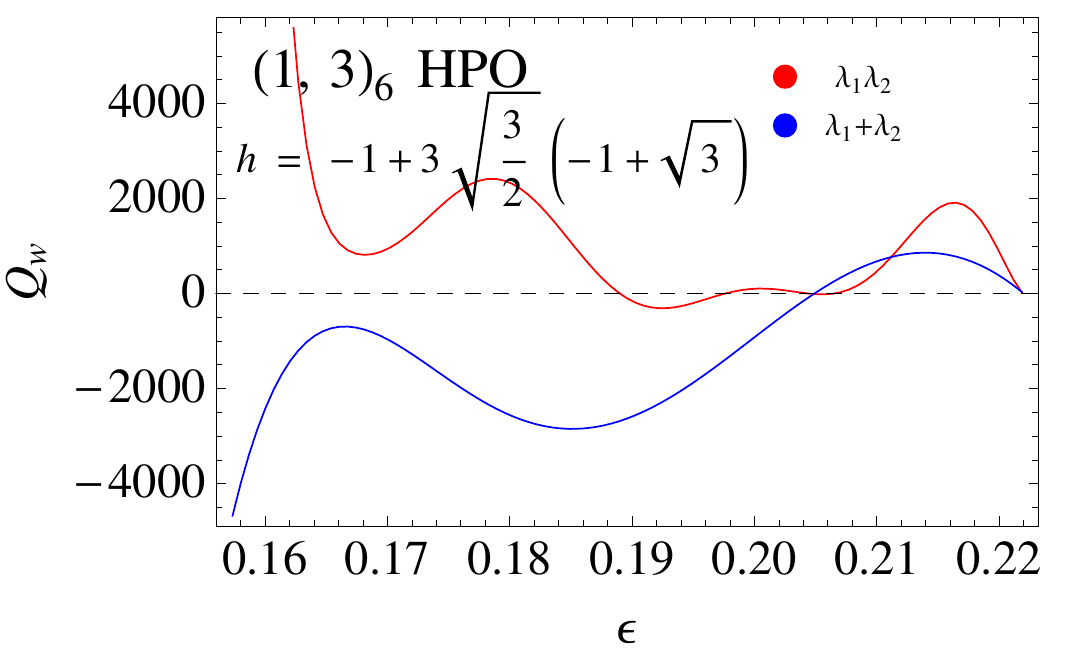}
  \hfill
  (f) 
  \includegraphics[width=0.45\textwidth]{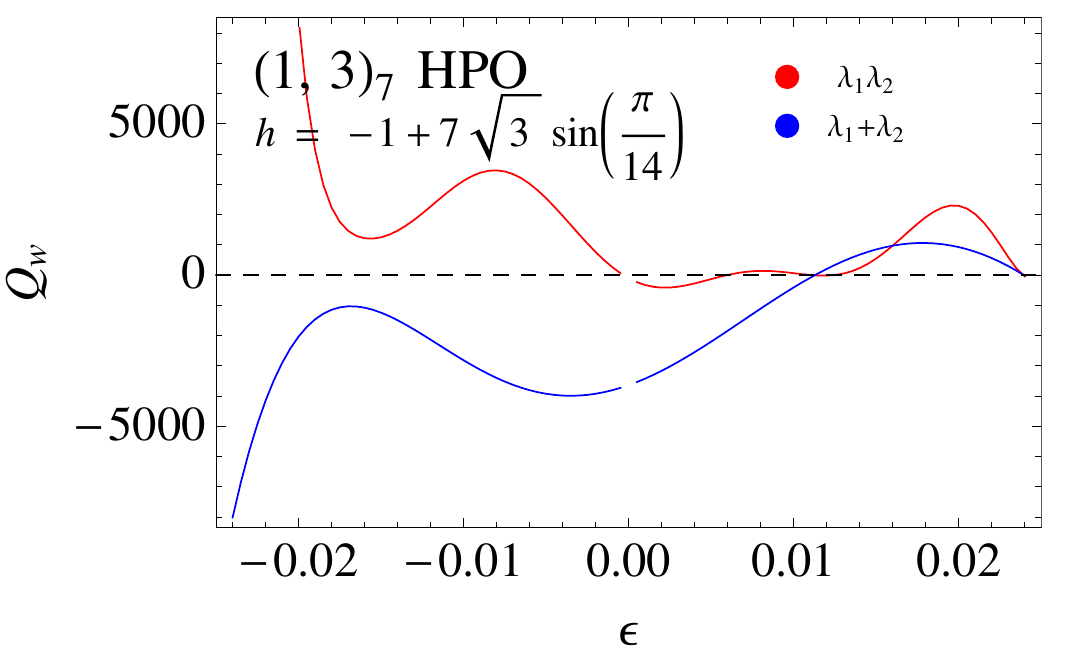}
  \caption{Nonlinear stability analysis of the families of $(1,3)_k$ HPOs
    shown in \fref{fig.hp1-3_k-family}.}
  \label{fig.hp1-3_k-nlstab}
\end{figure}

\begin{figure}[tbh]
  \centering
  (a) 
  \includegraphics[width=0.29\textwidth]{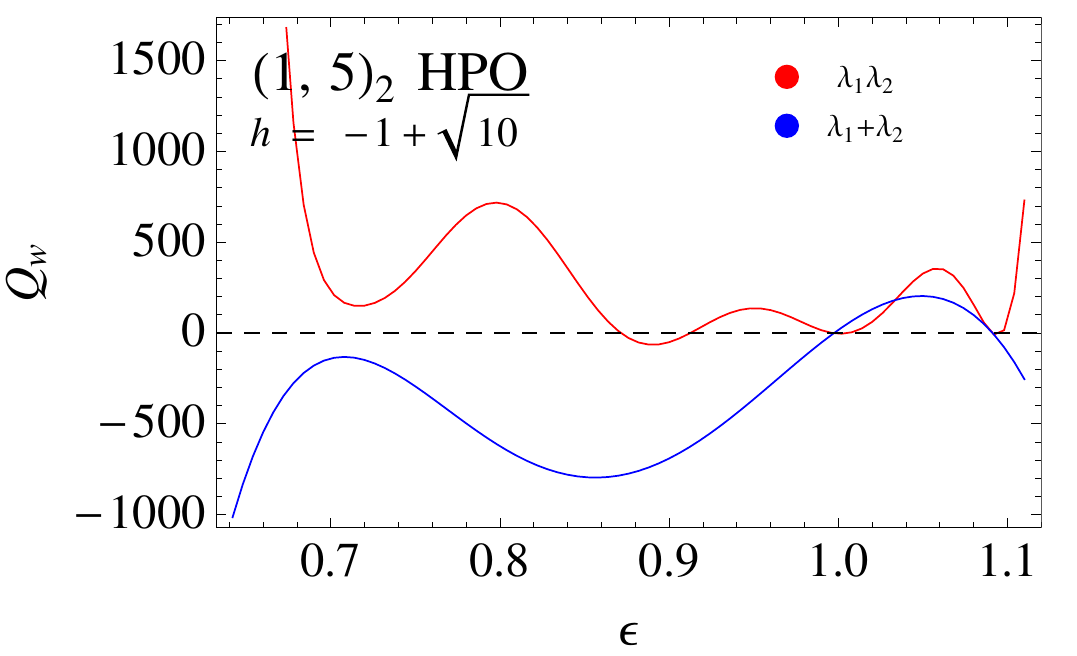}
  \hfill
  (b) 
  \includegraphics[width=0.29\textwidth]{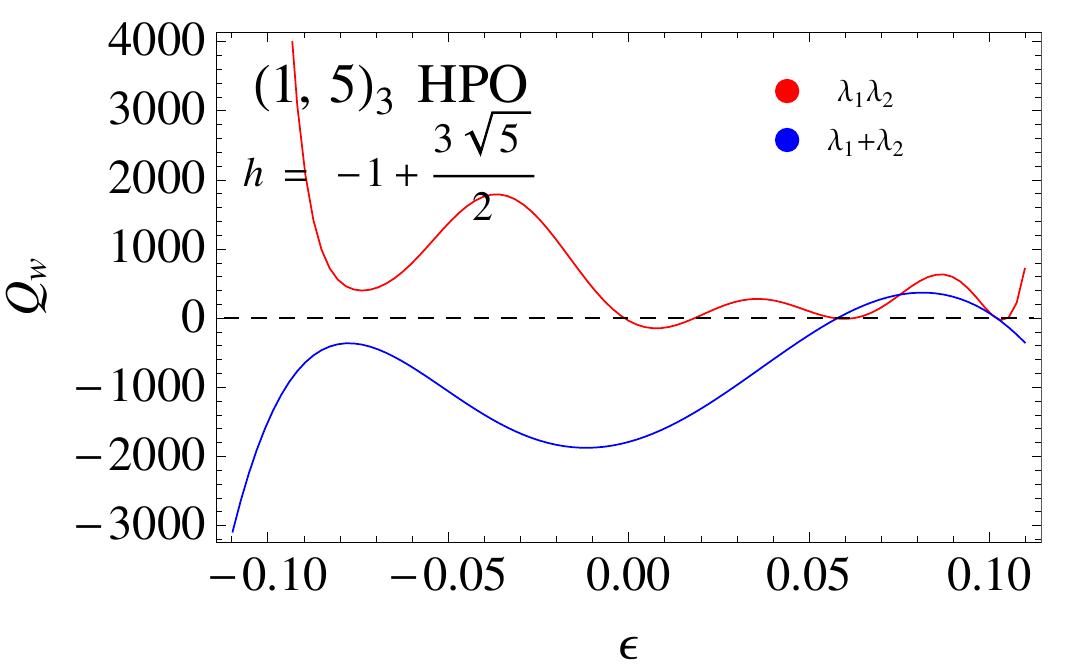}
  \hfill
  (c) 
  \includegraphics[width=0.29\textwidth]{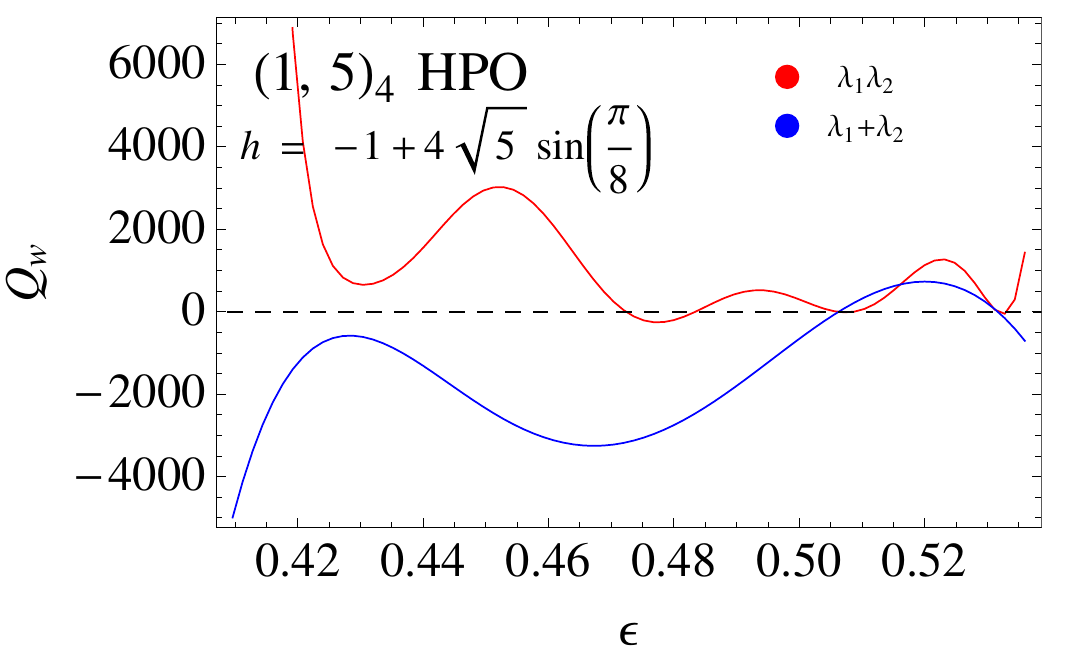}
  \caption{Nonlinear stability analysis of the families of $(1,5)_k$ HPOs
    shown in \fref{fig.hp1-5_k-family}.}
  \label{fig.hp1-5_k-nlstab}
\end{figure}

\begin{figure}[tbh]
  \centering
  (a) 
  \includegraphics[width=0.29\textwidth]{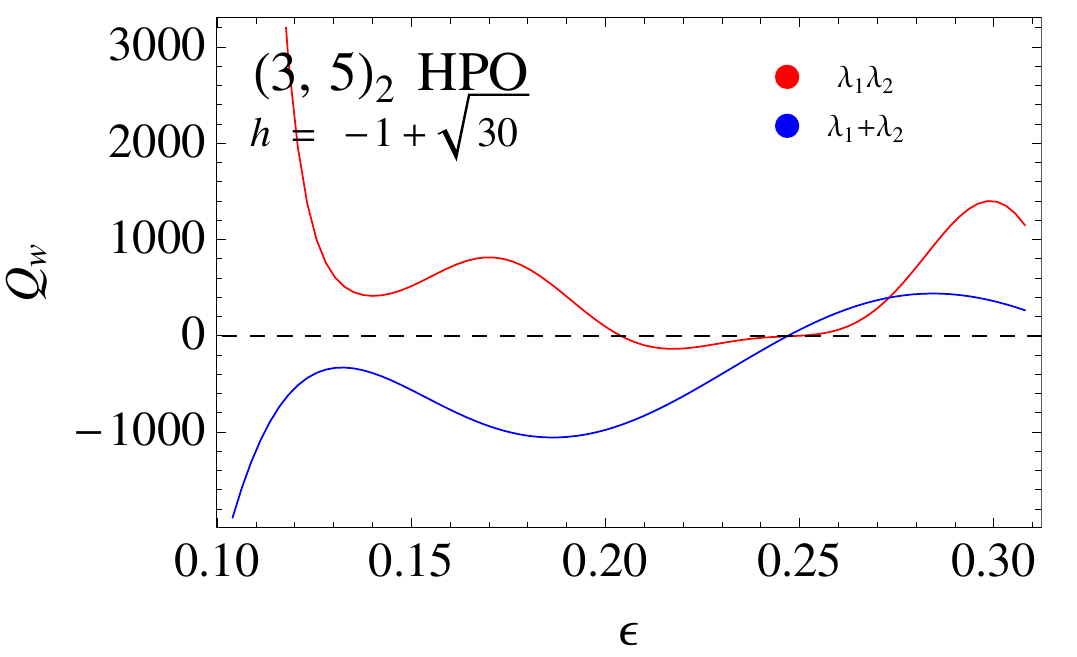}
  \hfill
  (b) 
  \includegraphics[width=0.29\textwidth]{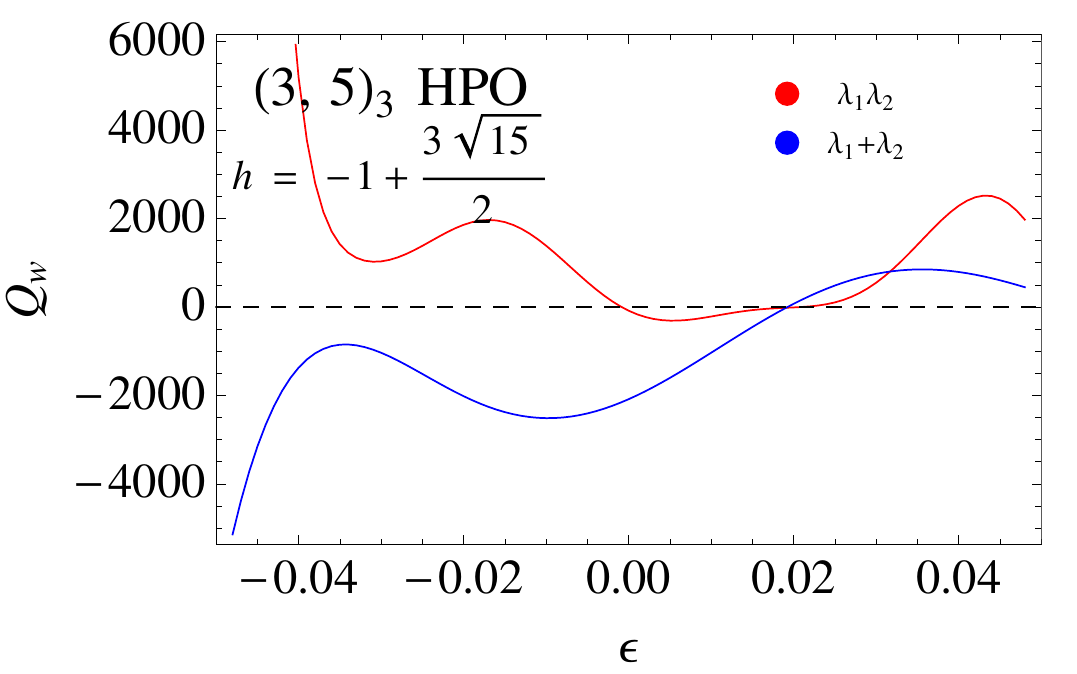}
  \hfill
  (c) 
  \includegraphics[width=0.29\textwidth]{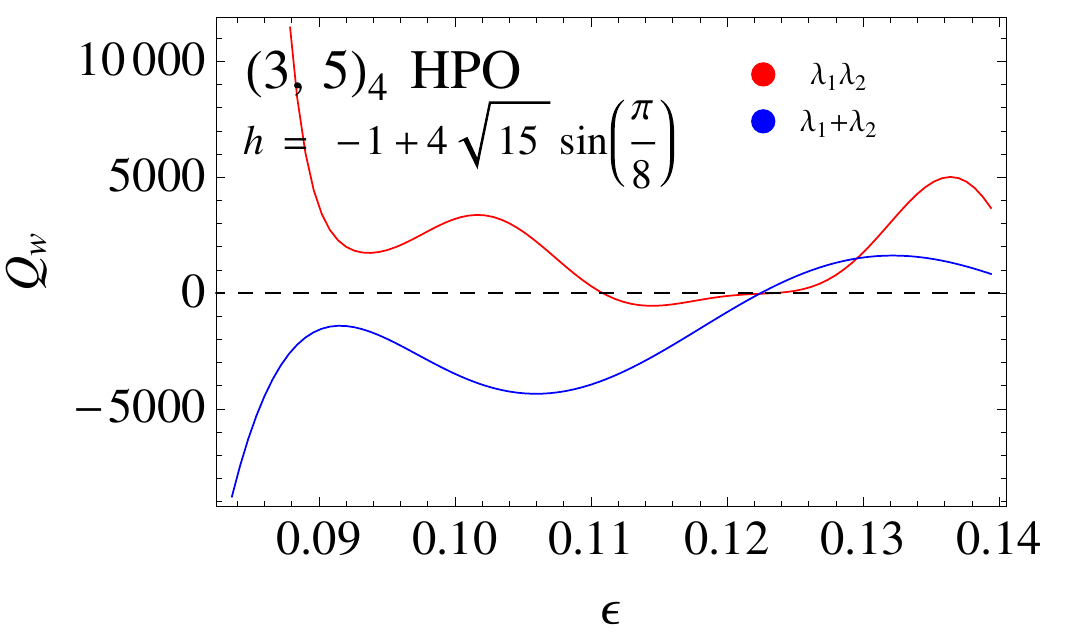}
  \caption{Nonlinear stability analysis of the families of $(3,5)_k$ HPOs
    shown in \fref{fig.hp3-5_k-family}.}
  \label{fig.hp3-5_k-nlstab}
\end{figure}

This computation can be carried out numerically for moderate values of the
overall period. The results are displayed in figures
\ref{fig.hp1-3_k-nlstab}, \ref{fig.hp1-5_k-nlstab} and
\ref{fig.hp3-5_k-nlstab}, respectively, where we plot the determinant
(=$\lambda_1 \lambda_2$) and 
trace (=$\lambda_1 + \lambda_2)$ of the quadratic 
form throughout the range of dynamical parameter $\epsilon$ values where
the eigenvalues of the linear stability analysis are found to be
elliptic. In all of these cases, we observe that the elliptic regime of 
dynamical parameter values is asymmetric with respect to $\epsilon = 0$ for
$k$ even and symmetric for $k$ odd. In these plots, candidate stable HPOs
would correspond to values of $\epsilon$ for which both
the determinant and the trace vanish, together with a negative derivative
of the 
trace, i.e.\ going from positive values on the left of this stable
point, to negative values on its right. This would ensure the existence of
an oscillating regime of the quadratic form, which might indeed stabilise
the given HPO.

Considering the stability analysis of the $(1,3)_k$ HPOS, shown in 
\fref{fig.hp1-3_k-nlstab}, we observe the opposite: namely, the points where
both the determinant and the trace of $Q_w$ vanish correspond to positive
values 
of the derivative of the trace. Therefore they are unstable, since
oscillations around the given HPO will tend to grow away from
it. Similar conclusions are drawn for the $(3,5)_k$ HPOs whose stability
analysis is shown in \fref{fig.hp1-5_k-nlstab}.

The case of the stability analysis of the $(1,5)_k$ HPOs, shown in 
\fref{fig.hp1-5_k-nlstab}, is different. There are now two  points where
both the determinant and trace of $Q_w$ vanish. The leftmost one is of the
same kind as the one observed for the $(1,3)_k$ HPOs. However, the
rightmost one is certainly a stable point, since it appears to correspond to a
minimum of the determinant of $Q_w$. It must, however, be noted that it
arises near the border of the bifurcation point of $\epsilon$ which separates
the elliptic region from the hyperbolic one. Stable oscillations around
this point must therefore be confined to a very small region of phase
space. We have not been able to obtain conclusive evidence of their
existence. 

\section{Summary and perspectives \label{sec.Con}}

The flat surface component of three-dimensional cylindrical billiards
complexifies the stability analysis of their periodic orbits. In the
presence of two elliptic eigenvalues which generate linear oscillations in
the plane transverse to the cylinder axis, nonlinear effects along the
neutral directions may act as a restoring force, allowing for stable
oscillations in all phase space directions.

Summarizing our results, the three-dimensional cylindrical stadium billiard
with a single oblique plane at angle $\pi/4$ with respect to its axis
displays a large range of different dynamical regimes, stable and
unstable, which can be analysed in terms of the spectral properties of
families of planar and helical periodic orbits of this billiard. We
classify these regimes in terms of the geometry of the expanded square
stadium billiard, which is parameterised by its height: 
\begin{list}{}{}
\item[\framebox{$-2<h<-1$}]
Stable oscillations can be observed in large regions of phase
space when the height is small enough that opposite circular arcs are
closer than their diameter. These oscillations take place near a number of
families of stable planar periodic orbits. Nonlinear effects act on these
oscillations as a restoring force 
which for some of these families may constrain the oscillations around the
periodic orbit located at the center of the family. Though most of the
periodic orbits are unstable in some regions of the parameter space in this
range, others remain stable so that there are no parameter values in this
region which correspond to a fully chaotic regime.
\item[\framebox{$-1<h<-1+1/\surd 2$}]
All but the $(1,1)_0$ planar periodic orbit are unstable in this parameter
range. Though the $(1,1)_0$ PPOs still display regions of elliptic
regimes, they are destabilized by nonlinear effects. Though the $(1,1)_2$
HPOs exist in this range, they are unstable. This range therefore
corresponds to a fully chaotic regime: the system is ergodic with all four
Lyapunov exponents non-zero. 
\item[\framebox{$-1+1/\surd 2 < h < 3\surd 2/4 -1$}]
The first familiy of helical periodic orbits, the $(1,1)_2$ HPOs, are
nonlinearly stable in this range, around the central orbit at $\epsilon =
0$. This gives rise to a small island of stability in an otherwise chaotic
phase space.
\item[\framebox{$3\surd 2/4 -1 < h < 1/6$}]
The $(1,1)_2$ HPOs are nonlinearly unstable in this range and the $(1,1)_3$
HPOs are unstable. This is another range of parameter values for which the
system is ergodic with all four Lyapunov exponents non-zero. 
\item[\framebox{$ 1/6< h < 1/3$}]
The $(1,1)_3$ HPOs are nonlinearly stable around $\epsilon = 0$. 
\item[\framebox{$ 1/3 \lnapprox h < \pi/2 - 1$}]
The sequence of $(1,1)_k$ HPOs, $k \geq 4$, follow each other closely in
this range of parameter values, displaying small regions of nonlinear
stability around $\epsilon = 0$. They culminate in the $(1,1)$ helix at
$h = \pi/2 - 1$.
\item[\framebox{$\pi/2 - 1 < h < 4/3$}]
The system is ergodic with all four Lyapunov exponents non-zero in this
range, which separates the last of the $(1,1)_k$ HPOs from the region of
nonlinear stability of the $(1,1)_1$ PPOs.
\item[\framebox{$4/3 < h < 5/3$}]
The $(1,1)_1$ PPOs are nonlinearly stable around $\epsilon = 0$ in this
range.
\end{list}

A similar succession of fully chaotic and ergodic regimes, and mixed,
non-ergodic, regimes where the phase space is separated into a single
chaotic region and small elliptic regions,  continues as we keep increasing
the height parameter $h$. Thus the next regions of nonlinear stability
correspond to the $(3,3)_k$ HPOs which start for $k=2$ near $h = 3$ and
culminate at $h = 3\pi/2-1$, followed, in the interval $18/5 < h < 19/5$,
by the $(1,1)_2$ PPOs, and so on. Let us also mention the possibility that
stable regimes of oscillations exist for isolated values of $h$, such as
with the $(1,5)_k$ HPOs. However, we have not been able to confirm their
existence numerically.

We note that the stable oscillation regimes described above are not typical
of all billiards in the larger class of convex, cylindrical stadium
billiards, as they can be easily destroyed by a change in the angle of the
oblique plane or by the insertion of another oblique plane. In particular,
these stable oscillations do not exist in the billiards obtained from
interacting-particle models confined by hard walls. The 
three-dimensional cylindrical stadium billiard with two perpendicular
planes at angle $\pi/4$ with respect its axis is thus ergodic and fully
chaotic in its whole range of parameter values. It is indeed easy to  see
that there are no stable planar periodic orbits and that helical periodic
orbits do not exist. A periodic orbit of this billiard always has at least
one segment long enough to induce defocusing.

Cylindrical stadium billiards are easily extended to higher-dimensional
billiards and provide a fertile ground for exploration of regular and
chaotic phenomena in higher-dimensional cavities.

\ack

The authors thank Carlangelo Liverani and Vered Rom-Kedar for helpful
discussions. This research benefitted from the joint support of FNRS
(Belgium) and CONACYT (Mexico) through a bilateral collaboration
project. The work of TG is financially supported by the Belgian Federal
Government under the Inter-university Attraction Pole project NOSY
P06/02. TG is financially supported by the Fonds de la Recherche
Scientifique F.R.S.-FNRS.  DPS acknowledges financial support from
DGAPA-UNAM grant  IN105209.

\begin{appendix}
  
  \section{Stability analysis of the Planar Periodic Orbits
    \label{app.PPO}}
  
  In this appendix we give the expressions for the coefficients
  of some of the linear and quadratic forms obtained in the stability
  analysis of planar periodic orbits.
  
  \subsection{$(1,1)_0$ PPOs \label{app.1-1_0-PPO}}
  
  We provide in this section the expressions of the coefficients of the
  quadratic forms
  \eref{pp1-1_0-quadraticformw}--\eref{pp1-1_0-quadraticformz} for the
  $(1,1)_0$ PPOs used  to obtained the results shown in
  \fref{fig.pp1-1_0_nlstab}. 

  \numparts
  \begin{eqnarray}
    \lo a_w =
    -16 \sqrt{2} h^6+32 \sqrt{2} h^5 \epsilon -80
    \sqrt{2} h^5+16 \sqrt{2} h^4 \epsilon ^2+144
    \sqrt{2} h^4 \epsilon -144 \sqrt{2} h^4
    \nonumber\\
    -64\sqrt{2} h^3 \epsilon ^3+32 \sqrt{2} h^3 \epsilon^2
    +224 \sqrt{2} h^3 \epsilon -112 \sqrt{2} h^3+16
    \sqrt{2} h^2 \epsilon ^4-160 \sqrt{2} h^2 \epsilon
    ^3
    \nonumber\\
    +136 \sqrt{2} h^2 \epsilon -36 \sqrt{2} h^2+32
    \sqrt{2} h \epsilon ^5+48 \sqrt{2} h \epsilon
    ^4-96 \sqrt{2} h \epsilon ^3-16 \sqrt{2} h
    \epsilon ^2
    \nonumber\\
    +28 \sqrt{2} h \epsilon -4 \sqrt{2}
    h-16 \sqrt{2} \epsilon ^6+16 \sqrt{2} \epsilon
    ^5+16 \sqrt{2} \epsilon ^4-8 \sqrt{2} \epsilon^3
    \nonumber\\
    -4 \sqrt{2} \epsilon ^2+2 \sqrt{2} \epsilon,
    \label{pp1-1_0-quadraticformaw}
  \end{eqnarray}
  \begin{eqnarray}
    \lo b_w =
    16 \sqrt{2} h^6-32 \sqrt{2} h^5 \epsilon +96 \sqrt{2}
    h^5-16 \sqrt{2} h^4 \epsilon ^2-160 \sqrt{2} h^4
    \epsilon +216 \sqrt{2} h^4
    \nonumber\\ 
    +64 \sqrt{2} h^3
    \epsilon ^3
    -64 \sqrt{2} h^3 \epsilon ^2-288
    \sqrt{2} h^3 \epsilon +224 \sqrt{2} h^3-16
    \sqrt{2} h^2 \epsilon ^4+192 \sqrt{2} h^2 \epsilon
    ^3
    \nonumber\\   
    -80 \sqrt{2} h^2 \epsilon ^2
    -224 \sqrt{2} h^2 \epsilon 
    +106 \sqrt{2} h^2-32 \sqrt{2} h \epsilon
    ^5-32 \sqrt{2} h \epsilon ^4+160 \sqrt{2} h
    \epsilon ^3
    \nonumber\\   
    -32 \sqrt{2} h \epsilon ^2
    -72 \sqrt{2} h \epsilon +20 \sqrt{2} h+16 \sqrt{2} \epsilon
    ^6-32 \sqrt{2} \epsilon ^5-8 \sqrt{2} \epsilon
    ^4+32 \sqrt{2} \epsilon ^3
    \nonumber\\   
    +2 \sqrt{2} \epsilon^2-8 \sqrt{2} \epsilon +\sqrt{2},
    \label{pp1-1_0-quadraticformbw}
  \end{eqnarray}
  \begin{eqnarray}
    \lo c_w =
    -16 \sqrt{2} h^6+32 \sqrt{2} h^5 \epsilon -112
    \sqrt{2} h^5+16 \sqrt{2} h^4 \epsilon ^2+176
    \sqrt{2} h^4 \epsilon -304 \sqrt{2} h^4
    \nonumber\\   
    -64
    \sqrt{2} h^3 \epsilon ^3+96 \sqrt{2} h^3 \epsilon
    ^2+352 \sqrt{2} h^3 \epsilon -400 \sqrt{2} h^3+16
    \sqrt{2} h^2 \epsilon ^4-224 \sqrt{2} h^2 \epsilon
    ^3
    \nonumber\\   
    +192 \sqrt{2} h^2 \epsilon ^2+312 \sqrt{2} h^2
    \epsilon -260 \sqrt{2} h^2+32 \sqrt{2} h \epsilon
    ^5+16 \sqrt{2} h \epsilon ^4
    \nonumber\\   
    -224 \sqrt{2} h
    \epsilon ^3+144 \sqrt{2} h \epsilon ^2+124
    \sqrt{2} h \epsilon -76 \sqrt{2} h-16 \sqrt{2}
    \epsilon ^6+48 \sqrt{2} \epsilon ^5
    \nonumber\\   
    -16 \sqrt{2}
    \epsilon ^4-56 \sqrt{2} \epsilon ^3+28 \sqrt{2}
    \epsilon ^2+22 \sqrt{2} \epsilon -8 \sqrt{2},
    \label{pp1-1_0-quadraticformcw}
  \end{eqnarray}
  \endnumparts
  
  \numparts
  \begin{eqnarray}
    \lo a_z =
    -32 h^8-288 h^7+128 h^6 \epsilon ^2+96 h^6 \epsilon
    -1040 h^6+864 h^5 \epsilon ^2+608 h^5 \epsilon
    \nonumber\\   
    -1904 h^5
    -192 h^4 \epsilon ^4-288 h^4 \epsilon^3
    +2096 h^4 \epsilon ^2+1440 h^4 \epsilon -1828
    h^4-864 h^3 \epsilon ^4
    \nonumber\\   
    -1216 h^3 \epsilon ^3 + 2144
    h^3 \epsilon ^2+1536 h^3 \epsilon -852 h^3
    +128 h^2
    \epsilon ^6+288 h^2 \epsilon ^5-1072 h^2 \epsilon
    ^4
    \nonumber\\   
    -1600 h^2 \epsilon ^3+744 h^2 \epsilon ^2+696
    h^2 \epsilon -158 h^2+288 h \epsilon ^6
    +608 h
    \epsilon ^5-240 h \epsilon ^4
    \nonumber\\   
    -640 h \epsilon ^3 - 44 h \epsilon ^2 + 120 h \epsilon 
    -10 h-32 \epsilon
    ^8-96 \epsilon ^7+16 \epsilon ^6+160 \epsilon
    ^5
    +60 \epsilon ^4
    \nonumber\\   
    -24 \epsilon ^3-34 \epsilon ^2+16
    \epsilon +4,
    \label{pp1-1_0-quadraticformaz}
  \end{eqnarray}
  \begin{eqnarray}
    \lo b_z =
    32 h^8+320 h^7-128 h^6 \epsilon ^2-64 h^6 \epsilon
    +1312 h^6-960 h^5 \epsilon ^2-448 h^5 \epsilon
    +2816 h^5
    \nonumber\\   
    +192 h^4 \epsilon ^4+192 h^4 \epsilon^3      
    -2752 h^4 \epsilon ^2-1216 h^4 \epsilon +3336
    h^4+960 h^3 \epsilon ^4+896 h^3 \epsilon ^3
    \nonumber\\   
    -3712
    h^3 \epsilon ^2-1600 h^3 \epsilon +2096 h^3
    -128
    h^2 \epsilon ^6-192 h^2 \epsilon ^5+1568 h^2
    \epsilon ^4
    \nonumber\\   
    +1408 h^2 \epsilon ^3-2288 h^2 \epsilon
    ^2-1048 h^2 \epsilon +608 h^2-320 h \epsilon
    ^6 -448 h \epsilon ^5
    \nonumber\\   
    +896 h \epsilon ^4+832 h
    \epsilon ^3-464 h \epsilon ^2-328 h \epsilon +64
    h+32 \epsilon ^8
    \nonumber\\   
    +64 \epsilon ^7-128 \epsilon^6
    -192 \epsilon ^5+104 \epsilon ^4+152 \epsilon
    ^3+40 \epsilon ^2-48 \epsilon,
    \label{pp1-1_0-quadraticformbz}
  \end{eqnarray}
  \begin{eqnarray}
    \lo c_z =
    -32 h^8-352 h^7+128 h^6 \epsilon ^2+32 h^6 \epsilon
    -1616 h^6+1056 h^5 \epsilon ^2+224 h^5 \epsilon
    \nonumber\\   
    -3984 h^5
    -192 h^4 \epsilon ^4-96 h^4 \epsilon^3
    +3440 h^4 \epsilon ^2+608 h^4 \epsilon -5636
    h^4-1056 h^3 \epsilon ^4
    \nonumber\\   
    -448 h^3 \epsilon ^3+5536
    h^3 \epsilon ^2+832 h^3 \epsilon -4508 h^3+128 h^2
    \epsilon ^6
    +96 h^2 \epsilon ^5-2032 h^2 \epsilon
    ^4
    \nonumber\\   
    -704 h^2 \epsilon ^3+4456 h^2 \epsilon ^2+680
    h^2 \epsilon -1862 h^2+352 h \epsilon ^6
    \nonumber\\   
    +224 h
    \epsilon ^5-1552 h \epsilon ^4-448 h \epsilon
    ^3+1564 h \epsilon ^2+408 h \epsilon -326 h-32
    \epsilon ^8-32 \epsilon ^7
    \nonumber\\   
    +208 \epsilon ^6+96
    \epsilon ^5-356 \epsilon ^4-136 \epsilon ^3+150
    \epsilon ^2+144 \epsilon -20.
    \label{pp1-1_0-quadraticformcz}
  \end{eqnarray}
  \endnumparts

  \subsection{$(1,2)_0$ PPOs \label{app.1-2_0-PPO}}

  The non-trivial coefficients of the linear form \eref{pp1-n_0-linearmap}
  for the $(1,2)_0$ PPO are found to be
  \numparts
  \begin{eqnarray}
    \lo 
    m_{\theta\theta} = 
    32 h^6+160 h^5-64 h^4 \epsilon ^2-64 h^4
    \epsilon +232 h^4-256 h^3 \epsilon ^2-256 h^3 \epsilon 
    \nonumber\\  
    +32
    h^2 \epsilon ^4+64 h^2 \epsilon ^3-272 h^2
    \epsilon ^2-272 h^2 \epsilon -192 h^2
    +96 h\epsilon ^4
    \nonumber\\  
    +192 h \epsilon ^3+32 h \epsilon ^2+32
    h \epsilon -40 h+72 \epsilon ^4+144 \epsilon
    ^3 +120 \epsilon^2
    \nonumber\\        
    +120 \epsilon +49
    ,
    \label{pp1-2_0-linearmap11}  
  \end{eqnarray}
  \begin{eqnarray}
    \lo 
    m_{\theta\xi} = 
    -32 h^6-224 h^5+64 h^4 \epsilon ^2-584 h^4+320 h^3
    \epsilon ^2-648 h^3-32 h^2 \epsilon ^4
    \nonumber\\        
    +560 h^2
    \epsilon ^2-164 h^2-96 h \epsilon ^4+392 h
    \epsilon ^2+212 h-72 \epsilon ^4+84 \epsilon
    ^2+120
    ,
    \label{pp1-2_0-linearmap12}
  \end{eqnarray}
  \begin{eqnarray}
    \lo 
    m_{\xi\theta} = 
    -32 h^6-96 h^5+64 h^4 \epsilon ^2-8 h^4+192 h^3
    \epsilon ^2+136 h^3-32 h^2 \epsilon ^4 
    \nonumber\\        
    +112 h^2 \epsilon ^2
    +4 h^2-96 h \epsilon ^4-72 h \epsilon
    ^2-68 h-72 \epsilon ^4-36 \epsilon ^2+20,
    \label{pp1-2_0-linearmap21}
  \end{eqnarray}
  \begin{eqnarray}
    \lo
    m_{\xi\xi} = 
    32 h^6+160 h^5-64 h^4 \epsilon ^2+64 h^4 \epsilon
    +232 h^4-256 h^3 \epsilon ^2+256 h^3 \epsilon 
    \nonumber\\        
    +32
    h^2 \epsilon ^4-64 h^2 \epsilon ^3-272 h^2
    \epsilon ^2+272 h^2 \epsilon -192 h^2+96 h
    \epsilon ^4
    -192 h \epsilon ^3
    \nonumber\\        
    +32 h \epsilon ^2-32
    h \epsilon -40 h+72 \epsilon ^4-144 \epsilon
    ^3+120 \epsilon ^2-120 \epsilon +49,
    \label{pp1-2_0-linearmap22}
  \end{eqnarray}
  \begin{equation}
    m_{z w} = 20 \sqrt{5} h+40 \sqrt{5},
    \label{pp1-2_0-linearmap43}
  \end{equation}
  \endnumparts

  The coefficients of the quadratic forms
  \eref{pp1-1_0-quadraticformw}--\eref{pp1-1_0-quadraticformz} 
  obtained for the $(1,2)_0$ PPOs are
  \numparts
  \begin{eqnarray}
    \lo a_w = -\frac{4 (2 h+3)}{\sqrt{5}} \Big[
    128 h^9-64 h^8 (4 \epsilon -11)-32 h^7 \left(4 \epsilon ^2+48
      \epsilon -33\right)
    \nonumber\\        
    +16 h^6 \left(32 \epsilon ^3-20 \epsilon
      ^2-182 \epsilon -27\right)
    \nonumber\\        
    -16 h^5 \left(8 \epsilon
      ^4-160 \epsilon ^3-42 \epsilon ^2+46 \epsilon
      +115\right)
    \nonumber\\        
    -8 h^4 \left(32 \epsilon ^5+120
      \epsilon ^4-472 \epsilon ^3-298 \epsilon ^2-395
      \epsilon +10\right)
    \nonumber\\        
    +8 h^3 \left(16 \epsilon ^6-128
      \epsilon ^5-292 \epsilon ^4+16 \epsilon ^3+112
      \epsilon ^2+233 \epsilon +159\right)
    \nonumber\\        
    +4 h^2
    \left(144 \epsilon ^6-280 \epsilon ^5-468 \epsilon
      ^4-780 \epsilon ^3-452 \epsilon ^2-353 \epsilon
      -11\right)
    \nonumber\\        
    +4 h \left(216 \epsilon ^6+24 \epsilon
      ^5+108 \epsilon ^4-186 \epsilon ^3-174 \epsilon
      ^2-171 \epsilon -95\right)
    \nonumber\\        
    +432 \epsilon ^6+504
    \epsilon ^5+864 \epsilon ^4+828 \epsilon ^3+540
    \epsilon ^2+347 \epsilon +100\Big]
    \label{pp1-2_0-quadraticformaw}
  \end{eqnarray}
  \begin{eqnarray}
    \lo b_w =
    \frac{4 (2 h+3)}{\sqrt{5}}
    \Big[128 h^9-64 h^8 (4 \epsilon
    -15)-32 h^7 \left(4 \epsilon ^2+56 \epsilon
      -81\right)
    \nonumber\\        
    +16 h^6 \left(32 \epsilon ^3-52 \epsilon
      ^2-286 \epsilon +155\right)
    \nonumber\\        
    -16 h^5 \left(8
      \epsilon ^4-192 \epsilon ^3+118 \epsilon ^2+282
      \epsilon +69\right)
    \nonumber\\        
    -8 h^4 \left(32 \epsilon ^5+88
      \epsilon ^4-824 \epsilon ^3+174 \epsilon ^2-47
      \epsilon +405\right)
    \nonumber\\        
    +4 h^3 \left(32 \epsilon
      ^6-320 \epsilon ^5-328 \epsilon ^4+1344 \epsilon
      ^3+192 \epsilon ^2+848 \epsilon -147\right)
    \nonumber\\        
    +2 h^2
    \left(288 \epsilon ^6-1136 \epsilon ^5-376
      \epsilon ^4-56 \epsilon ^3+656 \epsilon ^2+600
      \epsilon +685\right)
    \nonumber\\        
    +2 h \left(432 \epsilon ^6-816
      \epsilon ^5+168 \epsilon ^4-1040 \epsilon ^3+26
      \epsilon ^2-300 \epsilon +149\right)
    \nonumber\\        
    +432 \epsilon
    ^6-360 \epsilon ^5+360 \epsilon ^4-600 \epsilon
    ^3-282 \epsilon ^2-250 \epsilon
    -245\Big]
    \label{pp1-2_0-quadraticformbw}
  \end{eqnarray}   
  \begin{eqnarray}
    \lo c_w = -\frac{4 (2 h+3)}{\sqrt{5}} 
    \Big[128 h^9-64 h^8 (4 \epsilon
    -19)-32 h^7 \left(4 \epsilon ^2+64 \epsilon
      -145\right)
    \nonumber\\        
    +16 h^6 \left(32 \epsilon ^3-84
      \epsilon ^2-390 \epsilon +545\right)
    \nonumber\\        
    -16 h^5
    \left(8 \epsilon ^4-224 \epsilon ^3+342 \epsilon
      ^2+518 \epsilon -449\right)
    \nonumber\\        
    -8 h^4 \left(32
      \epsilon ^5+56 \epsilon ^4-1176 \epsilon ^3+1350
      \epsilon ^2+301 \epsilon +104\right)
    \nonumber\\        
    +8 h^3
    \left(16 \epsilon ^6-192 \epsilon ^5+28 \epsilon
      ^4+1328 \epsilon ^3-1232 \epsilon ^2+615 \epsilon
      -689\right)
    \nonumber\\        
    +4 h^2 \left(144 \epsilon ^6-856
      \epsilon ^5+668 \epsilon ^4+724 \epsilon ^3-396
      \epsilon ^2+949 \epsilon -611\right)
    \nonumber\\        
    +4 h \left(216
      \epsilon ^6-840 \epsilon ^5+924 \epsilon ^4-854
      \epsilon ^3+866 \epsilon ^2-137 \epsilon
      +205\right)
    \nonumber\\        
    +432 \epsilon ^6-1224 \epsilon ^5+1584
    \epsilon ^4-2028 \epsilon ^3+1740 \epsilon ^2-857
    \epsilon +600\Big]
    \label{pp1-2_0-quadraticformcw}
  \end{eqnarray}   
  \endnumparts
  
  \numparts
  \begin{eqnarray}
    \lo a_z = 
    -1024 h^{12}-10752 h^{11}+4096 h^{10} \epsilon
    ^2+4608 h^{10} \epsilon -44800 h^{10}
    +38400 h^9
    \epsilon ^2
    \nonumber\\  
    +42496 h^9 \epsilon -88000 h^9-6144 h^8
    \epsilon ^4-13824 h^8 \epsilon ^3+135424 h^8
    \epsilon ^2
    \nonumber\\  
    +152832 h^8 \epsilon -58464 h^8-50688
    h^7 \epsilon ^4-112640 h^7 \epsilon ^3+194112 h^7
    \epsilon ^2
    \nonumber\\  
    +250112 h^7 \epsilon +67728 h^7+4096
    h^6 \epsilon ^6+13824 h^6 \epsilon ^5-150272 h^6
    \epsilon ^4
    \nonumber\\  
    -352000 h^6 \epsilon ^3-11104 h^6
    \epsilon ^2+114592 h^6 \epsilon +125576 h^6+29184
    h^5 \epsilon ^6
    \nonumber\\  
    +97792 h^5 \epsilon ^5-153920 h^5
    \epsilon ^4-486656 h^5 \epsilon ^3-338288 h^5
    \epsilon ^2-184800 h^5 \epsilon 
    \nonumber\\  
    +15168 h^5-1024
    h^4 \epsilon ^8-4608 h^4 \epsilon ^7+73472 h^4
    \epsilon ^6+261376 h^4 \epsilon ^5
    \nonumber\\  
    +108704 h^4
    \epsilon ^4-165728 h^4 \epsilon ^3-268248 h^4
    \epsilon ^2-223648 h^4 \epsilon -74900 h^4
    \nonumber\\  
    -6144
    h^3 \epsilon ^8-27648 h^3 \epsilon ^7+61632 h^3
    \epsilon ^6+304384 h^3 \epsilon ^5+344816 h^3
    \epsilon ^4
    \nonumber\\  
    +278144 h^3 \epsilon ^3+103776 h^3
    \epsilon ^2+11664 h^3 \epsilon -25470 h^3-13824
    h^2 \epsilon ^8
    \nonumber\\  
    -62208 h^2 \epsilon ^7-40608 h^2
    \epsilon ^6+91872 h^2 \epsilon ^5+176824 h^2
    \epsilon ^4+240128 h^2 \epsilon ^3
    \nonumber\\  
    +171824 h^2
    \epsilon ^2+97624 h^2 \epsilon +22291 h^2-13824 h
    \epsilon ^8-62208 h \epsilon ^7
    \nonumber\\  
    -98064 h \epsilon
    ^6-95904 h \epsilon ^5-78432 h \epsilon ^4-22224 h
    \epsilon ^3+1262 h \epsilon ^2+10760 h \epsilon
    \nonumber\\  
    +6568 h-5184 \epsilon ^8-23328 \epsilon ^7-44712
    \epsilon ^6-61344 \epsilon ^5-67788 \epsilon
    ^4-52200 \epsilon ^3
    \nonumber\\  
    -33009 \epsilon ^2-16080
    \epsilon -3500
    \label{pp1-2_0-quadraticformaz}
  \end{eqnarray}   
  \begin{eqnarray}
    \lo b_z = 
    1024 h^{12}+12800 h^{11}-4096 h^{10} \epsilon ^2-2560
    h^{10} \epsilon +67328 h^{10}-44544 h^9 \epsilon
    ^2
    \nonumber\\  
    -28160 h^9 \epsilon +189120 h^9+6144 h^8
    \epsilon ^4+7680 h^8 \epsilon ^3-200960 h^8
    \epsilon ^2
    \nonumber\\  
    -129280 h^8 \epsilon +287456 h^8+56832
    h^7 \epsilon ^4+71680 h^7 \epsilon ^3-474944 h^7
    \epsilon ^2
    \nonumber\\  
    -313600 h^7 \epsilon +172720 h^7-4096
    h^6 \epsilon ^6-7680 h^6 \epsilon ^5+213760 h^6
    \epsilon ^4
    \nonumber\\  
    +275200 h^6 \epsilon ^3-588064 h^6
    \epsilon ^2-405280 h^6 \epsilon -125432 h^6-31232
    h^5 \epsilon ^6
    \nonumber\\  
    -58880 h^5 \epsilon ^5+410176 h^5
    \epsilon ^4+549120 h^5 \epsilon ^3-254416 h^5
    \epsilon ^2-198880 h^5 \epsilon
    \nonumber\\  
    -258240 h^5+1024
    h^4 \epsilon ^8+2560 h^4 \epsilon ^7-93952 h^4
    \epsilon ^6-180480 h^4 \epsilon ^5
    \nonumber\\  
    +396512 h^4
    \epsilon ^4+573920 h^4 \epsilon ^3+226136 h^4
    \epsilon ^2+135120 h^4 \epsilon -87280 h^4
    \nonumber\\  
    +6144
    h^3 \epsilon ^8+15360 h^3 \epsilon ^7-138176 h^3
    \epsilon ^6-277760 h^3 \epsilon ^5+120528 h^3
    \epsilon ^4
    \nonumber\\  
    +228800 h^3 \epsilon ^3+305072 h^3
    \epsilon ^2+213760 h^3 \epsilon +80300 h^3+13824
    h^2 \epsilon ^8
    \nonumber\\  
    +34560 h^2 \epsilon ^7-96480 h^2
    \epsilon ^6-217440 h^2 \epsilon ^5-96040 h^2
    \epsilon ^4-102000 h^2 \epsilon ^3
    \nonumber\\  
    +70200 h^2
    \epsilon ^2+55860 h^2 \epsilon +56074 h^2+13824 h
    \epsilon ^8+34560 h \epsilon ^7-21168 h \epsilon
    ^6
    \nonumber\\  
    -73440 h \epsilon ^5-83568 h \epsilon ^4-128880
    h \epsilon ^3-49004 h \epsilon ^2-37620 h \epsilon
    -6120 h
    \nonumber\\  
    +5184 \epsilon ^8+12960 \epsilon ^7+4104
    \epsilon ^6-4320 \epsilon ^5-16344 \epsilon
    ^4-34020 \epsilon ^3-20910 \epsilon ^2
    \nonumber\\  
    -18360
    \epsilon -8600
    \label{pp1-2_0-quadraticformbz}
  \end{eqnarray} 
  \begin{eqnarray}
    \lo c_z =
    -1024 h^{12}-14848 h^{11}+4096 h^{10} \epsilon ^2+512
    h^{10} \epsilon -93952 h^{10}+50688 h^9 \epsilon
    ^2
    \nonumber\\        
    +5632 h^9 \epsilon -337344 h^9-6144 h^8 \epsilon
    ^4-1536 h^8 \epsilon ^3+270592 h^8 \epsilon
    ^2+25856 h^8 \epsilon
    \nonumber\\        
    -742752 h^8-62976 h^7
    \epsilon ^4-14336 h^7 \epsilon ^3+809024 h^7
    \epsilon ^2+62720 h^7 \epsilon 
    \nonumber\\        
    -989424 h^7 +4096 h^6 \epsilon ^6
    +1536 h^6 \epsilon ^5-273152 h^6
    \epsilon ^4-55040 h^6 \epsilon ^3
    \nonumber\\        
    +1461664 h^6
    \epsilon ^2+81056 h^6 \epsilon -664408 h^6+33280
    h^5 \epsilon ^6+11776 h^5 \epsilon ^5
    \nonumber\\       
    -648000 h^5
    \epsilon ^4-109824 h^5 \epsilon ^3+1573648 h^5
    \epsilon ^2 
    +40096 h^5 \epsilon +64544 h^5
    \nonumber\\        
    -1024 h^4
    \epsilon ^8-512 h^4 \epsilon ^7+110336 h^4
    \epsilon ^6
    +36096 h^4 \epsilon ^5     -902752 h^4
    \epsilon ^4
    \nonumber\\
    -114784 h^4 \epsilon ^3+847784 h^4
    \epsilon ^2-24480 h^4 \epsilon +478612 h^4        
    -6144 h^3 \epsilon ^8
    -3072 h^3 \epsilon ^7
    \nonumber\\        
    +190144 h^3
    \epsilon ^6+55552 h^3 \epsilon ^5-727952 h^3
    \epsilon ^4-46080 h^3 \epsilon ^3
    -38176 h^3
    \epsilon ^2 
    \nonumber\\ 
    -34768 h^3 \epsilon +302514 h^3 -13824
    h^2 \epsilon ^8-6912 h^2 \epsilon ^7+178272 h^2
    \epsilon ^6       
    +43488 h^2 \epsilon ^5 
    \nonumber\\        
    - 299304 h^2
    \epsilon ^4+18816 h^2 \epsilon ^3-335344 h^2
    \epsilon ^2+1176 h^2 \epsilon +719 h^2
    -13824 h \epsilon ^8
    \nonumber\\        
    -6912 h \epsilon ^7+85104 h \epsilon^6
    +14688 h \epsilon ^5-32832 h \epsilon ^4+23184 h
    \epsilon ^3
    -176722 h \epsilon ^2 
    \nonumber\\        
    + 16920 h \epsilon
    -69272 h-5184 \epsilon ^8-2592 \epsilon ^7+15768
    \epsilon ^6+864 \epsilon ^5
    +9612 \epsilon ^4 
    \nonumber\\        
    +5400 \epsilon ^3-30309 \epsilon ^2
    +6480 \epsilon -21020
    \label{pp1-2_0-quadraticformcz}
  \end{eqnarray}   
  \endnumparts

  \section{Stability analysis of the Helical Periodic Orbits
    \label{app.HPO}}
  
  We provide below the expressions of the non-trivial
  coefficients of the linear stability matrix \eref{hp1-1_k-linearmap} and
  quadratic form \eref{hp1-1_k-quadraticform} for the $(1,1)_2$ and
  $(1,1)_3$ HPOs used to obtained the results shown in
  figures~\ref{fig.hp1-1_stab} and \ref{fig.hp1-1_nlstab}.
  
  \subsection{$(1,1)_2$ HPOs \label{app.1-1_2-HPO}}

  \numparts
  \begin{eqnarray}
    \lo
    m_{\theta\theta} = 256 h^4-704 \sqrt{2} h^3+1024 h^3-512 h^2
    \epsilon ^2+128 \sqrt{2} h^2 \epsilon -2112
    \sqrt{2} h^2
    \nonumber\\  
    +2928 h^2+704 \sqrt{2} h
    \epsilon^2-1024 h \epsilon ^2+256 \sqrt{2}
    h \epsilon -384 h \epsilon -2692 \sqrt{2}h
    \nonumber\\  
    +3808 h+256 \epsilon^4
    -128 \sqrt{2}
    \epsilon^3+704 \sqrt{2} \epsilon ^2-944
    \epsilon^2+264 \sqrt{2} \epsilon -384
    \epsilon 
    \nonumber\\  
    -1284 \sqrt{2}+1817, 
  \end{eqnarray}
  \begin{eqnarray}
    \lo
    m_{\theta\xi} = -768 h^4+1984 \sqrt{2} h^3-3072 h^3+1536 h^2
    \epsilon ^2-256 \sqrt{2} h^2 \epsilon +5952
    \sqrt{2} h^2
    \nonumber\\  
    -8304 h^2
    -1984 \sqrt{2} h
    \epsilon ^2+3072 h \epsilon ^2-512 \sqrt{2}
    h \epsilon +768 h \epsilon +7412 \sqrt{2}
    h
    \nonumber\\  
    -10464 h-768 \epsilon ^4+256 \sqrt{2}
    \epsilon ^3-1984 \sqrt{2} \epsilon ^2+2736
    \epsilon ^2-528 \sqrt{2} \epsilon +768
    \epsilon 
    \nonumber\\  
    +3444 \sqrt{2}-4872,
  \end{eqnarray}
  \begin{eqnarray}
    \lo
    m_{\xi\theta} = -256 h^4+832 \sqrt{2} h^3-1024 h^3+512 h^2
    \epsilon ^2+2496 \sqrt{2} h^2-3536 h^2
    \nonumber\\  
    -832
    \sqrt{2} h \epsilon ^2
    +1024 h \epsilon
    ^2+3548 \sqrt{2} h-5024 h-256 \epsilon
    ^4-832 \sqrt{2} \epsilon ^2
    \nonumber\\  
    +1168 \epsilon
    ^2+1884 \sqrt{2}
    -2664
  \end{eqnarray}
  \begin{eqnarray}
    \lo
    m_{\xi\xi} = 768 h^4-2368 \sqrt{2} h^3+3072 h^3-1536 h^2
    \epsilon ^2-128 \sqrt{2} h^2 \epsilon -7104
    \sqrt{2} h^2
    \nonumber\\  
    +10000 h^2
    +2368 \sqrt{2} h
    \epsilon ^2-3072 h \epsilon ^2-256 \sqrt{2}
    h \epsilon +384 h \epsilon -9788 \sqrt{2}
    h
    \nonumber\\  
    +13856 h+768 \epsilon ^4
    +128 \sqrt{2}
    \epsilon ^3+2368 \sqrt{2} \epsilon ^2-3280
    \epsilon ^2-264 \sqrt{2} \epsilon +384
    \epsilon
    \nonumber\\  
    -5052 \sqrt{2}+7145,
  \end{eqnarray}
  \begin{eqnarray}
    \lo
    m_{z\theta} = -128 \sqrt{2} h^4-512 \sqrt{2} h^3+704 h^3+256
    \sqrt{2} h^2 \epsilon ^2-128 h^2 \epsilon
    -1464 \sqrt{2} h^2
    \nonumber\\  
    +2112 h^2 + 512 \sqrt{2} h
    \epsilon ^2 -704 h \epsilon ^2 + 192 \sqrt{2}
    h \epsilon -256 h \epsilon -1904 \sqrt{2}
    h
    \nonumber\\  
    +2692 h-128 \sqrt{2} \epsilon ^4+128
    \epsilon ^3+472 \sqrt{2} \epsilon ^2-704
    \epsilon ^2+192 \sqrt{2} \epsilon -264
    \epsilon 
    \nonumber\\  
    -908 \sqrt{2}+1284,
  \end{eqnarray}
  \begin{eqnarray}
    \lo
    m_{z\xi} = 384 \sqrt{2} h^4+1536 \sqrt{2} h^3-1984
    h^3-768 \sqrt{2} h^2 \epsilon ^2+256 h^2
    \epsilon 
    \nonumber\\  
    +4152 \sqrt{2} h^2-5952 h^2-1536
    \sqrt{2} h \epsilon ^2+1984 h \epsilon
    ^2-384 \sqrt{2} h \epsilon +512 h \epsilon
    \nonumber\\  
    +5232 \sqrt{2} h
    -7412 h+384 \sqrt{2}
    \epsilon ^4
    -256 \epsilon ^3-1368 \sqrt{2}
    \epsilon ^2+1984 \epsilon ^2
    \nonumber\\  
    -384 \sqrt{2}
    \epsilon 
    +528 \epsilon +2436 \sqrt{2}-3444,
  \end{eqnarray}
  \begin{equation}
    m_{z w} = 8 \sqrt{2} h + 8 \sqrt{2}+16.
    \label{hp1-1_2-mzw}
  \end{equation}
  \endnumparts
  \numparts
  \begin{eqnarray}
    \lo
    a_w = 
    8192 \sqrt{2} h^6+16384 \sqrt{2} h^5 \epsilon
    +49152 \sqrt{2} h^5-77824 h^5-8192 \sqrt{2}
    h^4 \epsilon ^2
    \nonumber\\  
    +81920 \sqrt{2} h^4 \epsilon 
    -126976 h^4 \epsilon +275456 \sqrt{2}
    h^4-389120 h^4-32768 \sqrt{2} h^3 \epsilon
    ^3
    \nonumber\\  
    -32768 \sqrt{2} h^3 \epsilon ^2
    +57344 h^3
    \epsilon ^2+358400 \sqrt{2} h^3 \epsilon
    -507904 h^3 \epsilon +774144 \sqrt{2}
    h^3
    \nonumber\\  
    -1094144 h^3
    -8192 \sqrt{2} h^2 \epsilon
    ^4-98304 \sqrt{2} h^2 \epsilon ^3+155648
    h^2 \epsilon ^3
    \nonumber\\  
    -122880 \sqrt{2} h^2
    \epsilon ^2
    +172032 h^2 \epsilon ^2 +747520
    \sqrt{2} h^2 \epsilon -1056512 h^2 \epsilon
    \nonumber\\  
    +1220384 \sqrt{2} h^2
    -1725952 h^2
    +16384    \sqrt{2} h \epsilon ^5
    -16384 \sqrt{2} h
    \epsilon ^4+20480 h \epsilon ^4
    \nonumber\\  
    -219136 \sqrt{2} h \epsilon ^3
    +311296 h \epsilon^3
    -180224 \sqrt{2} h \epsilon ^2 +254464 h
    \epsilon ^2+775904 \sqrt{2} h \epsilon
    \nonumber\\  
    -1097216 h \epsilon 
    +1023552 \sqrt{2} h
    -1447536 h + 8192 \sqrt{2} \epsilon ^6+16384
    \sqrt{2} \epsilon ^5
    \nonumber\\  
    -28672 \epsilon
    ^5-13312 \sqrt{2} \epsilon ^4
    +20480 \epsilon ^4
    -153600 \sqrt{2} \epsilon^3 +216832 \epsilon ^3
    \nonumber\\  
    -98784 \sqrt{2}
    \epsilon ^2+139776 \epsilon ^2
    +321248 \sqrt{2} \epsilon 
    -454328 \epsilon 
    \nonumber\\  
    +356688 \sqrt{2}-504432,
  \end{eqnarray}
  \begin{eqnarray}
    \lo
    b_w = -24576 \sqrt{2} h^6-49152 \sqrt{2} h^5
   \epsilon -147456 \sqrt{2} h^5+225280
   h^5
   \nonumber\\  
   +24576 \sqrt{2} h^4 \epsilon ^2
   -245760
   \sqrt{2} h^4 \epsilon +372736 h^4
   \epsilon -794624 \sqrt{2} h^4+1126400
   h^4 
   \nonumber\\  
   +98304 \sqrt{2} h^3 \epsilon ^3
   +98304
   \sqrt{2} h^3 \epsilon ^2-155648 h^3
   \epsilon ^2-1050624 \sqrt{2} h^3 \epsilon
   \nonumber\\  
   +1490944 h^3 \epsilon -2195456 \sqrt{2} h^3
   +3103232 h^3+24576 \sqrt{2} h^2
   \epsilon ^4
   \nonumber\\  
   +294912 \sqrt{2} h^2 \epsilon^3
   -450560 h^2 \epsilon ^3+329728 \sqrt{2}
   h^2 \epsilon ^2
   -466944 h^2 \epsilon
   ^2
   \nonumber\\  
   -2168832 \sqrt{2} h^2 \epsilon +3065600
   h^2 \epsilon 
   -3397056 \sqrt{2}
   h^2+4804096 h^2
   \nonumber\\  
   -49152 \sqrt{2} h \epsilon
   ^5
   +49152 \sqrt{2} h \epsilon ^4-69632 h
   \epsilon ^4
   +632832 \sqrt{2} h \epsilon
   ^3-901120 h \epsilon ^3
   \nonumber\\  
   +462848 \sqrt{2} h
   \epsilon ^2-653824 h \epsilon ^2-2227040
   \sqrt{2} h \epsilon 
   +3149312 h \epsilon
   -2796416 \sqrt{2} h
   \nonumber\\  
   +3954768 h-24576
   \sqrt{2} \epsilon ^6-49152 \sqrt{2}
   \epsilon ^5+77824 \epsilon ^5
   +47104
   \sqrt{2} \epsilon ^4-69632 \epsilon
   ^4
   \nonumber\\  
   +436224 \sqrt{2} \epsilon ^3-616192
   \epsilon ^3+242112 \sqrt{2} \epsilon
   ^2
   -342528 \epsilon ^2-912224 \sqrt{2}
   \epsilon 
   \nonumber\\  
    +1290104 \epsilon-956562
   \sqrt{2}+1352784
   ,
  \end{eqnarray}
  \begin{eqnarray}
    \lo
    c_w = 73728 \sqrt{2} h^6+147456 \sqrt{2} h^5
   \epsilon +442368 \sqrt{2} h^5-651264
   h^5
   \nonumber\\  
   -73728 \sqrt{2} h^4 \epsilon ^2
   +737280
   \sqrt{2} h^4 \epsilon -1093632 h^4
   \epsilon +2292736 \sqrt{2} h^4
   \nonumber\\  
   -3256320
   h^4-294912 \sqrt{2} h^3 \epsilon^3
   -294912 \sqrt{2} h^3 \epsilon ^2+417792
   h^3 \epsilon ^2
   \nonumber\\  
   +3078144 \sqrt{2} h^3
   \epsilon -4374528 h^3 \epsilon +6221824
   \sqrt{2} h^3
   -8796672 h^3
   \nonumber\\  
   -73728 \sqrt{2}
   h^2 \epsilon ^4-884736 \sqrt{2} h^2
   \epsilon ^3+1302528 h^2 \epsilon
   ^3-880640 \sqrt{2} h^2 \epsilon^2
   \nonumber\\  
   +1253376 h^2 \epsilon ^2+6285312
   \sqrt{2} h^2 \epsilon -8884992 h^2
   \epsilon +9450720 \sqrt{2} h^2
   \nonumber\\  
   -13364736 h^2
   +147456 \sqrt{2} h \epsilon ^5-147456
   \sqrt{2} h \epsilon ^4+233472 h \epsilon
   ^4
   \nonumber\\  
   -1824768 \sqrt{2} h \epsilon ^3+2605056
   h \epsilon ^3
   -1171456 \sqrt{2} h \epsilon
   ^2+1657344 h \epsilon ^2
   \nonumber\\  
   +6379360 \sqrt{2}
   h \epsilon -9020928 h \epsilon +7637440
   \sqrt{2} h
   -10801008 h
   \nonumber\\  
   +73728 \sqrt{2}
   \epsilon ^6+147456 \sqrt{2} \epsilon
   ^5-208896 \epsilon ^5-158720 \sqrt{2}
   \epsilon ^4+233472 \epsilon ^4
   \nonumber\\  
   -1234944
   \sqrt{2} \epsilon ^3+1745664 \epsilon
   ^3-580896 \sqrt{2} \epsilon ^2+821760
   \epsilon ^2+2582368 \sqrt{2} \epsilon
   \nonumber\\  
   -3652056 \epsilon +2565300
   \sqrt{2}-3627888
   .
  \end{eqnarray}
  \endnumparts

  \subsection{$(1,1)_3$ HPOs \label{app.1-1_3-HPO}}

  \numparts
  \begin{eqnarray}
    \lo
    m_{\theta\theta} = 10368 h^4-13824 h^3-20736 h^2 \epsilon
    ^2-1728 h^2 \epsilon +6624 h^2+13824 h
    \epsilon ^2 \nonumber\\  
    +1152 h \epsilon -1344 h+10368
    \epsilon ^4+1728 \epsilon ^3-2016
    \epsilon ^2-168 \epsilon +97
  \end{eqnarray}
  \begin{eqnarray}
    \lo
    m_{\theta\xi} = -31104 h^4+36288 h^3+62208 h^2 \epsilon
    ^2-15120 h^2-36288 h \epsilon ^2+2664
    h \nonumber\\  
    -31104 \epsilon ^4+4752 \epsilon ^2-168
  \end{eqnarray}
  \begin{eqnarray}
    \lo
    m_{\xi\theta} = -3456 h^4+5184 h^3+6912 h^2 \epsilon ^2-2832
    h^2-5184 h \epsilon ^2+664 h \nonumber\\  
    -3456 \epsilon ^4+912 \epsilon ^2-56
  \end{eqnarray}
  \begin{eqnarray}
    \lo
    m_{\xi\xi} = 10368 h^4-13824 h^3-20736 h^2 \epsilon
    ^2+1728 h^2 \epsilon +6624 h^2+13824 h
    \epsilon ^2  \nonumber\\  
    -1152 h \epsilon - 1344 h+10368
    \epsilon ^4-1728 \epsilon ^3-2016
    \epsilon ^2+168 \epsilon +97
  \end{eqnarray}
  \begin{eqnarray}
    \lo
    m_{z\theta} = -5184 \sqrt{3} h^4+6912 \sqrt{3} h^3+10368
    \sqrt{3} h^2 \epsilon ^2+864 \sqrt{3} h^2
    \epsilon -3312 \sqrt{3} h^2 \nonumber\\  
    -6912 \sqrt{3}
    h \epsilon ^2-576 \sqrt{3} h \epsilon
    +672 \sqrt{3} h-5184 \sqrt{3} \epsilon
    ^4-864 \sqrt{3} \epsilon ^3 \nonumber\\  
    +1008 \sqrt{3}
    \epsilon ^2+84 \sqrt{3} \epsilon -48
    \sqrt{3}
  \end{eqnarray}
  \begin{eqnarray}
    \lo
    m_{z\xi} = 15552 \sqrt{3} h^4-18144 \sqrt{3} h^3-31104
    \sqrt{3} h^2 \epsilon ^2+7560 \sqrt{3}
    h^2+18144 \sqrt{3} h \epsilon ^2
    \nonumber\\     -1332
    \sqrt{3} h+15552 \sqrt{3} \epsilon
    ^4-2376 \sqrt{3} \epsilon ^2+84 \sqrt{3}
  \end{eqnarray}
  \begin{eqnarray}
    \lo
    m_{z w} = 8 \sqrt{2} h+20 \sqrt{2},
    \label{hp1-1_3-mzw}
  \end{eqnarray}
  \endnumparts
  \numparts
  \begin{eqnarray}
    \lo
    a_w = -746496 \sqrt{2} h^6+1492992 \sqrt{2} h^5
    \epsilon +1617408 \sqrt{2} h^5+746496
    \sqrt{2} h^4 \epsilon ^2 \nonumber\\  
    -2612736 \sqrt{2}
    h^4 \epsilon -1430784 \sqrt{2}
    h^4-2985984 \sqrt{2} h^3 \epsilon
    ^3-1244160 \sqrt{2} h^3 \epsilon
    ^2 \nonumber\\  
    +1783296 \sqrt{2} h^3 \epsilon +660096
    \sqrt{2} h^3+746496 \sqrt{2} h^2 \epsilon
    ^4+3234816 \sqrt{2} h^2 \epsilon
    ^3 \nonumber\\  
    +746496 \sqrt{2} h^2 \epsilon ^2-592704
    \sqrt{2} h^2 \epsilon -167184 \sqrt{2}
    h^2+1492992 \sqrt{2} h \epsilon ^5
    \nonumber\\  
    -373248
    \sqrt{2} h \epsilon ^4-1119744 \sqrt{2} h
    \epsilon ^3-190080 \sqrt{2} h \epsilon
    ^2+96048 \sqrt{2} h \epsilon  \nonumber\\  
    +22008
    \sqrt{2} h -746496 \sqrt{2} \epsilon
    ^6-622080 \sqrt{2} \epsilon ^5+20736
    \sqrt{2} \epsilon ^4+122688 \sqrt{2}
    \epsilon ^3  \nonumber\\  
    +17136 \sqrt{2} \epsilon
    ^2-6108 \sqrt{2} \epsilon -1176 \sqrt{2}
    ,
  \end{eqnarray}
  \begin{eqnarray}
    \lo
    b_w = 2239488 \sqrt{2} h^6-4478976 \sqrt{2} h^5
    \epsilon -4478976 \sqrt{2} h^5-2239488
    \sqrt{2} h^4 \epsilon ^2  \nonumber\\  
    +7464960 \sqrt{2}
    h^4 \epsilon +3639168 \sqrt{2}
    h^4+8957952 \sqrt{2} h^3 \epsilon
    ^3+2985984 \sqrt{2} h^3 \epsilon
    ^2  \nonumber\\  
    -4852224 \sqrt{2} h^3 \epsilon -1534464
    \sqrt{2} h^3-2239488 \sqrt{2} h^2
    \epsilon ^4-8957952 \sqrt{2} h^2 \epsilon
    ^3  \nonumber\\  
    -1430784 \sqrt{2} h^2 \epsilon
    ^2+1534464 \sqrt{2} h^2 \epsilon +353592
    \sqrt{2} h^2-4478976 \sqrt{2} h \epsilon
    ^5  \nonumber\\  
    +1492992 \sqrt{2} h \epsilon ^4+2861568
    \sqrt{2} h \epsilon ^3+290304 \sqrt{2} h
    \epsilon ^2-235872 \sqrt{2} h \epsilon
    \nonumber\\  
    -42192 \sqrt{2} h+2239488 \sqrt{2}
    \epsilon ^6+1492992 \sqrt{2} \epsilon
    ^5-217728 \sqrt{2} \epsilon ^4-290304
    \sqrt{2} \epsilon ^3  \nonumber\\  
    -20520 \sqrt{2}
    \epsilon ^2+14112 \sqrt{2} \epsilon +2037
    \sqrt{2}
    ,
  \end{eqnarray}
  \begin{eqnarray}
    \lo
    c_w = -6718464 \sqrt{2} h^6+13436928 \sqrt{2} h^5
    \epsilon +12317184 \sqrt{2} h^5+6718464
    \sqrt{2} h^4 \epsilon ^2  \nonumber\\  
    -21275136
    \sqrt{2} h^4 \epsilon -9144576 \sqrt{2}
    h^4-26873856 \sqrt{2} h^3 \epsilon
    ^3-6718464 \sqrt{2} h^3 \epsilon
    ^2  \nonumber\\  
    +13063680 \sqrt{2} h^3 \epsilon
    +3514752 \sqrt{2} h^3+6718464 \sqrt{2}
    h^2 \epsilon ^4+24634368 \sqrt{2} h^2
    \epsilon ^3  \nonumber\\  
    +2239488 \sqrt{2} h^2 \epsilon
    ^2-3872448 \sqrt{2} h^2 \epsilon -737424
    \sqrt{2} h^2+13436928 \sqrt{2} h \epsilon
    ^5  \nonumber\\  
    -5598720 \sqrt{2} h \epsilon ^4-7091712
    \sqrt{2} h \epsilon ^3-279936 \sqrt{2} h
    \epsilon ^2+553392 \sqrt{2} h \epsilon
    \nonumber\\  
    +80136 \sqrt{2} h-6718464 \sqrt{2}
    \epsilon ^6-3359232 \sqrt{2} \epsilon
    ^5+933120 \sqrt{2} \epsilon ^4
    \nonumber\\
    +637632
    \sqrt{2} \epsilon ^3+9072 \sqrt{2}
    \epsilon ^2-30564 \sqrt{2} \epsilon -3528
    \sqrt{2}
    .
  \end{eqnarray}
  \endnumparts

\end{appendix}

\section*{References}


\begin{thebibliography}{10}

\bibitem{Katok:2005p890}
Anatole Katok.
\newblock Billiard table as a playground for a mathematician.
\newblock {\em Cambridge University Press}, 321:216--242, Jan 2005.

\bibitem{Szasz:1993p996}
Domokos Sz\'asz.
\newblock Ergodicity of classical billiard balls.
\newblock {\em Physica A}, 194:86, Feb 1993.

\bibitem{Sinai:1970p970}
Yakov~G Sinai.
\newblock Dynamical systems with elastic reflections. ergodic properties of
  dispersing billiards.
\newblock {\em Russ. Math. Surv.}, 25(2):137, Oct 1970.

\bibitem{Bunimovich:1981p479}
Leonid~A Bunimovich and Yakov~G Sinai.
\newblock Statistical properties of lorentz gas with periodic configuration of
  scatterers.
\newblock {\em Commun. Math. Phys.}, 78(4):479--497, 1981.

\bibitem{Gaspard:2003p298}
Pierre Gaspard, Gregoire Nicolis, and J.~Robert Dorfman.
\newblock Diffusive lorentz gases and multibaker maps are compatible with
  irreversible thermodynamics.
\newblock {\em Physica A}, 323:294--322, 2003.

\bibitem{Bunimovich:1991p47}
Leonid~A Bunimovich, Yakov~G Sinai, and Nikolai Chernov.
\newblock Statistical properties of two-dimensional hyperbolic billiards.
\newblock {\em Russ. Math. Surv.}, 46(4):47--106, 1991.

\bibitem{Sinai:1991p978}
Yakov~G Sinai.
\newblock Hyperbolic billiards.
\newblock {\em Proceedings of the International Congress of 
  Mathematicians, Kyoto, Japan, August 21-29, 1990, ed. Ichiro Satake, The
  Mathematical Society of Japan}, page 242, Aug 1991.

\bibitem{Simanyi:1999p796}
N\'andor Sim\'anyi and Domokos Sz\'asz.
\newblock {\em Hard ball systems are completely hyperbolic}.
\newblock {\em Ann Math}, 149(1):35--96, 1999.

\bibitem{Szasz:2000book}
Domokos Sz\'asz (ed.).
\newblock {\em Hard Ball Systems and the Lorentz Gas}.
\newblock {\em Encyclopaedia of Mathematics (Springer)}, 101, 2000.

\bibitem{Szasz:2008p35}
Domokos Sz\'asz.
\newblock Some challenges in the theory of (semi)-dispersing billiards.
\newblock {\em Nonlinearity}, 21(10):T187--T193, Aug 2008.

\bibitem{Bunimovich:1974p930}
Leonid~A Bunimovich.
\newblock On ergodic properties of certain billiards.
\newblock {\em Functional Analysis and Its Applications}, 8(3):254, Dec 1974.

\bibitem{Bunimovich:1979p105}
Leonid~A Bunimovich.
\newblock On the ergodic properties of nowhere dispersing billiards.
\newblock {\em Commun. Math. Phys.}, 65(3):295--312, 1979.

\bibitem{Chernov:2006p683}
Nikolai Chernov and Roberto Markarian.
\newblock Chaotic billiards (mathematical surveys and monographs, 127).
\newblock {\em American Mathematical Society, Providence, RI}, 2006.

\bibitem{Stockmann:1999p979}
Hans-J{\"u}rgen St{\"o}ckmann.
\newblock Quantum chaos: An introduction.
\newblock {\em Cambridge University Press}, Aug 1999.

\bibitem{Draeger:1997p917}
Carsten Draeger and Mathias Fink.
\newblock One-channel time reversal of elastic waves in a chaotic 2d-silicon
  cavity.
\newblock {\em Phys Rev Lett}, 79:407, Jun 1997.

\bibitem{Friedman:2001p900}
Nir Friedman, Ariel Kaplan, Dina Carasso, and Nir Davidson.
\newblock Observation of chaotic and regular dynamics in atom-optics billiards.
\newblock {\em Phys Rev Lett}, 86:1518, Jan 2001.

\bibitem{Harayama:2003p901}
Takahisa Harayama, Peter Davis, and Kensuke~S Ikeda.
\newblock Stable oscillations of a spatially chaotic wave function in a
  microstadium laser.
\newblock {\em Phys Rev Lett}, 90:063901, Jan 2003.

\bibitem{Marcus:1992p977}
C.~M Marcus, A~J Rimberg, R.~M Westervelt, P.~F Hopkins, and A.~C Gossard.
\newblock Conductance fluctuations and chaotic scattering in ballistic
  microstructures.
\newblock {\em Phys Rev Lett}, 69:506, Jun 1992.

\bibitem{BunManyDimStadium}
Leonid~A Bunimovich.
\newblock Many-dimensional nowhere dispersing billiards with chaotic behaviour.
\newblock {\em Physica D}, 33(1--3):58--64, 1988.
\newblock Progress in chaotic dynamics.

\bibitem{Wojtkowski:1990p931}
Maciej~P Wojtkowski.
\newblock Linearly stable orbits in 3 dimensional billiards.
\newblock {\em Commun. Math. Phys.}, 129:319--327, Dec 1990.

\bibitem{Bunimovich:1997p729}
Leonid~A Bunimovich and Jan Rehacek.
\newblock Nowhere dispersing 3d billiards with non-vanishing lyapunov
  exponents.
\newblock {\em Commun. Math. Phys.}, 189(3):729--757, 1997.

\bibitem{Bunimovich:1998p421}
Leonid~A Bunimovich and Jan Rehacek.
\newblock On the ergodicity of many-dimensional focusing billiards.
\newblock {\em Ann. Inst. H. Poincar\'e}, 68(4):421--448, 1998.
\newblock Classical and quantum chaos.

\bibitem{Bunimovich:1998p164}
Leonid~A Bunimovich and Jan Rehacek.
\newblock How high-dimensional stadia look like.
\newblock {\em Commun. Math. Phys.}, 197(2):277--301, 1998.

\bibitem{Bunimovich:2006p213}
Leonid~A Bunimovich and Gianluigi~Del Magno.
\newblock Semi-focusing billiards: hyperbolicity.
\newblock {\em Commun. Math. Phys.}, 262(1):17--32, 2006.

\bibitem{Bunimovich:2008p1377}
Leonid~A Bunimovich and Gianluigi~Del Magno.
\newblock Semi-focusing billiards: Ergodicity.
\newblock {\em Erg. Theory Dyn. Sys.}, 28(05):1377--1417, 2008.

\bibitem{Papenbrock:2000p293}
Thomas Papenbrock.
\newblock Numerical study of a three-dimensional generalized stadium billiard.
\newblock {\em Phys Rev E}, 61:4626, Apr 2000.

\bibitem{Wojtkowski:2007p195}
Maciej~P Wojtkowski.
\newblock Design of hyperbolic billiards.
\newblock {\em Commun. Math. Phys.}, 273(2):283--304, 2007.

\bibitem{Bunimovich:1996p302}
Leonid~A Bunimovich, Giulio Casati, and Italo Guarneri.
\newblock Chaotic focusing billiards in higher dimensions.
\newblock {\em Phys Rev Lett}, 77:2941, Sep 1996.

\bibitem{Tanner:1997p950}
Gregor Tanner.
\newblock How chaotic is the stadium billiard? A semiclassical analysis.
\newblock {\em J Phys A Math Gen}, 30:2863, 1997.

\bibitem{Papenbrock:2000p765}
Thomas Papenbrock.
\newblock Collective and chaotic motion in self-bound many-body systems.
\newblock {\em Phys Rev C}, 61:34602, Mar 2000.

\bibitem{Papenbrock:2000p166}
Thomas Papenbrock.
\newblock Lyapunov exponents and kolmogorov-sinai entropy for a
  high-dimensional convex billiard.
\newblock {\em Phys Rev E}, 61:1337, Feb 2000.

\bibitem{Gilbert:2008p354}
Thomas Gilbert and Rapha{\"e}l Lefevere.
\newblock Heat conductivity from molecular chaos hypothesis in locally confined
  billiard systems.
\newblock {\em Phys Rev Lett}, 101:200601, Nov 2008.

\bibitem{Papenbrock:2000p196}
Thomas Papenbrock and Toma{\v z} Prosen.
\newblock Quantization of a billiard model for interacting particles.
\newblock {\em Phys Rev Lett}, 84:262, Jan 2000.

\bibitem{Dellago:1996p91}
Christoph Dellago, Harald~A Posch, and William~G Hoover.
\newblock Lyapunov instability in a system of hard disks in equilibrium and
  nonequilibrium steady states.
\newblock {\em Phys Rev E}, 53:1485, Feb 1996.

\end{thebibliography}

\end{document}